# Internal Length Gradient (ILG) Material Mechanics Across Scales & Disciplines

E.C. Aifantis

Aristotle University of Thessaloniki, Thessaloniki 54124, Greece
Michigan Technological University, Houghton, MI 49931, USA
ITMO University, St. Petersburg 197101, Russia

## Abstract

A combined theoretical/numerical/experimental program is outlined for extending the ILG approach to consider time lags, stochasticity and multiphysics couplings. Through this extension it is possible to discuss the interplay between deformation internal lengths (ILs) and ILs induced by thermal, diffusion or electric field gradients. Size-dependent multiphysics stability diagrams are obtained, and size-dependent serrated stress-strain curves are interpreted through combined gradient-stochastic models. When differential equations are not available for describing material behavior, a Tsallis non-extensive thermodynamic formulation is employed to characterize statistical properties. A novel multiscale coarse graining technique, the equation free method (EFM), is suggested for bridging length scales, and the same is done for determining ILs through novel laboratory tests by employing specimens with fabricated gradient micro/nano structures. The extension of ILG to consider fractional derivatives and fractal media is explored. Three apparently different emerging research areas of current scientific/technological/biomedical interest are discussed: (i) Plastic instabilities and size effects in nanocrystalline (NC)/ultrafine grain (UFG) and bulk metallic glass (BMG) materials; (ii) Chemomechanical damage, electromechanical degradation, and photomechanical aging in energetic materials; (iii) Brain tissue and neural cell modeling. Finally, a number of benchmark problems are considered in more detail. They include gradient chemoelasticity for Li-ion battery electrodes; gradient piezoelectric and flexoelectric materials; elimination of singularities from crack tips; derivation of size-dependent stability diagrams for shear banding in BMGs; modeling of serrated size-dependent stress-strain curves in micro/nanopillars; description of serrations and multifractal patterns through Tsallis q-statistics; and an extension of gradient elasticity/plasticity models to include fractional derivatives and fractal media.

# Table of Contents

# 1. Introduction

## *1.1 Background and Motivation*

The terms "gradient plasticity" and "material instabilities" were used in the mid 1980's [1] to denote load-induced spatiotemporal instabilities such as stress drops or strain bursts and shear bands, as well as structural defect clusters and dislocation patterns. In particular, the term "dislocation patterning" was introduced to denote the organization of dislocations in subgrain cell walls or persistent slip bands during monotonic or cyclic deformation. The author and his co-workers resorted to higher-order gradients (in the form of an additional Laplacian term multiplied by an internal length) of the key constitutive variables for modeling the evolution of deformation and fracture when homogeneous material states become unstable and the corresponding governing differential equations lead to pathological or unphysical behavior. Initial non-linear physics models were proposed, by viewing a deforming medium as a "far from thermodynamic equilibrium" driven system. Various theories of gradient plasticity and gradient damage were generated in the sequel to deal with shear band or damage zone thickness/spacings and mesh-size independence of finite element calculations in the material softening regime, as well as for interpreting size effects [2]. Soon afterwards (in the beginnings of 1990's), and partly motivated by the response of the material mechanics and physics communities, the author and his co-workers incorporated the Laplacian of strain into the standard constitutive equation of linear elasticity, and showed that this simple modification of Hooke's law leads to the elimination of singularities from dislocation (also disclination) lines and crack tips. An early review of such robust "gradient elasticity" theories can be found, for example, in [3] where related references are listed. The observation that these "strain gradient" elastic models are also capable for an effective interpretation of elastic size effects (noted, for example, in elastically deformed micro/nano beams, MEMS/NEMS devices and nanoindentation experiments in small depths), has led to a revival of nonlocal elasticity theories. Under suitable assumptions for the nonlocal Kernel, these theories reduce to "stress gradient" elastic models which are extensively being used in the current literature dealing with deformation and size effects observed (or simulated by molecular dynamics) at the nanoscale (nanotubes, nanobeams, nanoplates). To indicate the impact that these rather phenomenological gradient models had, reference is made to appropriate chapters in a number of recent books [4] and sections of dedicated reviews by leading authors in the field [5], as well as the articles quoted therein. More recently, the author and his co-workers suggested an extension of various deterministic ILG models for elasticity, plasticity, dislocation dynamics and diffusion processes by incorporating stochastic terms in the constitutive or governing differential equations. This conveniently accounts for the competition between deterministic gradients and stochastic effects due to randomly evolving micro/nano structures and corresponding internal stress fluctuations. Initial results are reported in [6], where a related discussion on advanced statistical and image analysis techniques is provided to describe power-law behavior and fractal characteristics when equations are not available.

Subsequently, or in parallel to the above developments other types of gradient models have been advanced by various leading researchers, such as the Fleck-Hutchinson and the Gao-Nix-Huang strain gradient theories, along with their crystal plasticity counterparts, as well as improved gradient theories taking into account surface effects (Gudmundson, K. Aifantis/Willis, Polizzotto, Voyiadjis et al). In addition, as an alternative to the initial Walgraef-Aifantis phenomenological model for dislocation patterning [6], a substantial effort has been initiated by prominent authors based on discrete dislocation dynamics (DDD) modeling (Kubin/Ghoniem/Bulatov/Zbib/Van der Giessen and coworkers). Due to computational limitations for obtaining dislocation patterns, alternative dislocation density based methods have also been pursued (Groma/El Azab/Zaiser/Hochrainer and coworkers) based on statistical mechanics considerations for line defects and rigorous extensions of the previously developed continuum dislocation theory of the Nye-Kroner-Bilby type. A comprehensive review of these very important developments is beyond the scope of the present article, but related references can be found in the bibliography listed in [1-6]. The ILG approach advocated herein, may be viewed as a compromise between the aforementioned dislocation density based continuum theories resting heavily on kinematics and the early much simpler dislocation kinetics and metal plasticity models of Taylor-Orowan-Gilman-Argon-Kocks type, which have been used successfully to model strain hardening. The proposed enhancement of such deformation and/or multi-defect kinetics models with gradient and stochastic terms enables to conveniently consider a variety of important problems of interdisciplinary science and technology by utilizing recently established powerful techniques and methods from nonlinear physics and non-equilibrium statistical thermodynamics to address pattern-forming instabilities and size effects in the presence of multiscale and multiphysics couplings. Continuous contact with both DDD simulations and dislocation density-based theories is necessary for tuning model parameters according to underlying physics, as well as with experiments for validating model predictions. In this connection, it should be pointed out that our gradient elasticity model has recently been successfully utilized by Ghoniem's group in UCLA to dispense with stress singularities of interacting dislocations which disable the computer codes in his 3D discrete dislocation dynamics simulations [7]. Moreover, our non-singular strain/stress crack tip solutions have been successfully used by Isaksson's group in Uppsala to interpret experimental measurements on crack-tip profiles in micro-heterogeneous materials such as solid foams and bone tissues [8].

### *1.2 Key Concepts and Techniques*

The above discussion points to the need for developing a robust interdisciplinary platform for considering deformation and fracture instabilities accounting for multiscale and multiphysics couplings. A promising possibility in this direction is to build on our previous work based on internal length gradient (ILG) material mechanics, by further elaborating on an effective incorporation of surface/interface energy terms to account for the interplay of *intrinsic* vs. extrinsic size effects; time-delay or time-lag terms to account for *incubation* effects of underlying micro/nanostructures; and stochastic terms to account for randomly evolving micro/nanostructures and internal stress/strain fluctuations. The resulting purely mechanical framework can then be further extended to include multiphysics processes (e.g. heat, mass, and electric charge transfer), as such thermomechanical, chemomechanical, and electromechanical couplings govern material stability and aging in a wide range of emerging technological and biomedical applications. To this end, we can use the standard approach of balance laws and constitutive equations of generalized continuum mechanics, enhanced by recently developed techniques in nonlinear physics, non-extensive statistical thermodynamics, and applied mathematics. In particular, methods for analyzing pattern-forming instabilities and self-organization phenomena can be utilized, as we have done in the past for dislocation patterning and shear banding problems. Such methods are particularly useful in conjunction with related atomistic (molecular dynamics/MD) and discrete element (cellular automata/CA, discrete dislocation dynamics/DDD) computer simulations, as well as with corresponding laboratory measurements that only recently have become possible, due to the development of new powerful numerical codes and high-resolution experimental probes. In particular, the so-called equation free method (EFM) and advanced multiscale finite element (FE) codes can be employed for coarse-graining procedures. In addition, electron microscopy (TEM/SEM) and nanoindentation/atomic force microscopy (NI/AFM) observations along with beam-based methods (synchrotron X-ray CT equipped with a micro-tensile stage and climate chamber) can be used to obtain statistical measurements on strength and surface patterning, as well as for validating non-singular solutions at crack tips in micro-heterogeneous materials.

When usual differential equations (with derivatives of integer order) are not available to describe the observed phenomena, we may resort to image processing and related algorithms for describing statistical properties of the processes involved. Our current research in this area focuses on analysing serrated stress-strain graphs for NC/UFG/BMG materials. It is shown that standard power law behavior based on Boltzmann-Gibbs-Shannon (B-G-S) entropy thermodynamics is not sufficient. Instead, Tsallis q-entropy statistics based on non-extensive thermodynamics should be used for interpreting the observed behavior, i.e. fitting the corresponding probability density functions (PDFs) and interpreting related fractal properties.

In concluding this subsection on new concepts and techniques, it is pointed out that another possibility based on fractional calculus (derivatives of non-integer order) and fractal media considerations is explored, for dealing with related intriguing peculiarities exhibited by the micro/nano deformation phenomena considered. This is motivated by very recent work on our fractional gradient elasticity and fractal elastic media [9] which is discussed here, along with a brief account on its extension to fractional gradient plasticity and fractal plastic media. Another new mathematical technique that can be employed is the so-called homotopy method [10] for obtaining analytical solutions of non-linear differential equations that are not possible to obtain with other (e.g. perturbation) methods. Some preliminary results on shear band profiles are in progress along these lines but reported elsewhere [10b].

## 1.3 *Relevance to Emerging Science/Technology/Biomedicine Research Areas*

We outline below three broad areas of emerging science, technology and biomedicine sectors that can benefit from the above described ILG theoretical/computational/experimental framework. One area is concerned with structural materials and pattern-forming instabilities, such as strain localization zones and multiple shear band networks, occurring in nanocrystalline/ultrafine grain (NC/UFG) materials and bulk metallic glasses (BMGs). A second area is concerned with energy materials and chemomechanical damage in next generation nanostructured battery electrodes, as well as electromechanical size effects in piezoelectric/flexoelectric materials, and photomechanical aging in semiconductor devices and light emitting diodes. The third area is concerned with modeling aspects of brain tissue and signal transmission through neural cells. A number of specific problems can be identified in this area where tissue growth and related physiological cell functioning induces internal strains which induce diffusion and electric field couplings leading to disease and its prevention or therapy.

### 1.3A *Structural NC/UFG/BMG and Micro/Nano-Heterogeneous Materials*

*(i) Shear Instabilities and Size Effects:* Adiabatic shear bands (ASBs), Portevin-Le Châtelier (PLC) bands and persistent slip bands (PSBs) in NC/UFG/BMG materials and micropillars exhibit substantial differences from their conventional or macroscopic counterparts, due to the small volumes available for the relevant plastic flow processes to evolve. It turns out that the proposed ILG approach allows for the description of micro/nanoshear bands and micro/nanonecks, as well as the nucleation and evolution of other types of micro/nanodeformation patterns. Size effects (both *extrinsic* related to specimen dimensions and *intrinsic* related to the substructure size) can also be interpreted in a robust manner. Size effects are especially pronounced at the micron and nano scales, where material heterogeneity and local gradients cannot be "averaged" or "smoothed out", and the surface-to-volume ratio increases. Typical experiments in the last fifteen years involved microtorsion and microbending tests, along with nanoindentation. More recent experiments have also shown size effects in tension, as well as in micro/nanopillar compression which, however, have not been attributed to strain gradients. Nevertheless, most recent work by the author and his coworkers [11] has shown that submicroscopic strain gradients (due to local heterogeneities and also observed experimentally) may be used to interpret this type of size effects, also consistently with atomistic simulations. This approach is in-line with our earlier treatment of size effects in tension and creep for macroscopically homogeneous specimens and deformation modes [12]. More work is needed, however, to elucidate this question and relate the interplay among substructure characteristics (grain size), specimen geometry (diameter, length), and surface conditions. In this connection, the combined effect of grain size (normal or inverse Hall-Petch behavior), temperature, and strain rate on flow curves has been investigated and the influence of specimen geometry and mode of loading has been examined [6a,13].

Existing ILG models for the above problems can be revisited by including time-lags, surface effects and stochastic terms aiming at capturing both stress-strain serrations and strain/defect density patterning characteristics. Our initial results show that when our earlier gradient models are revised to account for internal stress fluctuations, load drops or displacement bursts can be obtained and statistically interpreted through Tsallis non-extensive thermodynamics and corresponding q-probability density function (q-PDF) distributions [6e]. They also show that below a critical domain size, pattern-forming instabilities and/or stress/strain serrations may be suppressed [6g]. In general, this depends on the interplay between the time-lag parameter and the ratio of internal length vs. specimen size, as well as the stochastic contribution to the flow stress assumed. Related background on which such future developments can rest upon, can be found in [6].

***(ii) Multiphysics Couplings:*** The gradient modification of our previous purely mechanical models was based on the introduction of the Laplacian of strain and/or stress multiplied by an internal length. For elasticity, the extra gradient term is of the form $\ell_\varepsilon^2 \nabla^2 \left[ \lambda(\text{tr}\boldsymbol{\varepsilon})\mathbf{1} + 2G\boldsymbol{\varepsilon} \right]$, where $\ell_\varepsilon$ is an elastic internal length and the quantity in the brackets is the classical Hookean stress. For plasticity, the extra gradient term is of the form $\ell_p^2 \nabla^2 \gamma^p$, where $\ell_p$ is a plastic internal length and $\gamma^p$ is the equivalent plastic strain (second invariant of the plastic strain tensor). The inclusion of the Laplacian is not arbitrary. The simplest possible argument for its appearance is based on expressing an average field quantity over an elementary material volume as a nonlocal volume integral and then expanding in a Taylor series keeping terms up to the second order (Maxwell's physical interpretation of the Laplacian), taking also into account microstructure statistical correlation effects. The procedure also holds for the fractional Laplacian that enters into corresponding considerations for fractional elastic and plastic media. For multiphysics processes involving diffusion and/or heat transfer, it turns out that the above gradient formalism should be supplemented by an analogous generalization to include the Laplacian of the diffusion flux $\ell_{\mathbf{j}}^2 \nabla^2 \mathbf{j}$ and/or the heat flux $\ell_{\mathbf{q}}^2 \nabla^2 \mathbf{q}$, where $\left(\ell_{\mathbf{j}}, \ell_{\mathbf{q}}\right)$ denote respectively diffusional and thermal internal lengths [14]. Similar arguments may be used for introducing the Laplacian of the electric field $\ell_{\mathbf{E}}^2 \nabla^2 \mathbf{E}$ or the polarization rector $\ell_{\mathbf{P}}^2 \nabla^2 \mathbf{P}$. Details on this topic will be given later in Section 3.2.2, as well as at the end of the Chapter, in the "Concluding Remarks" Section. The implications of such generalizations could be quite significant to material fabrication and component design as they provide new criteria for the emergence of instabilities and the prediction of size effects. For example, in the case of adiabatic shear banding [15], it is shown that by manipulating the interaction between thermal and deformation-induced internal lengths can delay shear band formation for small specimens. Similarly, the interplay between diffusional and strain gradient internal lengths may suppress phase separation below a critical specimen size, leading to size-dependent spinodal gaps [6g]. Such type of multiphysics aspects with specific examples will be discussed in later sections. One specific area that has been entirely unexplored so far, is concerned with the effect of higher-order gradient couplings in photomechanics. In addition to possible re-interpretations of standard photoelasticity measurements, it is expected that such effects will be essential in describing light-matter interactions; in particular, the direct effect light has on inducing electronic strain, as well as the related effect light has on structural defects and the mobility of dislocations (photoplasticity, photodamage). Brief comments on this topic will be provided later in Section 3.2.3, as well as at the end of the Chapter in the "Concluding Remarks" Section.

*(iii) Non-singular Dislocation and Crack Solutions*: As already indicated, our earlier one-parameter simple gradient elasticity (GradEla) model has recently been used in three dimensional DD simulations [7], as well as in providing non-singular expressions for the crack tip "microstress" and the "incompatible" crack tip "microstrain". It is noted, in this connection, that some questions have risen recently by some authors in the literature [16], mainly due to the fact that these authors improperly identified "micro" with "macro" stress/strain fields. Related clarifications are included in [17], as well as in the present article where alternative forms of non-singular crack solutions are provided based on the concept of continuously distributed dislocations. Recent similar work on non-singular crack-tip solutions for heterogeneous materials (such as cellular solid foams or porous bone tissue) has also been conducted by Isaksson and co-workers [8]. These materials possess a large internal length as compared to metals and, therefore, experimental validation is more feasible. This was pursued in Isaksson's

Lab with the newest generation of a table-top X-ray CT scanner with submicron resolution. An initial comparison between theory and experiment was quite encouraging. It is proposed to further elaborate on this issue for such type of heterogeneous cellular materials (solid foams, paper, wood, bone) and pave the way for possible future analyses on metals in much more demanding synchrotron experiments requiring much higher beam intensity.

### *1.3B High-Energy Density Storage and Optoelectronic Materials*

The second case study area is concerned with next generation Li-ion and Na-ion battery (LiB and NaB) micro/nanostructured electrodes, micro/nano-electro-mechanical systems (MEMS/NEMS) and interconnect components, as well as optoelectronic material for light emitted diodes (LEDs).

*(i) LiB and NaB Electrodes:* Chemomechanical damage/cracking and capacity fade in next-generation battery materials, such as Li-ion and Na-ion battery electrodes, has extensively been considered for Li-ion nanostructured anodes, and an excessive number of papers on continuum modeling and simulation studies have been published in recent years (see, for example, [18] and references therein). In particular, chemomechanical degradation and fracture of nanocomposite anodes (Si or Sn active nanoparticles embedded in a carbon matrix) has been a problem of continuous interest. Upon Li insertion/de-insertion in the active sites, *colossal volume expansion* takes place (up to 300%) which cannot be accommodated by the surrounding matrix material during the electrochemical cycling, leading to damage/cracking. Understanding micro/nanocrack nucleation and pattern formation will help develop criteria for optimizing anode performance. Building on this experience, alternatives to Li-ion battery (LiBs) systems, such as Na-ion batteries (NaBs) can be considered by employing similar micro/nanomechanical models. In contrast to severe cracking observed in Li-ion anodes, it has been observed that Sn anodes in Na-cells do not experience fracture during electrochemical cycling despite the up to ~400% expansion/contraction during Na-ion insertion/deinsertion. Nevertheless, a significant capacity fade in pure Sn anodes has been observed despite of the lack of fracture. Preliminary SEM and TEM micrographs indicate "nanopore" formation which may be viewed as a possible mechanism to suppress fracture but not inhibit capacity fade. As in the case of coupled deformation-diffusion studies for Li-ion anodes, this problem can effectively be studied within the proposed multiphysics/multiscale thermo-chemo-mechanical ILG framework for advancing our understanding on nanopore-induced capacity fade in Na-Sn anodes.

*(ii) NEMS/MEMS and Interconnects Components:* A second topic in this area concerns the response and functionality of materials and structures used for energy-related applications in the presence of electromechanical couplings; in particular, size effects in *piezoelectricity/flexoelectricity* and *electromigration. Flexoelectricity* refers to strain gradient-induced electrical polarization, as opposed to *piezoelectricity* which refers to the coupling between uniform strain and polarization. Both of these electromechanical size effects (in particular, the flexoelectric effect) become increasingly significant as specimen dimensions are reduced down to the nanometer scale with the corresponding flexoelectric parameters increasing sometimes by three orders of magnitude. This can be detected and readily modeled through bending of micro/nano cantilever beams and nanoindentation as high strain gradients develop in these configurations. *Electromigration* refers to the stress gradient-induced mass transport due to the momentum transfer between the electron wind and the diffusing ions during the passage of electric current. This, in conjunction with the background stress field due to thermal expansion mismatch between conducting lines and their surroundings, leads to the development of *hillocks* (in regions of compressive stress) and *voids* (in regions of tensile stress), thus causing failure of the interconnects used to maintain electrical contact between neighboring devices on the chip. The demand for further miniaturization of electronic circuits requires additional fundamental understanding of the role of stress gradients and surface energy on related SEs leading to the determination of optimum line width/grain size ratio for reducing damage and prolonging lifetime.

***(iii) LED and LD Devices:*** Light emitted diodes (LEDs) and laser diodes (LDs) consist of energetic materials where the ILG approach can potentially be applied to model the effect that light has on inducing electronic/internal strain (Staebler-Wronsky effect), as well as the related light effect on structural defects and the mobility of dislocations (photoplastic effect). Gradient and size effects have not been considered in this field and it is pointed out that our unpublished research suggests that the ILG formalism can readily be employed to address degradation/aging in amorphous and crystalline semiconductors (chalcogenide glasses/a-Si:H, anthracene/GaAs) used in optoelectronic and light emitting diode technologies.

### *1.3C Brain Mechanics and Neuroelasticity*

The third case-study area is concerned with brain tissue gradient mechanics modeling and signal transmission in neural cells. This should not be viewed as an exotic distraction from the main focus of this report, as our earlier reaction-diffusion type models for the stress-induced nucleation and transport of strain and carriers of plastic deformation (structural defects) were partly inspired from biological and population dynamics models. The additional insight gained on the effect of strain gradients and internal stress fluctuations from the aforementioned models, can effectively be utilized to address related research problems in biomedicine; in particular, the micro/nano mechanics of brain tissue and neural cells. In this connection, it is noted that an alternative to Turing's interpretation of morphogenesis phenomena in biology is possible through the introduction of higher-order gradients of internal strain generated during the "crawling" of cells in the surrounding extracellular matrix (ECM) of living tissue – as shown in Murray's seminal work, which may here be referred as gradient bioelasticity; and, in relation to neurons, as neuroelasticity. A brief discussion on this topic is given in Section 3.3 and in the last section on "Concluding Remarks", where we point out the remarkable similarities between Murray's approach on internally generated strain gradient effects on living cells and the ILG approach on externally induced strain gradient effects on non-living deforming objects. Accordingly, some existing popular nonlinear signal propagation models can be revisited by incorporating gradient effects and internal stress fluctuations, as it has already been pursued for the case of propagating plastic instabilities in our earlier and also in our more recent studies. It is thus possible to explore the dynamics of neuro-transport within our internal length gradient multiscale/multiphysics mechanics framework. As an example, we may refer to reduced Huxley-Hodgkin type models, such as the Fitzhugh-Nagumo variant (enhanced with strain gradient effects and internal stress fluctuations) and its applicability to study signal propagation leading to neurological synapse formation due to decoherence in electrostatic pulse exchange. In this connection, it is noted that the dendritic structure of an axon (multi-branched tree like web), necessitates a finite time lapse in order for the signal from any node to propagate and fire another synapse. For axons in which this time delay is comparable to or larger than the system's relaxation time, the mechanism is largely controlled by the transport time. Using a stochastic version of the Fitzhugh-Nagumo model (enhanced with internal stress effects and a delay mechanism), we can address the strength of such neuronal impulses in orchestrating neural transport through neurological synapse firing. A similar situation may be envisioned for soliton-like deformation waves travelling through microtubules. A brief introductory discussion on neuro-transmission within the ILG framework is given later in Section 3.3, as well as at the end of the Chapter in the "Concluding Remarks" Section. Work along these lines, currently in progress and scheduled for publication elsewhere, is consistent with current views that biology and technology are both evolving toward more efficient methods of information processing (e.g. [19]: Mind/Tech Merger in the Nanoscale).

## 2. Methodology and Proposed ILG Platform

It follows from the discussion in the previous section that a combined theoretical-numerical-experimental framework, based on internal length and time scales gradient mechanics methodology, is desirable. The internal lengths emerge by averaging local micro/nano structural heterogeneities at the scale of the representative continuum element, while the internal times account for the time delay that these heterogeneities need in order to "mature" and affect the temporal behavior of the system. This can lead to a multiscale/multiphysics platform for addressing various seemingly different problems across space/time scales and disciplines. The currently available internal length gradient approach [6] can be used as a "starting point". However, it is necessary to extend it to include *surface/interface effects*, *internal time scales* and *stochastic heterogeneity*, with guidance from existing related experiments and simulations, as well as direct input from *novel multiscale computations and non-standard laboratory tests*. This would eventually lead to the desired comprehensive framework for transdisciplinary modeling (supported by related multiscale computations and experiments), within the overarching paradigm of continuum theory properly enhanced with features accounting for micro/nano scale phenomena recently observed in various study-areas in advanced technology, renewable energy and biomedical sectors.

Clearly, the above is not a trivial task: it is far beyond a mere "straight-forward" extension of existing gradient modeling which, in some aspects, has already reached a certain level of maturity. On the contrary, we are engaged in a "demanding" effort which, however, has the potential of producing new results and models with immediate "non-incremental" implications to a number of disciplines ranging from the established fields of mechanics and materials science to new emerging areas of nanosciences and nanotechnologies, including "nanoenergy", "nanobiology" and "nanobiomedicine".

### 2.1 Generic Theoretical Modeling and Numerical Issues

A major still unresolved issue of the ILG methodology and similar spinoff approaches is concerned with the development of an effective strategy for identifying (through physically-based micro/nanoscopic arguments) and calibrating (through corresponding multiscale simulations) the pertinent "bulk" and "surface" ILs – also in conjunction with especially designed for this purpose tests to be discussed in the next subsection. This should be done in connection with the underlying dominant deformation mechanisms, their dependence on the local state of stress, and their interaction with internal or external surfaces. It will enable us to model the interaction between bulk and surface/interface ILs [20] and assess the interplay of *extrinsic* (specimen dimensions) *vs. intrinsic* (microstructure dimensions) size effects [21] when small material volumes are considered. A novel (for material mechanics) mathematically inspired computational technique which can be employed here is the so-called *equation-free modeling* (EFM) [22]. This can be utilized to determine the right level of complexity and the degree of coarse graining required for gradient models by identifying low-dimensional representations of the pertinent essential variables through the use of modern data-mining techniques to analyze the high-dimensional data arising from related MD and DD (as well as CA and MC) simulations.

In the following we address the aforementioned outstanding theoretical modeling and numerical simulations issues which, when successfully resolved, can bring the ILG approach to a next level of development. Then, in Section 2.2 we discuss a number of experimental procedures and novel tests required for gaining further understanding and conclusively validating the theory.

*(i) Surface Effects:* The interaction of ILs of the above type with internal or external surfaces can be described [20] by introducing extra interface/surface energy terms $\int_S \Gamma_S dS$ in the material's gradient-dependent total energy functional ($S$ denotes surface and $\Gamma_S$ is a function depending on the deformation state and atomic defect configuration on $S$). In the case of elasticity, the form of $\Gamma_S$ may be deduced in connection with the atomic arrangements at the interface and related adhesion/binding properties of the adjacent materials. In the case of plasticity, the form of $\Gamma_S$ may be deduced from physical arguments and microscopic models concerning the structure of the interface (or free surface) and the interaction with arriving dislocations from the grain interior. In the context of interfaces one might envisage a "gradient multiscale" approach where on an atomistic "fine-scale" the interface is modeled as a continuous object allowing, for instance, to treat its interaction with discrete defects, and then on a "coarser-scale" one treats it as a surface of discontinuity with an associated energy "penalty" term.

*(ii) Internal Time Lags and Stability:* Deformation instabilities and the derivation of corresponding stability criteria depend not only on the introduced spatial gradients, but also on internal time-lags associated with the underlying evolving micro/nanostructures. Thus, in addition to ILs, time-lags or internal times (ITs) to account for the period elapsing between load application and nucleation of the dominant micro/nanostructures before they become "mature" for participating in the deformation process, need to be incorporated. This will enable us to critically examine the interplay between ILs and ITs and "tune" these new constitutive parameters for establishing desired structure-property relations and designing new protocols for optimum material/process performance. The inclusion of such time-lags or time delays and the use of the corresponding tool of delay differential equations (DDE) has not been explored so far in spatio-temporal pattern forming deformation instabilities, even though relevant physical mechanisms have been identified [23]. When a Taylor expansion in the time-delay variable is plausible, then an "internal inertia" term appears in the system. The concept of internal inertia has been used recently by us to account for coupling effects between ILs and ITs for elastic wave propagation in nanotubes with results consistent with related MD simulations [24]. This idea can be extended to plastic flow (Fig. 1a) and electromechanical processes to model front propagation in nanomaterials and neurons, respectively. It should be noted that while, in principle, discrete time-lags may not be needed if an internal variable-based formalism is introduced, the resulting phenomenological description may become quickly complex in view of uncertainties related to the optimum number of internal variables used

*(iii) Simulations:* Non-standard multiscale numerical codes, especially suited for considering IL gradient models, are available and involve molecular dynamics (MD), finite element (FE), discrete element (DE) and hybrid (MD-FE) or (DE-FE) codes as well as cellular automata (CA) and discrete dislocation dynamics/DDD codes. A robust FE code for gradient elasticity can be found in [24] and a hybrid code for bridging atomistic/nanoscopic and mesoscopic/macroscopic deformation within the ILG framework can be found in [25]. These codes can be extended to incorporate "time-lags" and then calibrated against analytical solutions for simple one-dimensional and radial symmetry configurations. They will be useful for validating new model predictions for benchmark problems involving stress concentrators around holes, dislocations, and cracks, as well as deformation patterns and size effects observed in tension/compression, bending, and indentation of micro/nano sized specimens. They will also provide critical (not currently available) feedback to the hypotheses adopted in the model development: in particular, to those pertaining to the microscopic mechanisms associated with specific gradient terms introduced at the continuum scale, in relation to the inhomogeneity measures used; the size of the representative volume element (RVE) used in specific applications, in relation to the internal length employed; and the form of the higher-order variationally consistent boundary conditions, in relation to the atomistic/dislocation configuration near internal or external surfaces. It is noted that such multiscale internal length-internal time (IL-IT) codes may pave the way for optimizing the academic and industrial impact of this research, through the extension of related solution algorithms to multi-field multiphysics formulations.

As already mentioned, a particularly novel computational aspect which has not been explored for solids (even though extensively used for fluids and reaction-diffusion systems), is the *equation-free modeling* (EFM) method [22]. While traditional material modeling starts by first formulating or deriving and then solving macroscopic evolution equations based on microscopic (atomistic, molecular, stochastic) models, EFM circumvents the step of obtaining accurate macroscopic descriptions, through the development and validation of a mathematically inspired computational enabling technology that allows to perform macroscopic tasks acting on the microscopic models directly. In this connection, it is pointed out that the Navier-Stokes equations were successfully used as a coarse-grained continuum description of the laminar flow in Newtonian fluids, before the corresponding approximate derivation from kinetic theory. In the case of complex material systems, the physics of processes occurring at fine micro/nano scales can be modeled on the basis of novel experiments and powerful simulation codes, but a reasonably accurate phenomenological "closure" system of equations is a challenge. Methods like EFM can help –with the aid of the computer and the underlying physical mechanisms – to test the suitability and effectiveness of gradient models as a phenomenological approximation at the level of the continuum element used. In particular, the role of EFM within the ILG framework may be: **(a)** To design computational experiments to test the nature of the proposed closures in the gradient constitutive models; **(b)** To use modern data-mining techniques/diffusion maps for analyzing the (very high-dimensional data arising from discrete) MD and DDD simulations, by identifying low-dimensional representations of the pertinent essential variables. This will determine the right level of complexity and the degree of coarse-graining required for the gradient-dependent models to be developed at a continuum level. In this connection, it should be pointed out that recent multiscale material mechanics models use generalized gradient (micromorphic) type arguments to interpolate between particle-level (DE) and continuum-level (FE) computations. While there is some conceptual resemblance between these methods and EFM, the latter is a quite novel approach resting on rigorous mathematical foundations which remains entirely unexplored for deformation problems. Moreover – being designed for complex multiscale problems where internal variables with diffusive transport may be *describable* by "fine scale" microscopic models, while system evolution is *observable* at a macroscopic "coarse scale" – it is particularly suitable to be implemented within the proposed IL-IT framework.

***(iv) Combined Gradient-Stochastic Models and Tsallis q-Statistics:*** A promising new area of research within the ILG framewok, concerns the development of robust models to consider the competition between *deterministic gradient terms* induced by the applied stress and *stochastic terms* associated with fluctuations of internal stress and randomly evolving micro/nano structures. This can be rigorously explored by adopting existing approaches of statistical physics leading to stochastic differential equations for the description of the material system under consideration However, a different procedure is adopted herein that is based on the introduction of phenomenological probability density functions (PDFs) for the random micro/nano structures, in conjunction with the deterministic gradient terms, leading to the construction of combined gradient-stochastic models [26,27]. This enables us to capture and theoretically interpret experimentally measured deformation features (spatio-temporal deformation avalanches, fractal patterns, power-law exponents, serrated stress-strain curves) that cannot be described by purely deterministic or stochastic models alone. Fractal characteristics and power-law exponents have been reported in recent work for plastic deformation by using statistical mechanics methods based on Boltzmann-Gibbs-Shannon entropy thermodynamics. A novel (for material mechanics/physics) technique to be employed here for analyzing experimental data for plastic deformation that cannot be fitted by conventional power-law relations, is the so-called Tsallis q-entropy statistics [28] inspired by non-extensive thermodynamics. Several examples will be given where Tsallis q-statistics provide a quantitative insight to the complex dynamics of plastic deformation in novel materials. Below we present a few more details on both of these theoretical modelling issues; i.e. the development of robust combined gradient-stochastic models and the interpretation of their corresponding statistical feature predictions through Tsallis q-statistics.

- ***Combined Gradient-Stochastic Models:*** Stochasticity can be incorporated through an enhancement of the aforementioned gradient deterministic models with the incorporation of stochastic terms to account for the heterogeneity of internal stresses and deformation-induced random micro/nanostructures. In the case of plasticity, a convenient way to account for the competition between deterministic gradient and random effects is to introduce (in analogy to Wiener processes in statistical mechanics) an additive stochastic term – of the form $h(\gamma)g(x)$; $\langle g(x)\,g(x')\rangle = l_{corr}\delta(x\text{-}x')$ with $l_{corr}$ denoting a correlation length and $\delta$ being the usual Dirac delta function – into the gradient expression of the flow stress ([26,27]; see also [6a-6e]). It is noted that this is not an arbitrary assumption but emerges generically if one aims at a description above the scale of the discrete substructure which defines the correlation length – i.e. within a continuum model. The delta function then simply emerges because the stress or strength fluctuations within individual volume elements of the continuum theory are effectively uncorrelated. The function $h(\gamma)$ also covers the limiting case where only the material parameters fluctuate while the evolution is deterministic (e.g. in the case of flow stress fluctuations due to fluctuating grain orientation). Recent preliminary work has shown (Fig. 1b) that such models can be effectively implemented using cellular automata (CA). They give access to the "discrete plasticity" events, generally observed in very small samples, and allow to understand the related spatio-temporal statistics of plastic deformation which is characterized by power-law distributions of the corresponding statistical events (defect avalanches, strain bursts, stress drops, and related irregularities in serrated stress-strain curves). Additional work along these lines is reported in later sections. Generalizations of this approach using kinetic Monte Carlo dynamics in conjunction with "gradient" and "stochastic" stresses can provide access to the time dependent deformation of very small systems and, thus, to the important question of durability and long-time performance of micro/nanomechanical components.

- ***Tsallis q-Statistics:*** Standard gradient deterministic models cannot provide any information on measured statistical aspects of plastic deformation. Fractal dimensions for deformation patterns, Hurst exponents for surface roughness determined through wavelet analysis, and power-law exponents for dislocation avalanches and strain bursts recorded during nanoindentation and micro/nanopillar compression tests [30] cannot be determined. But even the aforementioned enhanced gradient-stochastic models and existing Boltzmann-Gibbs-Shannon (B-G-S) entropy statistics are not always able to interpret statistical measurements of plastic deformation events for the whole spectrum of the available experimental data. Preliminary results suggest (Fig. 1c) that use of Tsallis q-statistics based on non-extensive thermodynamics can fit the whole spectrum of the experimental data for the probability density distribution of nanodeformation events (defect avalanches, strain bursts, stress drops). It turns out that near the "upper" and "lower" ends of the respective log-log plots for these events, the data "deviate" systematically from standard power-law exponent interpretations based on B-G-S entropy statistics. But, in fact, these "tails" which cannot be fitted by usual power-laws are the regimes of most interest, as they correspond either to "small magnitude events" with high probability or to "large magnitude events" with small probability. Additional results along these lines for serrated stress-strain curves and multiple shear band characterization are provided in later sections.

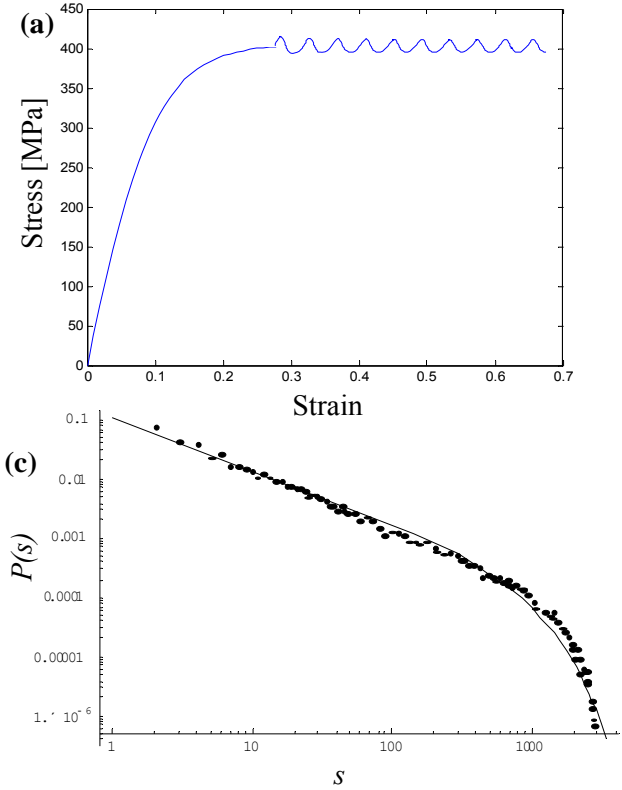

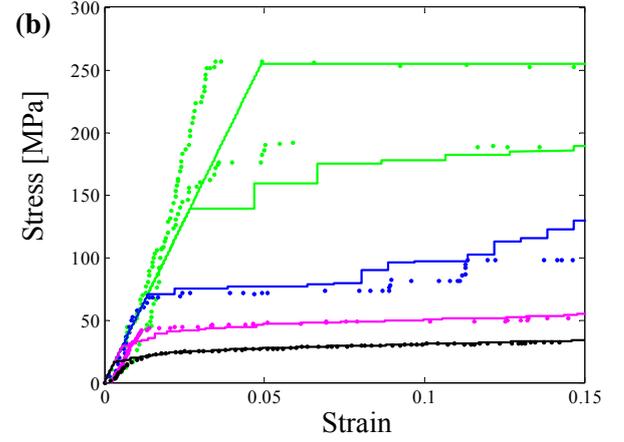

*Figure 1:* **(a)** *Qualitative stress-strain curves exhibiting oscillatory behavior by using a time-delay counterpart of Voce models (unpublished). Additional recent results on Tsallis q-statistics pertaining to plastic instabilities can be found in [6e, 6g].* **(b)** *Micro-pillar compression stress-strain curves (points) [29] modeled with CA simulations (lines) for 1, 2.4, 10.1 and 39.7 μm pillar diameters.* **(c)** *Application of Tsallis q-statistics on the probability density of slip avalanches (Reprinted from [28] with permission from Springer).*

*(v) Extension to Multiphysics:* The ILG approach for purely mechanical problems can readily be extended to consider novel "multiphysics" processes where the deformation field is coupled with temperature, diffusion or electromagnetic fields. Even though various types of thermo-mechanical, chemo-mechanical, and opto-electro-mechanical theories exist, higher-order internal length couplings have not been explored. Specific applications here include, for example, non-Fick diffusion in nanopolycrystals [6f, 31] and non-Fourier heat transfer during ultrashort-pulse laser treatment of surfaces and skin cancer radiation therapy [14], where higher-order mass and heat transport terms are necessary. Higher-order temperature gradients have also been thought as playing a role in adiabatic shear banding [32], an issue that may become even more important at the nanoscale in relation to pattern formation in nanocrystalline and amorphous solids [33]. Chemomechanical damage induced by the interaction of diffusional internal lengths and strain gradient-dependent internal lengths can be readily considered and its effect on capacity fade in rechargeable batteries during electrochemical cycling can be assessed. A discussion is given on extending our ILG framework to include electromechanical and photomechanical couplings, and some representative examples are considered. In particular, electromechanical size effects occurring in piezoelectricity and flexoelectricity phenomena observed in a variety of technological (MEMs/NEMs) and biological (neural cells) materials. Photomechanical instabilities and aging of photoplastic materials and light emitting diodes (LEDs) can be addressed in a similar way within our proposed framework, but this particular discussion is postponed until a future publication.

## *2.2 Generic Experimental Issues and Model Validation*

The novel micro/nano ILG models discussed in the previous section, should be developed in conjunction with available and ongoing measurements, recently being conducted through modern experimental probes. A key point for the successful integration of the proposed joined-up modeling and experimental approaches is to characterize the deformation and stress at the nanoscale (e.g. near dislocation pile-ups and crack tips). To accomplish this it is not sufficient to continue using the existing experimental tools. Rather, the development of new physically-based and reliable ILG models must go hand-to-hand with corresponding ILG experiments especially designed to reveal such local nanoscale features. The current "landscape" and associated techniques used in this area rely primarily on beam-based methods: X-ray diffraction (XRD) and imaging, including coherent diffraction imaging (CDI) and computed tomography (CT), as well as pair distribution function (PDF) analysis that is relevant to both nanocrystalline and amorphous materials. These are supplemented with advanced nanoindentation (NI)/atomic force microscopy (AFM) and transmission electron microscopy (TEM)/scanning electron microscopy (SEM) techniques, as well as focused ion beam (FIB) along with digital image correlation (DIC) analyses.

NI in particular, can be used to determine both elastic (small depths) and plastic (large depths) ILs in conjunction with the use of deterministic gradient elasticity and gradient plasticity models. Information on the random term introduced into the flow stress expression of gradient-stochastic models will also be obtained from multiple NI tests to determine the density probability function for the yield stress with the "mean" providing its average value and the "variance" its random counterpart. Such experiments can be conducted in two overlapping stages: First, an analysis of existing experimental data for model validation and IL calibration from local strain, spatial shear band characteristics (thickness/spacing) and size effect measurements should be conducted. Such measurements cannot be interpreted by classical deformation models. Second, a performance of a number of novel non-standard tests especially designed for direct measurements of ILs, in connection with benchmark model configurations involving, for example, non-singular stress/strain fields, dislocation patterns, and fabricated gradient microstructures, should be carried out.

**(*i*) IL Estimation from Local Non-Singular Strain/Stress Concentrators:** Recent work of the author and co-workers [34] used the non-singular strain field expressions for dislocations to determine experimentally the dislocation core size from high resolution TEM images using geometric phase analysis (GPA). The same was done for the calculation of the mean square strain $<\varepsilon_L{}^2>$ as obtained from the measured X-ray line profile analysis for a Cu nanopolycrystal. These preliminary results were in agreement with related MD simulations for dislocation cores and XRD data and were used to calibrate ILs for gradient elasticity. Further work should include the use of higher resolution for GPA analysis and X-ray profile analysis for other nanocrystalline materials.

Along similar lines, attention should focus on comparing our recently obtained non-singular crack solutions [6a,17,35] with corresponding measurements of strain [35b,c] near crack tips. Close to a crack-tip, the (elastic) non-singular strain field is controlled by two parameters: the usual LEFM stress intensity factor *K* and the IL parameter which eliminates the classical $1/\sqrt{r}$ stress singularity. The IL is strongly connected to the material's intrinsic length, typically ranging from nanometer (metals) to millimeter (paper). Such IL-determination experiments for fracture in cellulose nanofibril networks (for which the demands for the strain field resolution can be met by the new generation of table-top μCT scanners) are currently conducted by *Isaksson* and some of these preliminary results are shown in Fig. 2. These experiments can be used to validate (or otherwise) our non-singular IL-dependent solutions for crack tips obtained through gradient elasticity. Existing results for stress concentrators in holes obtained with the use of FIB assisted by DIC techniques will also be used to check our analytical gradient elasticity solutions, in relation to size-dependent estimates of elastic constants. Related experimental data are available from recent work at *Michel's* Micro-Nano Reliability Center in Berlin, by using the so-called fibDAC/microscopic hole method for residual stress determination.

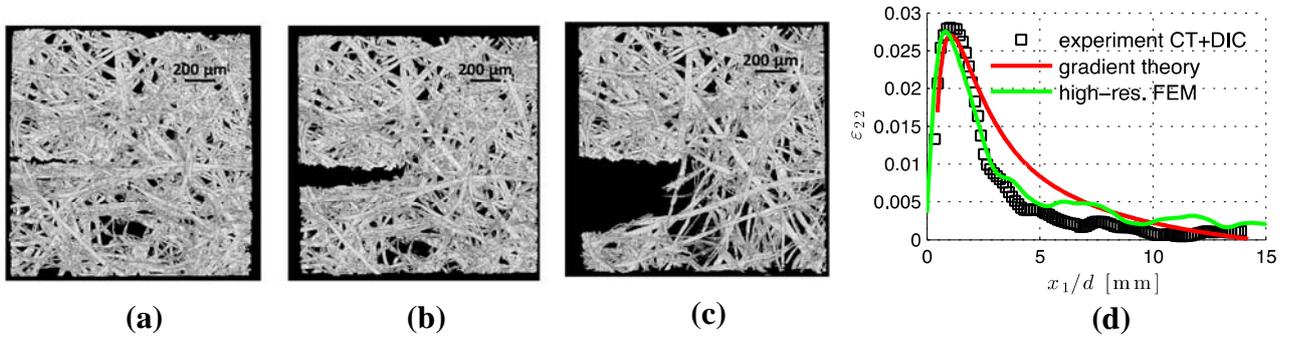

**Figure 2:** *Reconstructed cross section from a X-ray CT scan of a growing crack in a fibre material at ESRF (2a-c) (Reprinted from [8] with permission from Elsevier). CT/DIC-estimated strains along the crack plane in front of the tip. Observe the non-singular strain and that the maximum is located ahead of the tip at approximately the average cell diameter d. Also shown are the strains computed by a high-resolution FE model and the enhanced gradient theory (2d).*

***(ii) IL Estimation from Micro/Nano Indentation Tests:*** NI measurements have already been used to estimate ILs [36] for both elastic and plastic deformations. In particular, Yoffe's "blister" elasticity model and Johnson's "expanding cavity" plasticity model have recently been modified to include IL-gradient effects. The resultant IL-dependent expressions were successfully used to fit hardness vs. indenter's tip contact radius (or plastic zone size) data for a limited number of material systems and is now proposed to further elaborate on such analyses by using recently published data for bulk metallic glasses (bmg) and nanocrystalline (nc) materials.

***(iii) IL Estimation from Shear Band Characteristics:*** Shear band width and spacing measurements have already been used to estimate ILs for ultrafine grain (UFG) materials and relate them with grain size [37]. In fact, the ability of predicting shear band widths and spacings, as shown by the author's initial work on gradient elasticity [38], was one of the principal reasons for the interest shown by the mechanics and materials community in IL-gradient models. Most recently, the subject has attracted even more attention in relation to "shear-banding controlled" deformation in nanocrystalline (NC), bulk metallic glass (BMG) and nanoglass (NG) materials, along with the observed plateaus or serrations in the corresponding stress-strain graphs. Existing and new experimental data in this area can be used, in connection with estimating IL parameters of respective ILG models.

***(iv) IL Measurements for Gradient Microstructures:*** A set of non-standard novel experiments can be conducted by fabricating specimens with gradient microstructures, followed by subsequent mechanical testing. Specimens with controlled grain size gradients can be fabricated by expanding on our earlier MTU method [39], in which tapered plates were rolled to produce plastic strain gradients (Fig. 3a).

A modification of the above method utilizes rotary hammer swaging of conical rods to fabricate cylindrical materials with gradient microstructures for full-scale mechanical tests. Due to the initial variation in diameter, deformation gradients can be created in the material, and the resulting cylindrical specimens can be recrystallized to produce gradients in the grain size ranging from submicron to tens of microns. The samples may then be tested in tension, compression, or torsion. A high-resolution camera can be used to monitor cross-sectional area as a function of position during mechanical testing, which will enable real-time modeling of gradient effects on sample ductility. To our knowledge such type of tests for assessing size effects due to specimen dimensions, in conjunction with a controlled gradient in grain size, have not been conducted before.

*Wedge casting* (i.e. solidifying a metal inside a wedge-shaped copper mold) to produce gradient structures that could be machined into sub-sized cylinders for conventional compression testing (Fig. 3b) [40]. Mechanical behavior will be assessed with nanoindentation. Miniature mechanical test specimens will be fabricated by focused ion beam (FIB) milling that will be examined with transmission electron microscopy (TEM) using *in situ* nanoindentation or tensile deformation.

*Melt Spinning* is another method to produce nanoscale gradient materials for investigating gradient effects (Fig. 3c). The melt spinning solidification process depends on heat extraction from a stream of molten metal passing through a gas atmosphere and impinging on a spinning metal wheel. The heat extraction rate at any point is time dependent because of the varying shape of the puddle on the wheel and the difference in heat capacity between wheel, the melt and the atmosphere. The resulting solidified ribbon shaped piece may have a non-uniform microstructure due the time dependent heat flux at the wheel side and gas side of the ribbon. The ribbon can have a gradient in microstructure through the thickness because of the time dependent heat flow. The thickness of these types of ribbons depends on the wheel speed and the thickness of the molten stream, but can be developed in the 10 to 50 micron range with a width of 1 mm to 1 cm. An example of the gradient microstructure is shown in Fig. 3d.

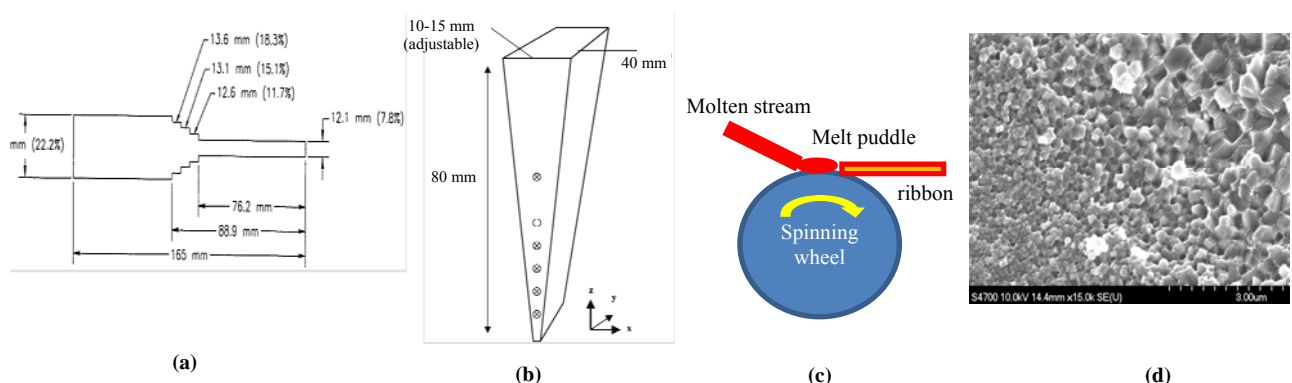

**Figure 3: (a)** *Geometry of tapered plate used for creating gradient microstructures via rolling [39]. The same type of profile in a cylindrical geometry will be deformed using rotary hammer swaging.* **(b)** *Wedge casting geometry showing approximate shape and location of temperature probes. The copper mold will water-cooled.* **(c)** *Geometry of melt spin produced gradient "grain size" materials.* **(d)** *Gradient in grain size through the thickness of a Fe-Nd-B hypo-eutectic composition. [unpublished results Hackney-Aifantis.] (reprinted from [6e] with permission from Springer).*

# 3. Emerging Research Case-Study Areas

*In this section we consider in some detail three topics* of current interest as case-studies to which the input from the theoretical and experimental developments can be fruitfully applied. The first topic is concerned with the development of a multiscale ILG framework for modeling the multiscale behavior of advanced structural materials such as *nanocrystalline/ultrafine grain (NC/UFG) polycrystals* and *bulk metallic glasses (BMGs)* – emerging multifunctional material classes sharing a common phenomenology. The second topic is concerned with chemomechanical, electromechanical and photomechanical degradation and aging of energetic materials, including next generation rechargeable battery nanostructured electrodes, micro/nano electromechanical materials (MEMs/NEMs), and interconnects, as well as optoelectronic materials used in semiconductor devices and light emitting diodes (LEDs). Finally, the last topic is concerned with the application of the ILG approach in addressing brain tissue modeling and neural cell transmission.

## 3.1 Structural Materials: Nanopolycrystals (NCs/UFGs) and Bulk Metallic Glasses (BMGs)

The applicability of the ILG approach to discuss deformation instabilities and shear banding in NC/UFG materials has been extensively discussed by the author and his co-workers in recent papers [37,41]. Some additional results pertaining to the emergence of shear bands in these materials and the occurrence of local deformation instabilities manifested as serrations in the corresponding stress-strain curves and describable by Tsallis q-statistics will be presented in detail in a later section. The same will be done for BMG materials, the response of which could also be conveniently modeled within the proposed ILG framework. This is due to the fact that all these materials may be viewed as sharing a common phenomenology, as discussed below. In this connection, we note that nanocrystalline and amorphous materials have in common that their deformation is not governed by the long-range motion of dislocations, but by localized and discrete deformation 'events': intergranular/grain boundary (GB) slip in nanocrystals, free volume/shear transformation zones (STZs) in amorphous materials. Despite the statistical homogeneity of the materials themselves above the atomic/crystallite scale, these events organize into complex spatial and temporal patterns from the nano up to the macroscale. For describing these patterns, it is possible to employ our multiscale ILG framework which comprises both *native* (GB size/STZ spacing) and *emergent* length scales (width/spacing of microshear bands), as discussed below:

• *On scales above the scale of the elementary deformation events*, i.e. intergranular/GB slip in nanocrystalline and free volume/STZ evolution in amorphous metals, deformation in both classes of materials can be modeled using an extension of the gradient-stochastic framework earlier applied to crystalline materials [27a]. Since its first publication, the same type of model has been used by several authors for providing a simplified description of deformation patterns in amorphous materials [42]. To develop the model into a physically realistic description of deformation in nanocrystalline and amorphous solids, we should use statistical distributions of local slip magnitudes and slip directions in deformation events parameterized by the densities of grain boundary dislocations/disclinations in nanocrystalline and of free volume/shear transformation zones in amorphous materials. The interaction between grains or STZs will be decomposed into a *long-range elastic interaction* and a *contact interaction* depending on the atomistic details between slip processes in adjacent grains/STZs. The former can, for an infinite solid, be evaluated in terms of the local eigenstrain and the elastic Green's function of the effective isotropic medium [27]. The 'contact interaction' can be obtained from atomistic simulation of the elementary slip processes and incorporated into the continuum framework by second-order gradient expansion of the finite-range interaction, leading to second-order gradient terms with an IL governed by the grain size/STZ spacing. To include temperature into this description, the CA dynamics can be replaced by a kinetic Monte Carlo formulation where elementary slips will be associated with stress dependent energy barriers. The discrete slip events (GB/intergranular slip for nanocrystalline, STZ/free volume evolution for amorphous solids) then occur at temperature-dependent rates which are governed by the corresponding activation barriers. In turn, the energy released through the heterogeneous slip events is converted into heat, which is propagated using the standard or a higher-order non-Fourier heat conduction equation [14,15,32a,b]. Heat conduction introduces additional spatio-temporal scales into the system and, thus, the proposed framework allows to study the transition between 'slow' deformation processes governed by the internal heterogeneity of the solid (slow on the time scale required for homogenization of the temperature field) and 'rapid' processes governed by the feedback between local slip and local heat generation.

• *On scales well above the scale of elementary slip events*, the deformation organizes into slip zones/microshear bands oriented along directions of maximum shear stress (or directions defined by a more general stability criterion accounting for the effect of hydrostatic pressure on a gradient-dependent yield stress [37]). These slip zones have finite extension both in the "forward" (maximum shear stress) and the perpendicular direction. The lower-scale model developed in the previous tasks will be used to establish the geometrical features of these microslip bands and the associated internal spatial scales (width/spacing/length). Unlike the *native* (or grown-in) ILs considered in the lower scale model, these significantly larger ILs constitute *emergent features* of the complex dynamics of microslip processes and themselves evolve in the course of deformation. To describe the spatial organization of deformation on the specimen scale, an effective formulation can be developed by taking up Escaig's idea for describing slip zone formation in amorphous materials in terms of the nucleation and propagation of Somigliana dislocations [43]. The collective evolution of these defects can be described in terms of nonlinear equations for their density, parameterized on the basis of model simulations from the previous tasks. Gradient and time lag terms in these equations are related to the interactions between the individual microshear bands, which can be envisaged as Somigliana dislocation dipoles and thus exhibit finite-range (dipolar) stress fields analogous to those associated with terminated shear bands in dislocation plasticity. The elastic interactions between these can be represented in terms of second-order gradients with an IL proportional to the dipole width (i.e. the shear band length). Thus, we can arrive at a multiscale gradient framework where the relevant elementary deformation events self-organize to produce new, emergent length scales which govern (on much larger scales) the ILG dynamics of macroscopic shear bands which, in turn, control the macroscopic deformation and failure behavior of these materials.

*(i) Micro/Nanoshear Banding and Micro/Nanodeformation Patterning:* Building on our previous work on shear banding and stationary plastic instabilities for meso/macroscopic specimens [38a-b, 44], attention will focus here on investigations of shear banding and deformation patterning phenomena at the micron and nanoscales. Our previous experimental observations for ultrafine grain size (UFG) materials (Fe-10% Cu alloys with grain size at ~150-1350 nm regime) have revealed [37] a new deformation plasticity mechanism i.e. massive shear band formation (or multiple shear banding), manifesting through the occurrence of a plateau in the stress-strain graph (i.e. perfectly plastic behavior). The orientation of shear bands was different in tension and compression and a yield asymmetry was also noted for the two types of loading. Application of a pressure-dependent yield criterion, also incorporating the Laplacian of effective plastic strain, provided some initial theoretical estimates [37] for the average orientation and thickness of shear bands, as well as for the yield asymmetry in agreement with the experiments, but several questions remained unanswered. They pertain to the correlation of grain size and grain rotation to statistical aspects of shear band characteristics (spacing, thickness, orientation), the role of grain size distribution, as well as the effect of inhomogeneous nucleation and evolution of shear bands on the form of overall stress-strain curves. The situation is reminiscent of the inhomogeneous deformation and multiple shear banding observed in foams and metallic glasses [45] for which satisfactory theoretical models to capture the available experimental data do not exist. A specific unresolved general question in this area is to correlate the evolution and interaction of the population of shear bands to the serrations (strain bursts or stress drops) observed in the corresponding size-dependent stress-strain curves, a problem of substantial current interest in the micro/nano pillar literature [46].

In an analogous manner, the nucleation and propagation of nanonecks can be considered along similar lines. While gradient theory has been adopted earlier (e.g. Barenblatt [47a], Coleman [47b]) to analyze necking in polymers, as well as by the author and his coworkers in metals at macroscopic scales [48], its implications to corresponding recently observed instability phenomena at the nanoscale [49] have not been considered. In particular, the critical instability conditions for the emergence of necking can be established (the analogue of the Considère condition for macroscopic specimens) and nanoneck propagation can be effectively studied within the ILG framework.

***(ii) Micro/Nanosize Effects and Micro/Nanoscale-Dependent Behavior:*** The problem of (extrinsic) size effect, i.e. the dependence of strength and mechanical behavior on specimen dimensions, can be traced back to Da Vinci and Galileo. The first who found experimentally that longer wires break easier than short ones for the same load; thus motivating a *probabilistic* "critical crack length" strength theory. The second found that thinner wires break easier than thicker ones for the same load; thus motivating a *deterministic* "maximum stress bearing capacity" theory. More recent treatments and related experiments for macroscopic specimens did not arrive at a widely accepted explanation (e.g. Bazant's size effect law and Carpinteri's multifractal size effect law, which are in conflict with each other [50]). An interesting strength of materials approach to interpret size effect experiments in torsion and bending of elastically deformed bone and polymeric foam specimens, as well as plastically twisted microwires and plastically bent microbeams, was presented by the author [51-52] on the basis of simple gradient elasticity and gradient plasticity models. Analogous results have been obtained by other authors using more sophisticated gradient or surface stress theories [53-55]. Atomistic MD simulations have also been employed to discuss size-dependent elastic and plastic properties at the nanoscale [56], without incorporating strain gradients. In the absence of macroscopically imposed strain gradients (tension/compression, creep configurations), the author and his co-workers [11-12,57], have introduced microscopic/submicroscopic gradients of strain or internal variables (nanovoid/dislocation density) to model the observed size effects, in agreement with experiments. This approach of internal variable theory with diffusive transport, enhanced with stochastic terms can be employed here for interpreting size effects in nano-objects (nanowires, micro/nanopillars).

Both *extrinsic* (associated with the "size" of the specimen) and *intrinsic* (associated with the "size" of the substructure) have been considered (e.g. Greer and De Hosson [21]). The most famous intrinsic size effect relationship (Hall-Petch/H-P) – stating that the yield/flow stress varies proportionally to the inverse square root of the grain size ($\sigma \sim d^{-1/2}$) – breaks down at the nanoscale where an abnormal or inverse H-P behavior has been observed and modeled either by revising standard dislocation arguments [33a,58] or using physically-based continuum modeling [59]. In this connection, it is noted that the ILG framework can result to different than the $\sim 1/\sqrt{d}$ dependence (e.g. $\sim 1/d$, in accordance with recent Bayesian analyses) of the yield stress, even in the conventional grain size strengthening regime. It all depends on the exponent used in the assumption introduced for relating the local flow stress to the magnitude of the local equivalent plastic strain ($\sim |\nabla \varepsilon|^{m/2}$; m = 1,2, ...) and its Laplacian. In fact, most recent work by the author [6a,f] has shown that simple-minded microscopic gradient arguments can effectively be used to deduce not only H-P standard and inverse behavior for the yield stress, but also grain size-dependent flow curves as a function of temperature and strain rate in agreement with existing experimental data for nanopolycrystals. Moreover, the grain size dependence of the pressure sensitivity and the activation volume parameters was obtained, in agreement with experimental trends [60] for both nanopolycrystalline and amorphous solids. These initial investigations have further been extended recently to consider normal and abnormal H-P behavior for nanocrystalline and nanotwinned metals, as observed experimentally [60, 61a-b], and provide convenient physically-based phenomenological models for data interpretation [61c-e].

Another issue that can readily be addressed is concerned with the modeling of *combined extrinsic-intrinsic size effects*. This amounts to simultaneously examining the interaction of the specimen dimensions with the size of the substructure as this manifests, in particular, in the dependence of flow curves on the ratios of (d/D, d/t, d/L) where d denotes grain size, D specimen diameter (for cylindrical specimens), t specimen thickness (for beam-like specimens) and L specimen length. An interesting account of such combined size effects has been provided recently by Ngan and coworkers [62] and certain experimental results for tensile specimens of varying grain size were reported, along with some empirical-like relations to fit the observed behavior. As indicated earlier, our recent results [11] in this direction, suggest that such combined extrinsic-intrinsic experimental size effect data can be interpreted through the aforementioned internal variable theory with diffusive transport, earlier used to interpret size effects in tension and creep of macroscale specimens. The new feature here that needs to be carefully examined is concerned with the assumed dependence of the yield stress on the grain size and the grain size-dependent nanoscale boundary conditions which have to be used, in order to determine the distribution of the inhomogeneously evolving internal variable (e.g. dislocation density). Along similar lines, it has been shown [63a] that introduction of microscopic strain gradients into the flow stress expression – due to local heterogeneities of the dislocation substructure, as observed experimentally [63b] – leads to a promising interpretation of size effect data for some micropillar testing realizations [46], also in agreement with simulations. More work is necessary, however, and stochastic effects need also to be accounted for in the flow stress expression, as discussed in later sections.

### 3.2 Energetic Materials: LiBs/Nabs, MEMs/NEMs, and LEDS

### 3.2.1 Gradient Chemomechanics and LiBs/NaBs

***(i) Li-ion Batteries (LiBs):*** Attention will focus here on chemically-induced damage/cracking that occurs in Li-ion rechargeable nanostructured anodes during electrochemical cycling. This problem is a major challenge for scientists and engineers due to recent needs for extending the lifetime of rechargeable batteries in order to increase the viability of the electric vehicle concept. Other important sectors to be directly affected by progress in this field include the portable electronic devices industry (laptops/cell-phones) and the biomedical industry (heart/brain pacemakers). A critical issue in determining the lifetime of a rechargeable battery is the interaction between the materials chemistry at the electrodes and the internal stress developed during operation [18,64]; i.e. combined expertise from mechanics, materials science and chemistry is required. During the charge/discharge cycle, the electrode materials will change shape and size, sometimes quite dramatically with up to a 300% volume expansion, due to chemical activity induced by the insertion/de-insertion process of charge carriers in the electrodes; i.e. upon Li-ion diffusion in Si or Sn nanoparticles (the active sites) encased in a brittle matrix. These *colossal strains* within the electrodes result in material damage, fracture and loss of electrical connection with the battery current collectors, thus leading to degradation of materials performance at large cycle numbers. The need for the development of a robust theoretical material mechanics framework has already been pointed out by workers in the field; e.g. the quotation [65]: "*In the lithium alloys studied here, enormous strain can be caused with zero applied stress. The strain is caused by the incorporation of interstitial Li atoms between the existing M atoms of the alloys. It is our opinion that the theories of elasticity in solids are not suited to describe the colossal volume changes described here. We invite theorists to take up the challenge to describe these phenomena.*" The approaches developed so far in considering "chemomechanical damage" seem to have neglected length scale and gradient effects due to nanoparticle or nanocrack interactions, as well as interfacial energy effects and couplings between deformation and diffusion fields at the nanoscale. All these are critical issues for interpreting the observed size effects and describe the corresponding size-dependent behavior. A different problem occurs in Li-ion cathodes where phase separation occurs also leading to capacity fade. In particular, the interaction between strain and diffusion fields along with their associated mechanical and diffusional internal lengths leads to size-dependent phase transformation diagrams and size-dependent spinodal gaps, controlling battery performance. A comprehensive review of various chemomechanical aspects pertaining to Li-ion electrodes can be found in [18] and some benchmark problems will be considered in a later section by also incorporating IL-gradient effects.

***(ii) Na-ion Batteries (NaBs):*** Recently, a renewed interest has been shown in alternative to Li-ion rechargeable battery systems for future energy storage. Na-ion rechargeable batteries is a promising example. The main advantage of this type of battery is the abundance of sodium, in contrast to the limited lithium resources. A most unique mechanics issue observed is these systems is that Sn anodes in Na-cells do not experience fracture during cycling, despite the up to ~400% expansions/contractions that occur during Na-ion insertion/deinsertion. This is very different than the severe fracture that Sn experiences upon the formation of lithium alloys. Despite the lack of fracture, a significant capacity fade in pure Sn anodes has been noted by embedding Sn nanopraticles in C. In fact, a careful examination of SEM and TEM micrographs [66] indicates the formation of "nanopores" in pure Sn anodes which, however, were not noticed (or not commented upon) by the authors. Pore formation is known as a mechanism by which materials respond to applied stress under mechanical creep conditions. This possibility has not been examined as a potential candidate mechanism to accommodate volume change stresses in electrochemical systems. This is important because pore formation in the Na-Sn is much less detrimental to electrochemical performance than the fracture mechanism observed in Li-Sn electrodes. The control of the transition between fracture and pore formation mechanisms could be a transformative concept for improving high capacity electrode lifetime in secondary batteries. It is therefore, necessary to perform a first systematic comparison of the mechanical effects that Na-ion insertion/de-insertion has on Sn-based anodes with the corresponding ones that Li-ion intercalation/de-intercalation has on such anodes.

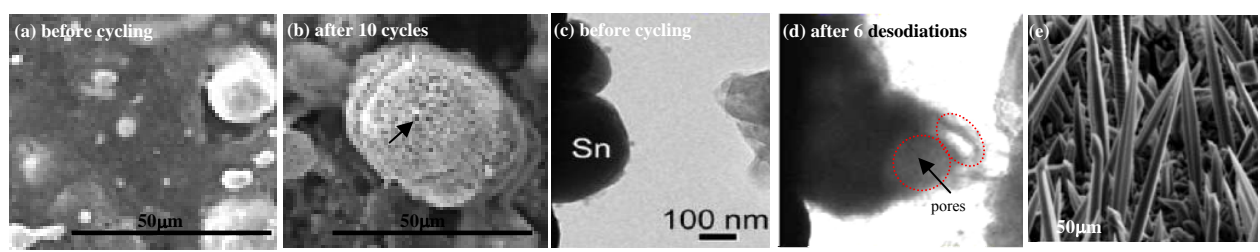

**Figure 4:** (a) SEM image of Sn anode prior Na-insertion (reprinted from [66a] with permission from Elsevier); (b) SEM image of Sn anode after Na-insertion indicating 'nanopores' (reprinted from [66a] with permission from Elsevier); (c) TEM image of Sn anode before cycling (reprinted from [66b] with permission from ACS); (d) TEM image of anode [$SnO_2$] (reprinted from [66b] with permission from ACS); (e) $SnO_2$ micro/nano structured anode to be cycled with respect to Li and Na [67].

One interesting topic that emerges for both Li-ion and Na-ion electrodes is to consider the implications of a stress-assisted diffusivity equation with stress-dependent diffusivity. A transport equation for the diffusion species of concentration $c$ can be adopted of the form $\dot{c} = \left( D + N\sigma_h \right)\nabla^2 c - M\nabla\sigma_h \cdot \nabla c$, where ($D$, $M$, $N$) are phenomenological constants and $\sigma_h$ is the hydrostatic stress determined by a purely mechanical problem (e.g. elasticity, or creep/damage) when certain uncoupling simplifications are assumed. Additional comments on this stress-assisted diffusion equation will be provided in a later section. In the general case, coupled diffusion-creep/damage equations need to be derived. Solutions of related boundary value problems will provide insight on chemomechanical damage and capacity fade in Na-ion anodes in analogy to corresponding studies in Li-ion anodes.

### 3.2.2 Gradient Electromechanics: MEMS/NEMS and Interconnects

*(i) Flexoelectricity:* In relation to flexoelectricity (observed in electronic and biological materials), the key constitutive assumption is $P_i = f_{ijk\ell}\varepsilon_{jk,\ell}$ where $P_i$ is the polarization vector, $f_{ijk\ell}$ is a fourth order flexoelectric tensor and $\varepsilon_{jk,\ell}$ is the strain gradient. For materials which are both piezoelectric and flexoelectric, the corresponding equation reads $P_i = d_{ijk}\varepsilon_{jk} + f_{ijk\ell}\varepsilon_{jk,\ell}$, where $d_{ijk}$ is the third order piezoelectric tensor. IL gradient effects can be incorporated by replacing the average quantities $\varepsilon_{jk}$ and $\varepsilon_{jk,\ell}$ with their local counterparts $\varepsilon_{ij} - \ell^2\nabla^2\varepsilon_{ij}$ and $\varepsilon_{ij,k} - \ell^2\nabla^2\varepsilon_{ij,k}$ and this may also account indirectly for surface effects. A more direct way to account for surface flexoelectricity (the *poling effect*) *is* to introduce a surface energy term $\int_S \eta P_i dS_i$ in the free energy of the system, where $\eta$ is a material parameter depending on the properties of the surface or interface (and the neighboring materials), with the integration being performed over the surface $S$. This is reminiscent of the practice of the IL gradient approach for plasticity where interfacial energy effects were accounted for through an "energy penalty" term $\int_S \phi(\varepsilon^p)dS$, where $\phi$ is a given function of the plastic strain $\varepsilon^p$ at the interface. Surface effects associated with electromechanical couplings were previously introduced indirectly [68] by including gradients of polarization, in the form $a_{ijk\ell}P_{i,j}P_{k,\ell}$ and $b_{ijk\ell}P_{i,j}\varepsilon_{k\ell}$, in the expression for the free energy of the system. The above variants of introducing "bulk" and "surface" flexoelectric terms will be the subject of this part of the proposed work. In addition to theoretical work, corresponding nanoindentation tests may be carried out by using, for example, titanate films of varying thicknesses (25, 50, 100, 500nm) in order to determine the effect of strain gradients on piezoelectric and flexoelectric parameters in a manner similar to that earlier employed to determine the indentation size effects (ISE) for purely mechanical deformations (e.g. [36] and references quoted therein).

*(ii) Electromigration:* In relation to electromigration (observed in interconnects), the central transport equation used to model void growth and shape changes is given by the equation $\boldsymbol{J}_S = D_S\delta_S\left(-q_S^*\boldsymbol{E}_S + \nabla_S\mu\right)/(\Omega kT)$ where $\boldsymbol{J}_S$ is the total mass flux on the void surface, $D_S$ is the surface atomic diffusivity, $\Omega$ is the atomic volume, $\delta_S$ denotes effective surface thickness, $q_S^*$ denotes surface effective charge, $k$ is the Boltzmann constant, and $T$ is the temperature. The quantity $\boldsymbol{E}_S = (\boldsymbol{E}\cdot\boldsymbol{s})\boldsymbol{s}$ is the tangential component of the electric field $\boldsymbol{E}$ on the void surface with unit tangent $\boldsymbol{s}$, and the quantity $\mu$ is the bulk chemical potential ($\nabla_S$ denotes surface gradient) given by $\mu = \mu_0 + \Omega(\gamma_s \kappa - \sigma_n + w)$. In this last expression, $\mu_0$ is the intrinsic chemical potential; $\gamma_s$ is the surface energy (energy per unit area in the stress free configuration); $\sigma_n = \sigma_{ij} n_i n_j$ is the normal stress on the void surface with unit outnormal $n_i$; $\kappa = n_{i,i}$ is the local curvature, and $w$ is the elastic strain energy density $w = \frac{1}{2}\varepsilon_{ij} L_{ijkl}\varepsilon_{kl}$ with $L_{ijkl}$ being the elastic matrix and $\varepsilon_{ij}$ the elastic strain. In this formulation, internal length scale effects are not accounted for. Within the ILG formulation, such effects can be introduced through the inclusion of $\frac{1}{2}\ell_\varepsilon^2 \varepsilon_{ij,k}\varepsilon_{ij,k}$ in the expression for $w$ and by replacing the $\sigma_n$ term with $\sigma_n - \ell_\sigma^2 \nabla^2 \sigma_n$. The effect of this gradient generalization on the void's shape can be obtained through the equation expressing the surface mass conversion, i.e. $\partial u_n / \partial t = -\Omega \nabla_s \cdot \boldsymbol{J}_s$, which suggests that the temporal change of local displacement (or void velocity) normal to the void's surface is balanced by the surface mass divergence, and the term $\partial u_n / \partial t$ may be replaced by $\partial h / \partial t$ where $h$ denotes the thickness of deposited/redeposited material. The electrostatic field in the standard equations of electromigration is derivable from a potential $\varphi$, i.e. by setting $\boldsymbol{E} = -\nabla\varphi$ where $\nabla^2\varphi = 0$. Within a gradient electromechanics generalization, it turns out that the potential $\varphi$ obeys the higher-order equation $\nabla^2\varphi - \alpha^*\nabla^4\varphi = 0$ where $\alpha^*$ accounts for heterogeneity effects exclusively associated with the transport of the charged carriers. Electromechanical coupling effects can be incorporated in the above considerations through extra terms of the form $\sigma_{ij}E_{i,j}$ and $\sigma_{ij,k}E_{i,jk}$. These are all aspects that can be conveniently explored within the proposed ILG framework. Finally, it is noted that hillock formation and growth is traditionally modeled [69] through a stress-diffusion equation which in one-dimension reads $\partial\sigma/\partial t = \Lambda \partial\sigma/\partial x$, where $\sigma$ is the inhomogeneous stress field along the interconnect and $\Lambda$ is a material constant. This equation results by assuming that the divergence $dJ_V / dx$ of vacancy flux $J_V$ along grain boundaries, (which is taken to be proportional to $\partial\sigma/\partial x$) in one dimension (along the interconnect) is directly related to stress relaxation $\partial\sigma/\partial t$, i.e. $\partial\sigma/\partial t \sim \Omega \partial J_V/\partial x$. If additional diffusion paths are assumed (dislocation cores for pipe diffusion, triple grain boundary junctions, internal void surfaces), they may all be lumped together into another effective diffusion path which coupled with the grain boundary (and bulk) diffusion path could fit within the framework of the author's double diffusivity framework [6f] which was successfully used recently to model diffusion in nanopolycrystals. It turns out that the final modified diffusion equation in this case reads $\partial\sigma/\partial t = \Lambda\partial^2\sigma/\partial x^2 + M\partial^3\sigma/\partial x^2\partial t + N\partial^4\sigma/\partial x^4$ where $\Lambda$, $M$, $N$ are constants depending on the two aforementioned diffusivities and mass exchange between the two diffusion paths.

### 3.2.3 Gradient Photomechanics and LEDs

The challenge here is to model the direct effect that light has on inducing electronic strain, as well as on the related effect of photo-induced configurational changes of structural defects and the mobility of dislocations. Such photo-induced strain effects have been observed in amorphous, as well as crystalline semiconductors (e.g. chalcogenide glasses and α-Si:H, as well as anthracene and GaAs) as reported in the physics literature [70]. Even though in the above works the mechanisms of photo-induced deformation (both reversible and irreversible) have been elucidated and the associated effects have been measured experimentally, a continuum gradient nanomechanics framework with internal variables (to represent photogenerated electron-hole pairs and reversible anisotropy, physical dimmers and dislocations) is not available. Gradients and size effects, defect self-organization and stochastic considerations on nanodeformation patterning through instability analysis, have also not been explored so far in this field. Such type of theoretical developments may be pursued within our ILG framework, also guided and confirmed experimentally through complementary AFM bending measurements of microcantilever beams. The photo-induced internal strain (Staebler-Wronski/SW) effect (e.g. [70a-c]) under various illumination and specimen size conditions for amorphous and crystalline semiconducting films, as well as the spatial characteristics of the dark-line defects (DLDs) generated through photo-induced dislocation glide [70d], can be investigated. In this connection, the remarkable analogy between the photoplastic (PhP) effect and the PLC effect (as convincingly discussed but not further explored by [70e] for anthracene crystals) will be considered as a representative case-study where nonlinear instability analysis and power-law statistics can be used to capture the observed behavior. Similarly the "giant" photo-softening effect [70f] observed in chalcogenide films without corresponding changes in photo-induced optical properties will be another case-study problem. Finally, the pattern forming DLD's instabilities observed in III-V semiconducting materials during device operation causing degradation and aging of light emitting diodes (LEDs) is a third example that can effectively be considered within this case-study area.

### 3.3 Brain ILG Mechanics and Neuroelasticity

*The third case-study area* is concerned with the application of the expertise and tools developed in the earlier discussed areas, to consider open questions related to *signal transmission in neural cells* (Fig. 5).

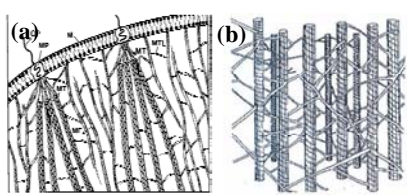

**Figure 5: (a)** *Schematic of cellular cytoskeleton/ membrane. M: cell membrane, MP: membrane protein, GP: glycoprotein extending into extra-cellular space, MT: microtubules, MF: microfilaments (actin filaments or intermediate filaments), MTL: microtrabecular lattice.* **(b)** *Interior of neuron showing cytoskeletal network. Straight cylinders are microtubules, 25 nm, in diameter. Branching interconnections are microtrabecular lattice filaments.*

**(i) Signal Transmission through Neuron's Membrane:** Existing studies based on the Fitzhugh-Nagumo (F-N) model [71] can be extended to consider the implications of ILs and ITs along with flexoelectric effects induced by the deformation of the cell membrane hosting the (nano)channels that $Na^+$ and $K^+$ ions diffuse through. The cellular membrane has the structure of a liquid crystal consisting of electric dipoles, the lipids. By applying flexion deformation on the membrane, the direction of the dipoles changes and electric polarization between the surfaces of the membrane is created. As a result, a voltage $V_f \sim f H$ appears, with $f$ denoting the flexoelectric coefficient and $H$ being the mean curvature of the membrane [72]. The discovery of mechano-sensitivity as a property of ion channels led to the study of the effect of mechanical deformation of the cellular membrane on the kinetics of voltage-sensitive domain (VSD) of potassium channels. The stretching deformation of the membrane could cause a time delay in the movement of VSD [73]. Time-delayed F-N equations have already been used to consider delayed feedback mechanisms associated with the influence of alcohol and coupling effects at the synapses [74]. However, such equations do not account for the effect of mechanical deformation on the action potential, an aspect that can readily be explored within our ILG framework

*(ii) Signal Transmission through Neuron's Microtubules:* A related problem is concerned with the propagation of signal through the microtubules of the neuron's axon. Each microtubule consists of 13 parallel protofilaments to form a hollow tube with length usually in the μm scale, and diameter in the nm scale. Each protofilament consists of a long series of tubulin dimmers: α- and β-tubulin are slightly different momoners that compound to form a dimer with certain dipolar moment in the electric field E of the microtubule. During GTP hydrolysis, the axis of each monomer shifts (initially perpendicular to the longitudinal axis) by $29^o$ so that the axis of the monomer has a projection u on the longitudinal axis, with consequences for the dipolar moment of the dimer. Furthermore, the microtubules (MTs) form a matrix bound together by the microtubule associated proteins (MAPs). It has been suggested [75] that the motion of the monomers may be modeled by the equation $M\ddot{u} - KR_0^2\partial_x^2 u - Au + Bu^3 + \gamma\dot{u} - qE = 0$ : M is the mass of a dimer; $K$ is the elastic modulus; $R_0$ is the equilibrium spacing between adjacent dimmers; *(A, B)* are constants; $\gamma$ is a damping coefficient; and *(q,E)* denote effective charge of the dimer and electric field, respectively. Thus, the microtubule is capable to transfer an electromechanical excitation (soliton) along its longitudinal axis. However, if two or more microtubules are connected with a MAP, an electrical excitation in one of them might influence the others. Preliminary research shows that this may be taken into account through an additional term proportional to $u(t-\tau)$ where $\tau$ denotes a time-delay parameter depending on the MAP properties. Similarly, gradient elasticity considerations lead to another extra term proportional to $\partial_x^4 u$ with the constant of proportionality related to an IL. For $\tau = E = 0$, the linear version of the resulting equation is similar to the one used in [24] to model wave propagation in nanotubes. Variants of the above nonlinear equation can be studied within our ILG framework also accounting for internal strain and flexoelectric effects. Such studies may also be useful in understanding various aspects of Alzheimer's disease [76], an important issue currently under consideration.

## 4. Benchmark Problems

### 4.1. Gradient Chemoelasticity: Size-Dependent Damage and Phase Separation in LiBs

#### 4.1.1 LiB Anodes and Size-Dependent Chemomechanical Damage

Based on experiments, initial mechanics considerations on LiB anodes employed simple elastic models to interpret deformation and fracture [18,64a,77-79] at the micro/nanoscale during lithiation/delithiation. In particular, this preliminary work aimed at predicting: **(a)** The swelling internal stress profiles that develop during maximum Li-insertion in Sn or Si spherical particles (active sites), embedded in an inactive, with respect to Li, matrix. **(b)** The critical stable crack length, as well as the energy released during cracking, thus allowing for the determination of the most promising anode configurations/active site volume fractions (e.g. nanospheres vs. nanofibers vs. nanodiscs) and material selections (for both the active site and the matrix) that would limit damage. In order to understand the experimentally observed enhanced stability during Li-insertion in active micro/nano islands or micro/nano pillars patterned on a less active matrix (as compared to thin films), the critical delamination compressive stress concept was employed [18,64c] to show that detachment of the active sites is inhibited by keeping their aspect ratio (height/length) larger than ~0.3. Direct experimental measurements [80-81] confirmed that the capacity fade with cycle number in fabricated Sn/C nanocomposite anodes was due to severe fracture that occurred when the particle size was greater than 100nm, whereas below 20nm both mechanical and electrochemical stability were observed allowing a capacity retention of 100% for 400 cycles. Detailed area analysis of the corresponding transmission electron microscopy (TEM) and field emission scanning electron microscopy (SEM) images provided an empirical relationship between capacity fade and initial Sn particle size/Sn surface area. This suggested that it is necessary to use an improved theoretical model that could explicitly account for this size effect that our previous conventional mechanics models could not capture. As a result, an initial gradient-dependent damage model was developed [82] which allowed for the prediction of the active site diameter, as well as the interparticle spacing, that would limit damage growth.

***(i) Active Particle Embedded in a Gradient Elastic Matrix:*** Since classical elasticity, used in a first attempt to estimate internal stress development during lithiation, is not able to capture size effects, we suggest here to use gradient elasticity theory for obtaining a more accurate size-dependent estimation of the stresses. We start with an elastic circular Si/Sn particle embedded in an anode matrix made from a gradient elastic material. The governing equations for the displacements $\boldsymbol{u}$ and stresses $\boldsymbol{\sigma}$ take the form (see, for example, [83])

$$u_k^g - \ell^2 u_{k,mm}^g = u_k^c \quad ; \qquad \sigma_{ij}^g - \ell^2 \sigma_{ij,mm}^g = C_{ijk\ell} u_{k,\ell}^c , \qquad (4.1.1)$$

where the superscript "g" denotes gradient (or nonlocal) quantities and the superscript "c" denotes local quantities predicted from classical elasticity theory. The quantity $C_{ijk\ell}$ is the elastic tensor, whereas the quantity $\ell$ denotes the internal length. In principle, both of these constitutive parameters can be adjusted through proper material selection and proper fabrication or manipulation of the processing conditions. The problem analyzed in [77] can then be revisited by adopting Eqs. (4.1.1), implemented with a finite element methodology by discretizing the continuum displacements $\boldsymbol{u}^c$ and stresses $\boldsymbol{\sigma}^g$ for the unit cell of Fig. 6 (for details, see [24, 83]).

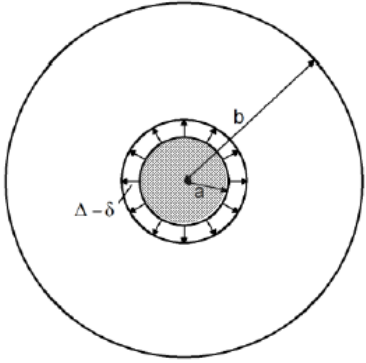

**Figure 6:** *Idealized geometry of the anode's unit cell, where the inner shaded circular area of radius* r = a *represents the active site and the surrounding white annular area, of inner radius* a *and outer radius* b, *is the glass/ceramic matrix. The quantity* Δ *denotes the unconstrained expansion of the active site and δ the distance the matrix pushes back during Li-intercalation [77].*

The size effect analysis is performed by studying the stress ratio defined as $stress\ ratio = \left| \sigma_{ref} \right| / \max \left| \sigma_{rr}^g \right|$. [ $\sigma_{ref}$ is the reference stress, assumed to be equal to the maximum compressive radial stress obtained by applying classical elasticity, while $\sigma_{rr}^g$ is the non-local radial stress.] This stress ratio is an indicator for the spare capacity of the battery towards mechanical failure. For typical values of the constants [77] and various values of the internal length $\ell$, the stress ratio is plotted against the ratio between the internal radius a and the length scale $\ell$ in Fig. 7.

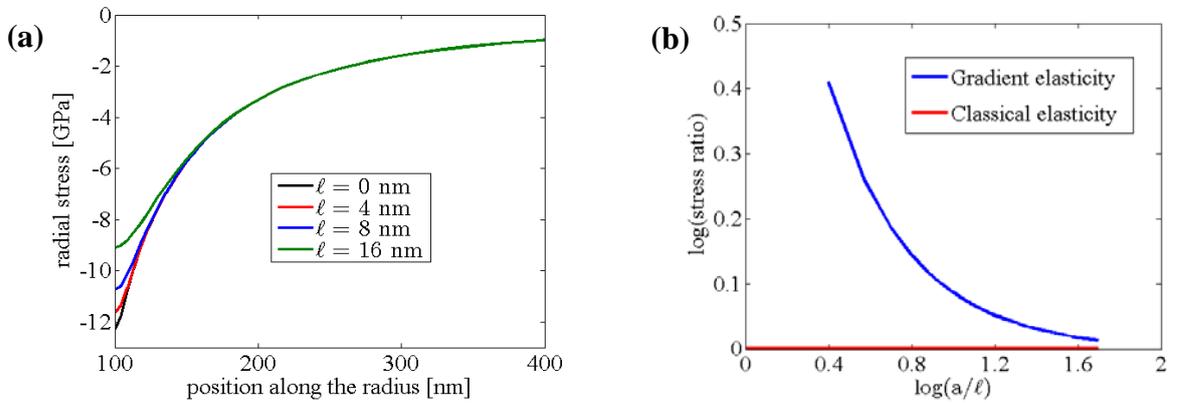

**Figure 7:** *(a) Profile of the radial stress for different values of internal length* $\ell$ *; (b) size effect curves.*

It is evident that while the local radial stress determined through classical elasticity does not show any size effect, gradient elasticity produces a stress ratio that increases for decreasing dimensions of the unit cell. For increasing size of the unit cell, the stress ratio exhibits a horizontal asymptote, coinciding with the solution of classical elasticity. This means that the maximum (compressive) non-local radial stress decrease for decreasing size of the unit cell, while it tends to the value determined according to classical elasticity for increasing dimensions.

It may be concluded that by using gradient elasticity (with a limited additional computational cost with respect to classical elasticity) it is possible to take into account size effects, and obtain more accurate estimations of the stresses experienced by the matrix. Furthermore, this feature of gradient elasticity enables a more accurate choice of the material to use for the matrix (different materials have different length scales), and allows to select the most appropriate dimensions of the unit cell.

***(ii) Gradient Elasticity for Both Active Particle and Inactive Matrix:*** The previous exercise may be repeated by assuming that both active site and the surrounding matrix (glass) are gradient elastic materials, with internal lengths $\ell_s$ and $\ell_g$, respectively. In Fig. 8a the profile of the radial stress along the radius is plotted for a portion of the unit cell for different values of internal lengths, along with the results previously obtained by modeling the surrounding matrix only as a gradient elastic material. It can be observed (for details, see [83]) that there are no significant differences between the radial stress values in the matrix, obtained by modeling both the active site and the surrounding matrix as gradient elastic materials. But it can also be noticed that by modeling the entire unit cell as a gradient elastic composite, the maximum compressive radial stress is not experienced at the interface between active site and matrix, but at a certain distance inside the matrix. This distance depends on the relative values of the adopted length scales. Furthermore, by focusing attention on the radial stress profile in the active site, it can be observed that an increase in the value of its internal length $\ell_s$, leads to a smoothing effect on the radial stress profile itself. The size effect analysis is performed by studying the stress ratio defined as before and the results are plotted in Fig. 8b.

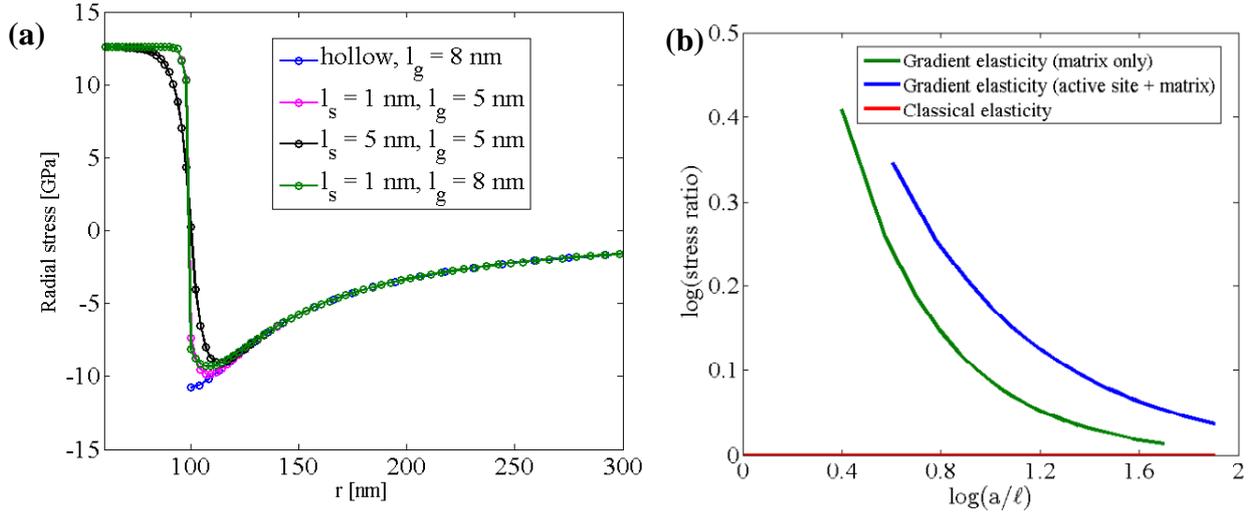

***Figure 8: (a)*** *Profile of the radial stress (left) for different values of internal length and comparison with the profile obtained by modeling only the surrounding matrix (hollow cylinder) as gradient elastic materials.* ***(b)*** *Size effect with* $\ell_g = 5\,\text{nm}$ *for the case where both active site and matrix are gradient elastic, and with* $\ell_g = 8\,\text{nm}$ *for the case where only the matrix is gradient elastic. It is seen that in this case the size effect becomes slightly less accentuated as compared to the case where only the matrix is modeled through gradient elasticity.*

In a similar way, gradient elasticity can be employed to consider size effects for the benchmark problems considered earlier [18] to obtain input for the most favorable configurations (spheres vs. cylindrical fibers) and material selection for preventing fracture. Some interesting results are also expected by considering the parameters $\Delta$ and $\delta$ (see Fig. 6 for the definition of these symbols) as being dependent on the concentration $c$ of the diffusing Li-ions. As already mentioned, the concentration $c$ of the diffusing species may be determined from a stress-assisted diffusion equation of the form

$$\dot{c} = \left(D + N\sigma_h\right)\nabla^2 c - M\nabla\sigma_h \cdot \nabla c \,, \qquad (4.1.2)$$

where $\sigma_h$ is the hydrostatic stress and $(D,M,N)$ are phenomenological coefficients. It is noted that Eq. (4.1.2) has been derived and used by the author [84] to model hydrogen embrittlement. The hydrostatic stress $\sigma_h$ $(\sigma_h = 1/3\,\sigma_{ii})$ can be obtained either from linear elasticity or (more accurately) gradient elasticity theory, as outlined above.

In concluding this section, it is noted that internal stress development in LiB anodes has been considered recently by a number of leading mechanics researchers (see, for example, [85-86]) but internal length gradient effects were not considered and the dependence of diffusivity on the hydrostatic stress in Eq. (4.1.2) was neglected ($N = 0$).

### 4.1.2 LiB Cathodes and Size-Dependent Phase Transformations

A central problem in LiB cathodes – as, for example, in those based on $Li_xFePO_4$ materials – is the occurrence of phase separation and pattern formation of new nucleating phases which cause electrochemical degradation and capacity fade. This effect is more detrimental than the volume expansion which for these cathode materials is about 15%. A number of investigators (see, for example, [87-89]) have recently addressed this problem of phase separation by using Cahn-Hillard [90] type arguments with or without strain effect considerations. It was found that confinement of the active particles in small volumes can suppress the occurrence of Li-induced phase changes, thus improving electrochemical performance and prolonging cathode life. However, strain gradient effects were not taken into account, an issue that will be undertaken in the IL chemomechanical couplings as discussed earlier. Recent results [91] show that spinodal and miscibility gaps are size-dependent, as a result of the "small" volumes available for both Li-ion concentration and concentration gradients to adjust, as well as for internal stress and associated stress gradients to develop. Below we briefly sketch these preliminary findings which can further complement the results recently obtained in this topic by other researchers without considering strain gradient IL effects.

We start with the following gradient-dependent expression for the free energy density of the system $\psi$

$$\psi = \hat{\psi}\left(c, \nabla c, \boldsymbol{\varepsilon}, \nabla e\right) = f(c) + \frac{1}{2}\nabla c \cdot \boldsymbol{K} \nabla c + \frac{1}{2}\boldsymbol{\varepsilon}^E \cdot \boldsymbol{\mathcal{C}}\boldsymbol{\varepsilon}^E + \frac{1}{2}\nabla e^E \cdot \boldsymbol{H}\nabla e^E \,, \qquad (4.1.3)$$

where $\boldsymbol{\varepsilon}^E = \boldsymbol{\varepsilon} - \bar{c}\boldsymbol{M}$ is the elastic strain tensor, with $\boldsymbol{M}$ denoting the lattice mismatch tensor of the two phases, which in its simplest form reads $\boldsymbol{M} = M_o\boldsymbol{1}$. The fourth-order elasticity tensor $\boldsymbol{\mathcal{C}}$ is assumed to be constant, and for isotropic materials we have $\boldsymbol{\mathcal{C}}\boldsymbol{\varepsilon}^E = 2G\boldsymbol{\varepsilon}^E + \lambda e^E\boldsymbol{1}$, with $(G, \lambda)$ being the Lamé parameters. The second-order tensors $\boldsymbol{K}$ and $\boldsymbol{H}$ are symmetric and positive definite and, for simplicity, they are taken to be of the form $\boldsymbol{K} = \kappa\boldsymbol{1}$ $(\kappa = \text{const} > 0)$ and $\boldsymbol{H} = \ell_\varepsilon^2\left(\lambda + 2G/3\right)\boldsymbol{1}$, where $\ell_\varepsilon$ is the internal length associated with strain gradients, whereas the diffusional internal length is associated with the parameter $\kappa$ (not to be confused with the same symbol used for the curvature in Section 3.2.2(ii)).

The function $f(c)$ represents the free energy per unit volume of a homogeneous and strain-free system of uniform concentration $c$. An appropriate expression for it is

$$f(c) = \mu^0 c + R\vartheta c_{\max}\left[\bar{c}\ln\bar{c} + (1-\bar{c})\ln(1-\bar{c})\right] + R\vartheta c_{\max}\left[\alpha\bar{c}(1-\bar{c})\right], \qquad (4.1.4)$$

where $\bar{c} = c/c_{\max}$ $(0 \le \bar{c} \le 1)$ is a normalized concentration, $R$ is the gas constant, and $\vartheta$ denotes temperature. The quantity $\mu^0$ is a reference value of the chemical potential $\mu$, the gradient of which gives the flux $\boldsymbol{j}$ of the diffusing species, i.e. $\boldsymbol{j} = -D(c)\nabla\mu$; $D(c) = B_o c\left(1 - \bar{c}\right)$, where $B_o$ is the mobility constant. In what follows, the value $\alpha = 4$ is used; only for $\alpha > 2$ the chemical free energy $f(c)$ is non-convex, as required for phase separation through spinodal decomposition.

It then turns out (for details see [91]) that within a variational formulation based on Eq. (4.1.3), the constitutive equations for $(\mu, \boldsymbol{\sigma})$ are $(\bar{M}_0 = M_0/c_{\max})$

$$\left.\begin{array}{l} \mu = f'(c) - \kappa\nabla^2 c + (2G+3\lambda)\overline{M}_o\big[3\overline{M}_o(c-\ell_\varepsilon^2\nabla^2 c) - \big(\mathrm{tr}\,\boldsymbol{\varepsilon} - \ell_\varepsilon^2\nabla^2\,\mathrm{tr}\,\boldsymbol{\varepsilon}\big)\big], \\[2mm] \boldsymbol{\sigma} = 2G\boldsymbol{\varepsilon} + \lambda e\mathbf{1} - \ell_\varepsilon^2\left(\dfrac{2G+3\lambda}{3}\right)\nabla^2 e\mathbf{1} - (2G+3\lambda)\overline{M}_o(c-\ell_\varepsilon^2\nabla^2 c)\mathbf{1}, \end{array}\right\} \quad (4.1.5)$$

and the resulting governing differential equations for the displacement and concentration fields $(\boldsymbol{u},c)$ read

$$\left.\begin{array}{l} G\nabla^2\boldsymbol{u} + (\lambda+G)\nabla(\nabla\cdot\boldsymbol{u}) - \left(\dfrac{2G+3\lambda}{3}\right)\ell_\varepsilon^2\nabla^2\big[\nabla(\nabla\cdot\boldsymbol{u})\big] - (2G+3\lambda)\overline{M}_o\nabla\big(c-\ell_\varepsilon^2\nabla^2 c\big) = 0, \\[2mm] \dot{c} = g_1(c)\nabla c\cdot\nabla c + g_2(c)\nabla^2 c - g_3(c)\nabla^4 c - g_4(c)\nabla c\cdot\big[\nabla\cdot(\nabla^2 c)\big] - \\[2mm] \qquad - g_5(c)\nabla c\cdot\nabla\big[\nabla\cdot\boldsymbol{u} - \ell_\varepsilon^2\nabla^2(\nabla\cdot\boldsymbol{u})\big] - g_6(c)\big[\nabla^2(\nabla\cdot\boldsymbol{u}) - \ell_\varepsilon^2\nabla^4(\nabla\cdot\boldsymbol{u})\big], \end{array}\right\} \quad (4.1.6)$$

where the functions $g_i(c)$ are defined in [91]. It is noted that Eqs. $(4.1.6)_{1,2}$ result by inserting Eqs. $(4.1.5)_{1,2}$ respectively into the force balance equation for the stress $(\nabla\cdot\boldsymbol{\sigma} = 0)$ and the mass balance for the flux $(\dot{c} + \nabla\cdot\boldsymbol{j} = 0)$, where the definition $\boldsymbol{\varepsilon} = 1/2[\nabla\boldsymbol{u} + (\nabla\boldsymbol{u})^T]$ is also used.

Next, we employ linear stability analysis, by considering perturbations of the form $c(\boldsymbol{x},t) = c_o + \varepsilon\tilde{c}(\boldsymbol{x},t)$, $\boldsymbol{u}(\boldsymbol{x},t) = \boldsymbol{u}_o(\boldsymbol{x}) + \varepsilon\tilde{\boldsymbol{u}}(\boldsymbol{x},t)$, where $\varepsilon$ is a small parameter. Then the system of the first-order approximation of Eqs. (4.1.6) reads

$$\left.\begin{array}{l} G\nabla^2\tilde{\boldsymbol{u}} + (\lambda+G)\nabla(\nabla\cdot\tilde{\boldsymbol{u}}) - (2G+3\lambda)(\ell_\varepsilon^2/3)\nabla^2\big[\nabla(\nabla\cdot\tilde{\boldsymbol{u}})\big] - (2G+3\lambda)\overline{M}_o\nabla\big(\tilde{c}-\ell_\varepsilon^2\nabla^2\tilde{c}\big) = 0, \\[2mm] \dot{\tilde{c}} = g_2(c_o)\nabla^2\tilde{c} - g_3(c_o)\nabla^4\tilde{c} - g_6(c_o)\big[\nabla^2(\nabla\cdot\tilde{\boldsymbol{u}}) - \ell_\varepsilon^2\nabla^4(\nabla\cdot\tilde{\boldsymbol{u}})\big]. \end{array}\right\} \quad (4.1.7)$$

*(i) Infinite Domains:* For an infinite system, we seek solutions of the form $\tilde{c}(\boldsymbol{x},t) = \hat{c}\exp\big(\omega t + i\boldsymbol{q}\cdot\boldsymbol{x}\big)$ and $\tilde{\boldsymbol{u}}(\boldsymbol{x},t) = \hat{\boldsymbol{u}}\exp(\omega t + i\boldsymbol{q}\cdot\boldsymbol{x})$. The resulting dispersion equation (with $q = \sqrt{\boldsymbol{q}\cdot\boldsymbol{q}}$ ) reads

$$\frac{\omega(q)}{q^2 D(c_o)} = -f''(c_o) - \kappa q^2 - \frac{12G\overline{M}_o^2\big(1+\ell_\varepsilon^2 q^2\big)}{3\left(\dfrac{2G+\lambda}{2G+3\lambda}\right)+\ell_\varepsilon^2 q^2}. \qquad (4.1.8)$$

It follows that $\omega(q)$ is positive and thus, a uniform concentration $c_o$ is unstable, if and only if

$$f''(c_o) < -\kappa q^2 - \frac{12G\overline{M}_o^2\big(1+\ell_\varepsilon^2 q^2\big)}{3\left(\dfrac{2G+\lambda}{2G+3\lambda}\right)+\ell_\varepsilon^2 q^2}. \qquad (4.1.9)$$

In view of Eq. (4.1.4), the instability condition in Eq. (4.1.9) can be solved with respect $\overline{c}_o$ giving the concentration range of the spinodal regime as

$$\frac{1-\sqrt{1-4/b}}{2} < \overline{c}_o < \frac{1+\sqrt{1-4/b}}{2}, \quad \text{for } b > 4 \qquad (4.1.10)$$

with the parameter $b$ given by the relation

$$b = 2\alpha - \ell_c^2 q^2 - \frac{12GM_o^2}{R\vartheta c_{\max}} \frac{1 + \ell_\varepsilon^2 q^2}{3\left(\frac{2G+\lambda}{2G+3\lambda}\right) + \ell_\varepsilon^2 q^2}, \qquad (4.1.11)$$

where $\ell_c^2 = \kappa c_{\max} / R\vartheta$ defines an internal length associated with the presence of the concentration gradient in the free energy density. As follows from Eq.(4.1.10) the spinodal gap width ($w$) − not to be confused with the same symbol used for the strain energy density in Section 3.2.2(ii) − is given as $w = \sqrt{1 - 4/b}$. Thus, the critical value $q_{cr}$, i.e. the maximum wavenumber that can induce instability, corresponds to $b = 4$ in Eq.(4.1.11), while $q \geq q_{cr}$ is equivalent to $b \leq 4$.

***(ii) Finite Domains and Size Effects:*** The dependence of the instability regime on the wavenumber implies the occurrence of size effects. In particular, for a finite system the boundary conditions will constrain the set of allowable wave vectors $\boldsymbol{q}$. For example, for three-dimensional systems where the phase boundary is perpendicular to the $x$-direction and whose concentration field is uniform in the other two directions, the general solution of the perturbed Eqs.(4.1.7) is a sum of terms of the form

$$\tilde{c}(x,t) = \hat{c}e^{\omega t}\cos\left(\frac{n\pi}{L}x\right), \qquad \tilde{\boldsymbol{u}}(x,t) = \hat{u}e^{\omega t}\sin\left(\frac{n\pi}{L}x\right)\hat{\boldsymbol{e}}_x, \qquad (4.1.12)$$

where $\hat{\boldsymbol{e}}_x$ is the unit vector along the positive $x$-direction and $n \neq 0$ an integer, such that proper zero flux and higher-order stress conditions (deduced from the aforementioned variational formulation) are fulfilled at the boundary, as discussed in [91]. Then, the dispersion relation of Eq. (4.1.8) still holds, but with $q = n\pi / L$. Uniform concentrations are unstable when $\omega(n\pi / L) > 0$ for at least one value of $n$, which gives again Eq. (4.1.9) but with $q = \pi / L$. Accordingly, the same replacement for $q$ holds in Eq.(4.1.11) for the parameter $b$, in Eq. (4.1.11) is now given by

$$b = 2\alpha - \pi^2\left(\frac{\ell_c}{L}\right)^2 - \frac{12GM_o^2}{R\vartheta c_{\max}} \frac{1 + \pi^2\left(\frac{\ell_\varepsilon}{L}\right)^2}{3\left(\frac{2G+\lambda}{2G+3\lambda}\right) + \pi^2\left(\frac{\ell_\varepsilon}{L}\right)^2}, \qquad (4.1.13)$$

rendering a size effect on the spinodal gap width $w = \sqrt{1 - 4/b}$, which is illustrated in Figure 9.

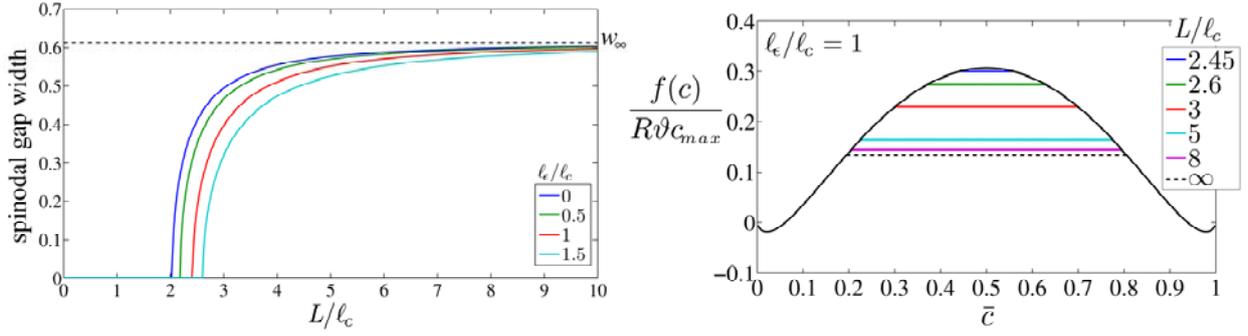

***Figure 9:*** *Size effects on the spinodal gap width.*

It is noted that the spinodal range is narrower as the crystal size decreases. Moreover, there is a critical length $L_{cr} = 2\pi / q_{cr}$ such that when $L \le L_{cr}$ the spinodal region ceases to exist, i.e. in quite small crystals all possible uniform concentrations are stable to any fluctuation. Similarly with $q_{cr}$, this critical length value corresponds also to $b = 4$ in Eq.(4.1.13), while $L \le L_{cr}$ is equivalent to $b \le 4$. Furthermore, due to the aforementioned stabilizing role of strain gradients, the spinodal gap width decreases while $L_{cr}$ becomes larger as the internal length scale $\ell_\varepsilon$ increases. On the other hand, for very large specimens, i.e. as $L \to \infty$, Eq.(4.1.13) combined with $w = \sqrt{1 - 4/b}$ yields

$$w_\infty = \sqrt{1 - 4\left(2a - \frac{4GM_o^2}{R\vartheta c_{\max}}\frac{2G+3\lambda}{2G+\lambda}\right)^{-1}} \qquad (4.1.14)$$

which, for typical values of the parameters (see [91]) gives $w_\infty \cong 0.611$.

### 4.2 Gradient Electroelasticity and Size Effects

As in the case of gradient elastic chemomechanics, we outline here a corresponding formulation for elastic gradient electromechanics by also taking into account surface effects. Two example problems are considered: one for a gradient piezoelectric hollow specimen subjected to shear loading, and another for a flexoelectric cantilever beam with an end point load.

### 4.2.1 Gradient Piezoelectric Perforated Plate under Shear

For elastic piezoelectric materials, gradient effects can be considered through a gradient-dependent free energy density function $\psi$ which, in analogy to the previous case of gradient chemoelasticity, may be expressed in the following form

$$\psi = \hat{\psi}\left(\varepsilon_{ij}, \varepsilon_{ij,k}, E_i, E_{i,j}\right) = \frac{1}{2}\varepsilon_{ij}c_{ijkl}\varepsilon_{kl} + \frac{1}{2}\varepsilon_{ij,m}B_{ijmkln}\varepsilon_{kl,n} - \varepsilon_{ij,m}D_{ijmkl}E_{k,l}$$
$$-\varepsilon_{ij}e_{ijk}E_k - \frac{1}{2}E_i\lambda_{ij}E_j - \frac{1}{2}E_{i,j}F_{ijkl}E_{k,l}, \qquad (4.2.1)$$

where an indicial notation is adopted here for convenience. The quantity $\varepsilon_{ij}$ is the strain tensor defined as $\varepsilon_{ij} = \frac{1}{2}(u_{i,j} + u_{j,i})$, with $u_i$ denoting displacement, and $E_i$ is the electric field vector defined as $E_i = -\phi_{,i}$, with $\phi$ denoting electric potential. The quantities $c_{ijkl}$, $e_{ijk}$ and $\lambda_{ij}$ are respectively, the elastic, piezoelectric and dielectric permittivity material tensors, while the higher-order new material tensors $B_{ijklmn}$, $D_{ijmkl}$ and $F_{ijkl}$ are new material constants owing to the introduction of strain gradient ($\varepsilon_{ij,k}$) and electric field gradient ($E_{i,j}$). Under appropriate symmetry assumptions [92-93], it turns out that the corresponding constitutive equations for the stress ($\sigma_{ij}$) and the electric displacement ($D_i$) are of the form

$$\sigma_{ij} = c_{ijkl}(\varepsilon_{kl} - l^2\varepsilon_{kl,mm}) - e_{ijk}(E_k - \gamma^2 E_{k,mm}),$$
$$D_i = \lambda_{ij}(E_j - \rho^2 E_{k,jj}) + e_{ijk}(\varepsilon_{jk} - \gamma^2\varepsilon_{kl,jj}), \qquad (4.2.2)$$

where $(\ell, \gamma, \rho)$ denote internal lengths, and the relations $B_{ijmkln} = c_{ijkl}l^2\delta_{mn}$, $F_{ijkl} = \chi_{ik}\rho^2\delta_{jl}$ and $D_{ijmkl} = e_{ijk}\gamma^2\delta_{ml}$ have been used.

Next, we focus attention on an infinite piezoelectric perforated plate containing a hole of radius $R$ (Fig. 10).

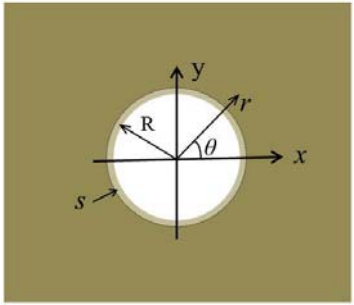

**Figure 10.** *An infinitely extended circular hole embedded in a piezoelectric matrix under uniform remote anti-plane shear and in-plane electric field.*

The piezoelectric material is transversely isotropic and polarized along the *z*-direction. Therefore, the *xy*-plane is a plane of isotropy. The in-plane displacement components are zero ($u_1 = u_2 = 0$), whereas the out-of-plane displacement and the in-plane electric potential are denoted by $u_3 = w(x, y)$ and $\phi = \phi(x, y)$. Then, for the simple problem under consideration, the constitutive equations can be rewritten as

$$\begin{Bmatrix} \sigma_{31} \\ \sigma_{32} \end{Bmatrix} = c_{44}[\nabla w - l^2\nabla(\nabla^2 w)] + e_{15}[\nabla\phi - \gamma^2\nabla(\nabla^2\phi)], \qquad (4.2.3)$$

$$\begin{Bmatrix} D_1 \\ D_2 \end{Bmatrix} = e_{15}[\nabla w - \gamma^2\nabla(\nabla^2 w)] + \lambda_{11}[-\nabla\phi + \rho^2\nabla(\nabla^2\phi)]. \qquad (4.2.4)$$

In the absence of body force and free volume charge, the final governing equations are obtained by introducing the above constitutive equations into the equilibrium equations for the stress and the electric field, as

$$\left.\begin{matrix} c_{44}(\nabla^2 w - l^2\nabla^4 w) + e_{15}(\nabla^2\phi - \gamma^2\nabla^4\phi) = 0, \\ e_{15}(\nabla^2 w - \gamma^2\nabla^4 w) - \lambda_{11}(\nabla^2\phi - \rho^2\nabla^4\phi) = 0. \end{matrix}\right\} \qquad (4.2.5)$$

The above set of coupled differential equations can be solved by separation of variables for the displacement and the electric potential in the matrix ($w^{(m)}, \phi^{(m)}$). The details are given in [93], where a discussion on the form of the appropriate boundary conditions is also provided. Here, we assume that the displacement and electric potential in the hole are given by $w^{(h)} = 0$, $\phi^{(h)} = m_1 r \cos\theta$, whereas the remaining boundary and continuity conditions are as follows:

$$\phi^{(m)}\big|_{r=R} = \phi^{(h)}\big|_{r=R}, \ \sigma_{zr}^{(m)}\big|_{r=R} = \frac{1}{R}\frac{\partial \sigma_{z\theta}^s}{\partial \theta}, \ D_r^{(m)}\big|_{r=R} = \frac{1}{R}\frac{\partial D_\theta^s}{\partial \theta},$$

$$\frac{\partial^2 w}{\partial r^2}\Big|_{r=R} = 0, \frac{\partial^2 \phi}{\partial r^2}\Big|_{r=R} = 0, E_\theta^m = E_\theta^s, \varepsilon_{z\theta}^m = \varepsilon_{z\theta}^s. \tag{4.2.6}$$

The superscripts $m$ and $h$ refer respectively to the matrix and the hole, whereas the superscript $s$ refers to the surface layer of the hole where the corresponding surface stress and surface electric displacement constitutive equations are assumed to be of the form [93-94]

$$\sigma_{z\theta}^s = \sigma_{z\theta}^0 + c_{44}^s \varepsilon_{z\theta}^s - e_{15}^s E_\theta^s, \ \ D_\theta^s = e_{15}^s \varepsilon_{z\theta}^s + \lambda_{11}^s E_\theta^s. \tag{4.2.7}$$

To proceed further and provide numerical results on size effects, we assume that the piezoelectric matrix is PZT-4, and the bulk material parameters are taken as $c_{44}^M = 25.6$ GPa, $e_{15}^M = 13.44$ C/m$^2$, $\lambda_{11}^M = 6 \times 10^{-9}$ C/V·m. The applied uniform far field anti-plane strain and electric fields are assumed to be given as $\varepsilon_{xz} = S_0 = 0.02$ and $E_x = E_0 = 10^6$ V/m. The surface material parameters are assumed as $c_{44}^s = l_s c_{44}^M$, $e_{15}^s = l_s e_{15}^M$ and $\lambda_{11}^s = l_s \lambda_{11}^M$ with $l_s$ being the surface layer thickness. The size effect results are shown in Figs. 11-12. Figure 11 suggests that the surface layer properties have a significant effect on the electric and stress fields around the hole. Only when the surface layer is very thin, the surface effect may be neglected. Figure 12 shows the effect of internal length scales on the electric and stress fields. For the electric field, it turns out that the size effect is more pronounced when the surface effect is taken into account. When $l=R/15$, the distribution of the electric field component $E_r$ follows nearly a straight line. It may be concluded that for the values of the parameters chosen and the boundary conditions assumed, the effect of surface layer and internal lengths on the electric field may be neglected. Both of the stress components do not seem to change significantly as $l$ takes different values, but when the surface effect is considered, $\sigma_{zr}$ decreases.

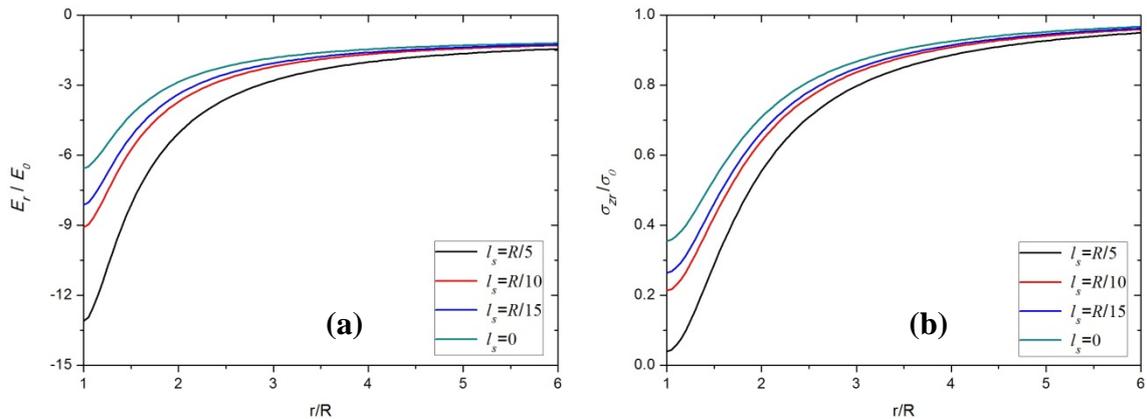

**Figure 11.** *Variation of normalized (a) electric field $E_r$, and (b) stress $\sigma_{zr}$ for different surface layer thickness ( $l = R/5$, $\rho = R/10$, $\gamma = R/20$).*

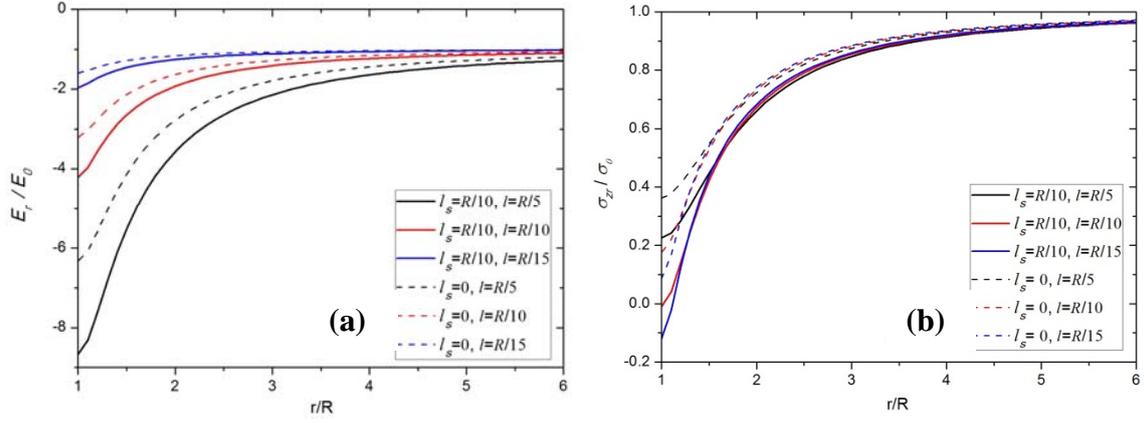

**Figure 12.** *Variation of normalized **(a)** electric field $E_r$, and **(b)** stress $\sigma_{zr}$ for different internal lengths $(l = 3\rho = 4\gamma)$ .*

### 4.2.2 Gradient Piezoelectric Beam with Flexoelectric and Surface Effects

For surface treated micro/nano-electro-mechanical beams, the constitutive equation for the surface stress and electric fields can be written as

$$\sigma_{\alpha\beta}^s = \sigma_{\alpha\beta}^0 + c_{\alpha\beta\gamma\delta}^s \varepsilon_{\gamma\delta}^s + d_{\alpha\beta\gamma}^s P_\gamma^s, \quad E_\alpha^s = a_{\alpha\beta}^s P_\beta^s + d_{\alpha\beta\gamma}^s \varepsilon_{\alpha\beta}^s, \quad (4.2.8)$$

where $\sigma_{\alpha\beta}^s$ is the surface stress tensor and $E_\alpha^s$ is the surface electric field vector, while $\varepsilon_{\alpha\beta}^s$ and $P_\alpha^s$ denote the surface strain tensor and surface polarization vector. For the piezoelectric bulk, strain gradient and flexoelectric effects can be accounted for through the strain energy density function given by

$$U_b = \frac{1}{2} c_{ijkl} \varepsilon_{ij} \varepsilon_{kl} + d_{ijk} \varepsilon_{ij} P_k + \frac{1}{2} a_{kl} P_k P_l + f_{ijkl} \varepsilon_{ij,k} P_l + \frac{1}{2} g_{ijmkln} \varepsilon_{ij,m} \varepsilon_{kl,n}, \quad (4.2.9)$$

where $\varepsilon_{ij}$ is the bulk strain tensor and $P_k$ is the bulk polarization vector; $c_{ijkl}$, $d_{ijk}$, $a_{ij}$, $f_{ijkl}$ and $g_{ijmkln}$ denote elastic, piezoelectric, reciprocal dielectric susceptibility, flexoelectric and strain gradient tensor material coefficients, respectively.

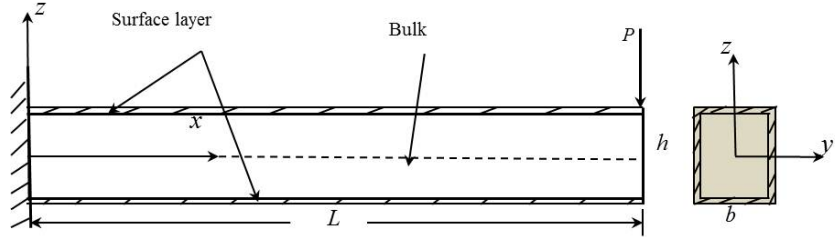

**Figure 13.** *Schematic of a piezoelectric beam with a surface layer.*

Next, we focus on the problem of a slender cantilever beam subjected to a static concentrated force $P$ at the beam end (see Fig. 13). The constitutive equations for the usual (partial) elastic stress $\sigma_{ij}^e$, the hyperstress $\tau_{ijk}$ and the electric field $E_i$ for the simple configuration considered, read

$$\left.\begin{array}{c}
\sigma_{ij}^{e} = c_{11}\varepsilon_{11} + d_{31}P_3, \\
\tau_{111} = g_{111111}\varepsilon_{11,1}, \ \tau_{113} = g_{113113}\varepsilon_{11,3} + f_{13}P_3 \\
E_3 = a_{33}P_3 + d_{31}\varepsilon_{11} + f_{13}\varepsilon_{11,3}.
\end{array}\right\} \quad (4.2.10)$$

The corresponding variationally-consistent boundary conditions [93] require the prescription of the following quantities at the beams ends

$M^h$ or $\dfrac{\mathrm{d}^2w(x)}{\mathrm{d}x^2}$; $\quad M - \dfrac{\mathrm{d}M^h}{\mathrm{d}x}$ or $\dfrac{\mathrm{d}w(x)}{\mathrm{d}x}$; $\quad \dfrac{\mathrm{d}}{\mathrm{d}x}\left(M - \dfrac{\mathrm{d}M^h}{\mathrm{d}x} + M^s\right)$ or $w(x)$,

$M = \int_A (\sigma_{11}^e z + \tau_{113})\mathrm{d}A = \left[\left(\dfrac{\epsilon_0 d_{31}^2}{1+\epsilon_0 a_{33}} - c_{11}\right)I + \left(\dfrac{f_{13}^2}{a_{33}} - l_2^2 c_{11}\right)A\right]\dfrac{\mathrm{d}^2w}{\mathrm{d}x^2} - \dfrac{f_{13}V}{a_{33}h}A,$

$M^h = \int_A \tau_{111}z\mathrm{d}A = -l_1^2 c_{11}I\dfrac{\mathrm{d}^3w}{\mathrm{d}x^3}; \quad M^s = \int_c \sigma_{11}^s z\mathrm{d}c = \left(\dfrac{\epsilon_0 d_{31}d_{31}^s}{1+\epsilon_0 a_{33}} - c_{11}^s\right)\dfrac{\mathrm{d}^2w}{\mathrm{d}x^2}\left(\dfrac{h^3}{6} + \dfrac{bh^2}{2}\right).$

with the various symbols (not earlier defined, e.g. $I = bh^3/12, A = bh,$ etc.) denoting standard quantities in beam theory. The final governing equation for the beam's displacement turns out to be of the form

$$g_{111111}I\dfrac{d^6w}{dx^6} - \left[\left(c_{11} - \dfrac{\epsilon_0 d_{31}^2}{1+\epsilon_0 a_{33}}\right)I + \left(g_{113113} - \dfrac{f_{13}^2}{a_{33}}\right)A + \left(c_{11}^s - \dfrac{\epsilon_0 d_{31}d_{31}^s}{1+\epsilon_0 a_{33}}\right)I_s\right]\dfrac{d^4w}{dx^4} + 2b\sigma^0\dfrac{d^2w}{dx^2} = 0. \quad (4.2.11)$$

Even though this equation is of similar form as that obtained by others (e.g. [95-96]), it is more general as all three effects (piezoelectric, flexoelectric, surface) are taken into account, as explicitly shown in the expressions for the coefficients. For the numerical evaluation of size effects, we assume a BaTiO$_3$ material for which parameter values are available in the literature [97-98], as follows: $c_{11}$=131GPa, $d_{31}$=1.87×10$^8$Vm$^{-1}$, $a_{33}$=0.79×10$^8$VmC$^{-1}$, $f_{13}$=5V, $c_{11}^s$ = 9.72N/m, $d_{31}^s = -0.056$C/m. The residual surface stress is taken as $\sigma^0 = 0.5$N/m, while the high-order material parameters are given as $g_{111111} = l_1^2 c_{11}$ and $g_{113113} = l_2^2 c_{44}$ [98]. The difference between the results of this and a previous model is illustrated in Fig. 14(a). Figure 14(b) shows the influence of each term (gradient, flexoelectric, surface) on the beam deflection. One thing to be noted here is that we assumed $l_1$=$l_2$=1$nm$ in Fig. 14(b).

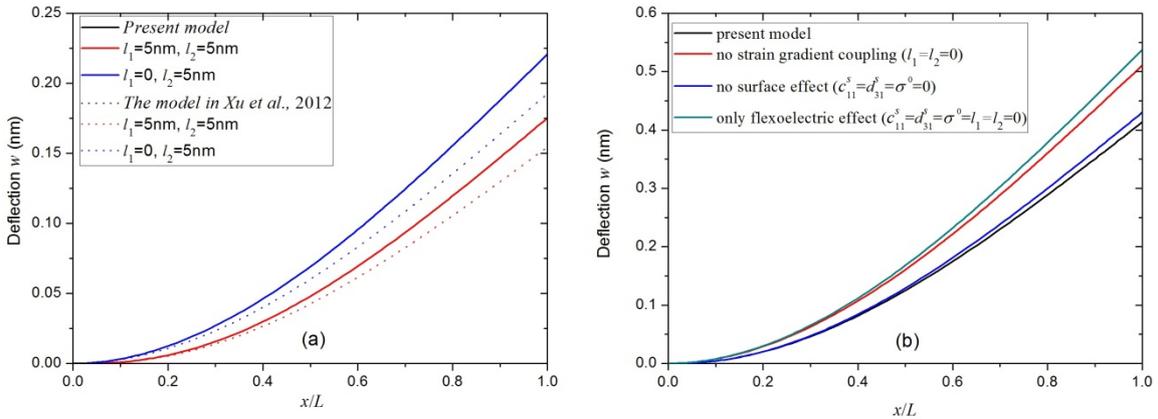

**Figure 14.** The deflection of a BaTiO$_3$ cantilever beam for various values of the electromechanical internal lenght parameters.

More specifically, Fig. 14(a) indicates that the deflection predicted by the present model is larger than that predicted by a previous gradient model without surface effects. This suggests that the surface and flexoelectric effects can reduce the rigidity of the beam. Figure 14(b) shows that when only the flexoelectric effect is considered, the deflection attains the largest values. For the evaluation of the surface effect, we test the influence of $\sigma^0$ on the deflection and electric field. The results are shown in Fig. 15. We can conclude that when strain gradient couplings are considered, the residual stress has a small effect on the electromechanical fields, while when strain gradient couplings are neglected, the residual stress effect becomes quite important. There is a significant difference for the value of the electric field at the fixed end of the cantilever for the two cases, i.e. with and without strain gradient effect. With the strain gradient effect, $E_3$ is zero near the fixed end. However, $E_3$ assumes its largest value when the strain gradient is neglected. This interesting phenomenon, which relates to the boundary conditions assumed, may be used for model validation in conjunction with experiments.

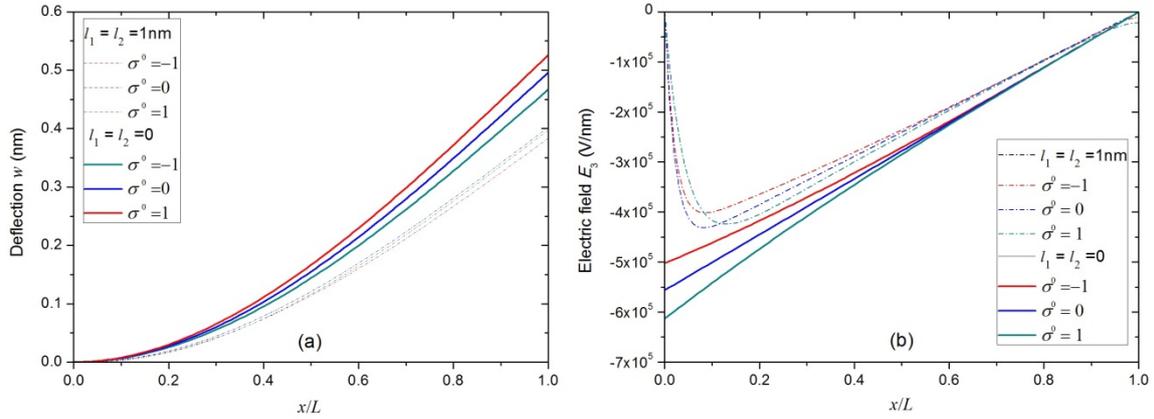

**Figure 15.** *Variation of **(a)** deflection and **(b)** electric field along the cantilever, for different values of the residual surface stress.*

### *4.3 Gradient Elastic Fracture Mechanics*

### *4.3.1 Gradient Elasticity (GradEla) Non-singular Crack Fields*

In this section we show the implications of our previously proposed (GradEla) model [3,6a,24] in classical linear elastic fracture mechanics (LEFM). Focusing on the determination of the stress field ($\sigma_{ij}$) near the crack tip, we list the non-standard equilibrium equation that it satisfies i.e.

$$[\sigma_{ij} - \ell^2 \nabla^2 \sigma_{ij}]_{,j} = 0 \ , \tag{4.3.1}$$

suggesting that the gradient enriched quantity $\sigma_{ij} - \ell^2 \nabla^2 \sigma_{ij}$ is divergence free, as the singular Hookean stress $\sigma_{ij}^0$ of classical elasticity is, when the gradient-enriched partial differential equation $\sigma_{ij} - \ell^2 \nabla^2 \sigma_{ij} = \sigma_{ij}^0$ is adopted to remove singularities from dislocation and crack problems. In other words, the quantity $\sigma_{ij}$ may not be considered as a standard "macrostress" field but rather as a non-standard "microstress" field which satisfies a nonlocal-type equilibrium equation of the form $\sigma_{ij,j} = f_i$, where $\sigma_{ij}$ denotes the bulk stress and $f_i$ is an internal body force taking into account the bulk-surface interaction. This was, in fact, the physical basis for the author's proposal for a continuum with microstructure [99], which pointwise exchanges mass, momentum, and energy with external or internal surfaces. In this connection, it should be pointed out that the aforementioned "microstress" quantity and its corresponding conjugate "microstrain" counterpart discussed in the next paragraph may not be necessarily associated or identified in a straight-forward manner with other analogous quantities introduced within the framework of generalized continua (micropolar, microstretch, micromorphic, etc.) as reviewed, for example, by Eringen [100]. Thus, in the present case of the GradEla model $\sigma_{ij}$ may be identified with "bulk" stress, and $\ell^2 \nabla^2 \sigma_{ij,j}$ with the internal body force $f_i$, such that the overall macroscopic stress $\sigma_{ij}^0$ satisfies the usual equilibrium equation. The overall macroscopic stress remains singular (as discussed in [101]), whereas the "bulk" or microstress $\sigma_{ij}$ is non-singular. Analogous is the situation for the non-singular stresses in modes I and II obtained in [35a,102]. Such expressions were also listed in [8c,103] and in Eqs. (47)-(49) of [16]. It was further stated in [16] that the stress $\sigma_{12}$ ($t_{xy}$ in their terminology) does not vanish on the crack surface. The appropriate expression for this shear stress component which vanishes on the crack surface and, thus, satisfies the relevant boundary condition is given in [102]. It follows that the criticism in [16] that the aforementioned expressions for the microstress $\sigma_{ij}$ do not satisfy equilibrium is misleading, as the discussion of Eringen's nonlocal elasticity or stress gradient theory [100b] is misleading, which was the first attempt to eliminate stress (but not strain) singularities at dislocation lines and crack tips. It should be noted, however, that Eringen ([100b]; p. 100) explicitly states that "... the real stress is not $\sigma_{k\ell}$ but $t_{k\ell}$ ..." and, thus, the Hookean stress ($\sigma_{ij}^0$ in the notation of [16]) may not be enforced, in general, to satisfy a balance law ($\sigma_{ij,j}^0 = 0$) as suggested by these authors. In fact, strictly speaking, the statement given by Eq. (5) of [16] may be viewed as an artifact of a number of assumptions, and it is not a fundamental equation of Eringen's theory. Only under suitable assumptions for the nonlocal kernel (Green's function of a linear differential operator with constant coefficients, or a Taylor series expandable quantity with the local Hookean stress being divergent free), the argument embodied in Eqs. (1)-(7) of [16] holds, and the nonlocal integral theory can be reduced to a "stress gradient" theory. In this connection, it should be pointed out that in the GradEla terminology the term "stress" or "Cauchy stress" is used for the stress satisfying the classical equilibrium equation, which is designated as "total stress" in [16]. This contains both the Hookean stress ("Cauchy-like stress" in Eq. (12) of [16], "Cauchy stress" in [104]) defined as the energy conjugate of the elastic strain, as well as the divergence of a higher-order stress ("double stress" in [104]). The terminology problem has also been addressed recently in [105], but the remarks on the author's work are not appropriate. In particular, there is no problem with the solutions listed in [52] for the components $(\varepsilon_{23}^G, \varepsilon_{32}^G)$ – or $(e_{23}^1, e_{32}^1)$ in the notation of [16] – since the components of $\boldsymbol{\varepsilon}^0$ (or $\boldsymbol{e}^0$ in that notation) are harmonic functions for the antiplane problem considered (Note 4 of [105]), and Note 5 of [105] does not apply to the reference [53] listed in [105]. The notation $(\boldsymbol{\sigma}, \boldsymbol{\sigma}^0)$ used in Section 4 of that article (listed as [53] in [105] and as [6a] herein) is different than that used in [105] and the meaning of these stress quantities should not be identified with that of $(\boldsymbol{\sigma}, \boldsymbol{\sigma}^c)$ listed in Section 2 of the same paper.

Details on the comparison between the GradEla model and Eringen's nonlocal or stress gradient theory can be found in an overview article [24]. In the same overview article, the so-called Ru-Aifantis (R-A) theorem that [16] comments upon is utilized to facilitate the solution of boundary value problems for static and dynamic configurations. The conditions for its validity are clearly stated in the original work of Ru and Aifantis [101]. Its principal aim was to express (by adopting suitable assumptions) solutions of gradient elasticity in terms of solutions of classical elasticity by reducing the fourth-order governing partial differential equation (pde) for the displacement field to a much easier treatable second order pde of the (modified) Helmholtz type. Thus, the claim in [16] that "…in the presence of boundary conditions, the Ru-Aifantis theorem is no longer valid and can lead to erroneous solutions" is not appropriate. For example, in the case of straight boundaries (dislocations, cracks), it turns out that the variationally consistent boundary conditions can be replaced or approximated by simpler ones with more clear physical meaning and experimentally attainable for which the Ru-Aifantis procedure can effectively be applied [106]. This procedure, in fact, has been used extensively [107-108] by the first author of [16]. The Ru-Aifantis procedure is an efficient mathematical tool for directly extending basic formulas of classical dislocation theory to the GradEla case by simply replacing the linear elasticity Green's function in the classical formulas by its GradEla counterpart which is well-known in mathematical physics for the Helmholtz operator. This is precisely the approach utilized in [105] for illustrating that the well-known analogy between classical electrostatics and classical dislocation theory can be directly extended to the GradEla case. Such straightforward extension is due to the use of the R-A theorem which holds for GradEla but not for the general Mindlin's theory [104]. And, in fact, the Ru-Aifantis observation was also used by the authors of [107] in line with our earlier work [108-109] for solving dislocation and disclination problems within the structure of the GradEla model. A generalization of the R-A theorem for fractional gradient elasticity has been recently presented in [110].

In view of the above discussion, it is also reasonable to expect that the non-singular strains conjugate to the above non-singular stresses (i.e. the strain expressions given by Eqs. (33)-(34) of [16]) are not compatible. Again, these non-singular strains should be viewed as microstrains, not necessarily associated with (or derivable from) a macroscopic displacement field. These incompatible microstrains are, in this respect, of a similar character as the incompatible strain fields emerging in Eringen's 3M (micromorphic, microstretch, micropolar) theories, and for which generalized compatibility conditions can be derived ([100a]; p. 21). In fact, the strains derived by Lazar and Maugin [107a] and listed in Eq. (46) of that paper for the case of a screw dislocation (as well as other cases of dislocation/disclination fields) are not compatible, whereas their classical elasticity textbook counterparts are. In this connection, it should be pointed out that the insistence for both their $\sigma_{ij}$ (second-order) and $\tau_{ijk}$ (third-order) tensor quantities to be divergence free (p. 1161 of [107a]) could be overly restrictive for general boundary value problems. Only by identifying $\tau_{ij}$ with a divergence free (self-equilibrated Hookean) stress, such practice may be justified. As mentioned earlier, they refer to "their quantity" $\sigma_{ij}$ as being the Cauchy stress, whereas it is a Hookean stress (given by the usual linear relationship in terms of strain) and may not be taken, in general, as the divergence free quantity that relates linearly the traction on a surface with the unit normal acting upon it. The same argument holds for their hyperstress quantity $\tau_{ijk}$ for which there is also no physically substantiated mathematical argument or other justification for assuming it as being divergence free. This, of course, does not preclude the fact in some instances (e.g. for GradEla dislocations), it turns out that these conditions are fulfilled.

Another point that should be addressed is concerned with the aforementioned authors' clearly stated preference [16] for other crack-tip gradient elasticity solutions (e.g. those obtained in [111]) which produce not only infinite but also stresses changing sign at the tip of a mode III crack, as well as compressive infinite stresses at the tip of a mode I crack under tensile loading. Why these "unphysical" hypersingular stress solutions (of the order of $r^{-3/2}$) should be favored over the classical non-singular solutions (of the order of $r^{-1/2}$), remains unclear. Similar hypersingular asymptotic and full-field analytical results for GradEla crack-tip solutions (which were also checked against corresponding finite element calculations), were derived in [112] published at the same time. It may only be remarked that the effect of strain/stress gradients is expected to be important near the crack tip region only, whereas far away from it, classical elasticity may suffice to describe elastic behavior. Moreover, if non-singular expressions for the crack-tip stresses and strain can be potentially useful for engineering applications, they should be simple and easy-to-use for reformulating or improving previous fracture mechanics arguments (e.g. failure criteria), commonly based on linear elastic fracture mechanics (LEFM). In this connection, it has recently been pointed out [8c,103,113] that using a classical continuum LEFM approach to describe macroscopic singular-dominated deformation fields in heterogeneous cellular or porous materials (nonwoven felts, solid foams, wood, textiles or paper) can lead to erroneous results, especially in materials having low or moderate degrees of density. The LEFM description is not sufficient to capture the essential mechanical behavior for this class of materials because a structural defect may significantly alter the displacement field, resulting in severe elastic blunting of the crack-tip [8d, 114]. Initial efforts by Isaksson and co-workers to establish contact between their experimental results and theoretical predictions are in-line with the aforementioned "complexity vs. simplicity" approach advocated by the author, and recent results on this issue will be reported elsewhere. In this connection and on the basis of experimental observations, it should be remarked that the displacement field describing the crack profile may not be smooth for heterogeneous materials. In fact, the use of Fourier transform in the Helmholtz equation for dislocation and crack problems should account for both possibilities, i.e. "smooth" and "non-smooth" source terms. Then, within the framework of generalized functions, it can been shown that the solution for a screw dislocation given by Gutkin and Aifantis [115] satisfies the equation $(1 - \ell^2 \Delta)u_z = u_z^o - \ell^2 b(x)\delta'(y)$. Thus, another criticism stated in [107b] based on the insistence of "smooth" displacements up to the dislocation line, is not founded. Similar arguments hold for a mode III crack with a blunted tip, as well as for other crack configurations (e.g. modes I, II).

### 4.3.2 Dislocation-based Gradient Elastic Fracture Mechanics

To provide further support on the efficiency of strain gradients in providing non-singular stress fields and smooth closure crack profiles, we list below recent results obtained in [116] for a mode III crack modeled by convolution of screw dislocations within an "incompatible" framework. Since the notation adopted here departs slightly from the one used in the earlier subsection, we first list below the governing equations

$$\left.\begin{array}{l} u_{i,j} = \beta_{ij}^T = \beta_{ij} + \beta_{ij}^P, \qquad \varepsilon_{ij} = \frac{1}{2}\left(\beta_{ij} + \beta_{ji}\right), \\ \alpha_{ij} = e_{jkl}\beta_{ik,l} = -e_{jkl}\beta_{ik,l}^P \Rightarrow \alpha_{ij,j} = 0; \\ \sigma_{ij} = \lambda\delta_{ij}\varepsilon_{mm} + 2\mu\varepsilon_{ij}, \quad \sigma_{ijk} = \ell^2\sigma_{ij,k}; \\ \sigma_{ij,j} - \sigma_{ijk,kj} = 0; \end{array}\right\} \qquad (4.3.2a)$$

and associated variationally-consistent boundary conditions

$$t_i = \left( \sigma_{ij} - \partial_k \sigma_{ijk} \right) n_j - \partial_j \left( \sigma_{ijk} n_k \right) + n_j \partial_l \left( \sigma_{ijk} n_k n_l \right), \Big\}$$
$$q_i = \sigma_{ijk} n_j n_k .$$

(4.3.2b)

The kinematic quantities $(u_i, \varepsilon_{ij}, \alpha_{ij})$ denote the displacement vector, the elastic strain tensor, and the dislocation density tensor. The total distortion tensor $\beta_{ij}^T$ is curl-free, while the elastic $\beta_{ij}$ and plastic $\beta_{ij}^P$ distortions are not curl-free within an incompatible framework. The constitutive quantities $\sigma_{ij}$ and $\sigma_{ijk}$ denote the elastic stress and the double stress tensor with the total (or real Cauchy) stress $\sigma_{ij} - \sigma_{ijk,k}$ satisfying the standard equilibrium equation. The symbols $\delta_{ij}$ and $e_{ijk}$ denote respectively the Kronecker delta and the Levi–Civita tensors, while $\ell$ is the internal length and $(\lambda, \mu)$ the usual Lamé constants. The "natural" boundary conditions for the nonclassical traction vector $t_i$ and the double traction vector $q_i$ to be used in the sequel, result from a variational formulation, with $n_i$ denoting unit outnormal.

The dislocation density $B_{zj}(t)$ is determined by solving a system of integral equations [116] and is depicted in Fig. 16. To shed light on the effect of the internal length gradient parameter, the dislocation density distribution for $l = 0.05a$, $0.1a$ and $0.5a$ is compared to the one corresponding to classical elasticity. It is observed that by decreasing the gradient parameter, the results of gradient elasticity converge to those of classical elasticity everywhere (even in the vicinity of the crack tips).

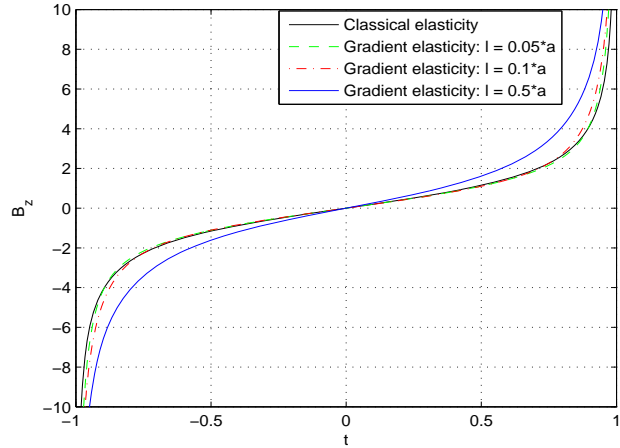

***Figure 16:*** *Dislocation density of a crack.*

Having determined the dislocation density, the crack opening displacement (CoD) can be obtained, as demonstrated in Fig. 17, which depicts the crack opening displacement (CoD) in classical and gradient elasticity.

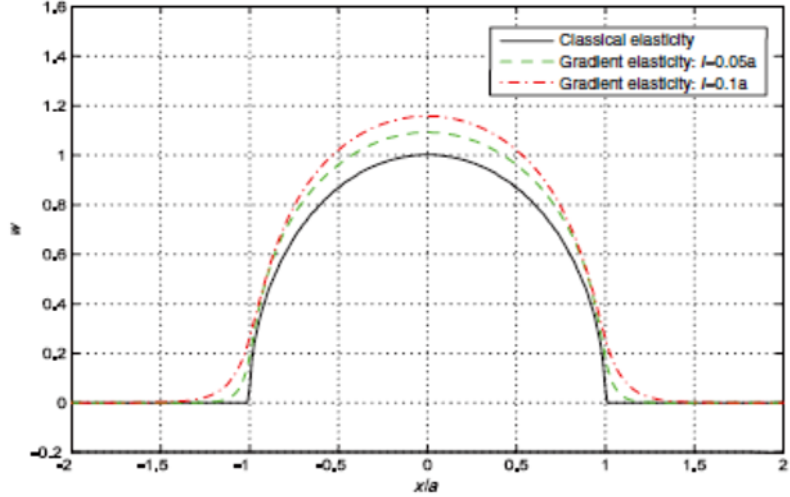

*Figure 17: Crack opening displacement in classical and gradient elasticity.*

The stress field along the x-axis reads [116]

$$\sigma_{yz}^{crack}(x,0) = \sigma_{yz}^{\infty} + \frac{a\mu}{2\pi}\int_{-1}^{1}\frac{1}{X}\left[1 - \frac{R}{l}K_1\left(\frac{|X|}{l}\right)\right]B_z\mathrm{d}t, \qquad (4.3.3)$$

while $\sigma_{xz}^{crack}(x) = 0$. It is interesting to compare the present results with those obtained by adopting Eringen's nonlocal elasticity theory, by also using a similar (incompatible) dislocation-based approach [117]. Figure 18 depicts the Cauchy-like stress tensor within classical, dislocation-based gradient, and dislocation-based nonlocal frameworks. The gradient parameter is taken as $l = 0.2a$. It is noted that since the non-classical traction-free boundary condition is employed, it is the traction vector which vanishes at the crack faces, whereas the stress component $\sigma_{yz}$ is non-zero there (Fig. 18).

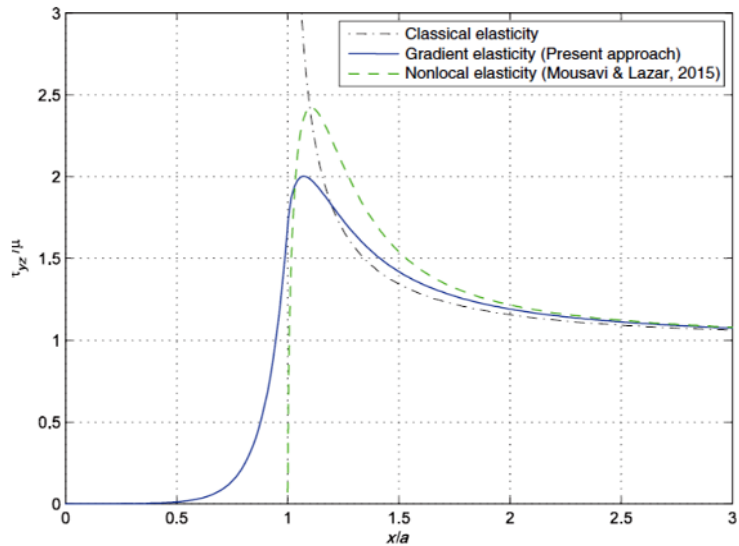

*Figure 18: Cauchy-like stress field around the crack tip, l = 0.2a, $\sigma_{yz}^{\infty} = \mu$.*

This result for the full-field non-singular crack solution for mode III, can be contrasted with the corresponding asymptotic non-singular solution given, for example, in derived in [31,52] which reads

$$\sigma_{yz}^{crack}\left(x\right)=\frac{K_{III}}{\sqrt{2\pi r}}\left(1-e^{-r/l}\right),\tag{4.3.4}$$

where $K_{III}\left(=\sigma_{yz}^{\infty}\sqrt{\pi a}\right)$ is the stress intensity factor, and $r=\left|x-a\right|$ denotes the polar coordinates centered at the crack tip. Figure 19 depicts the asymptotic stress component $\sigma_{yz}$, along with its classical counterpart ($l$=0).

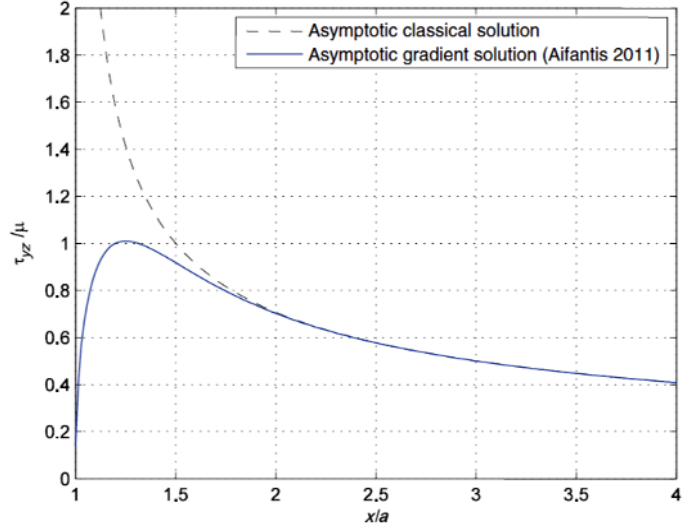

**Figure 19:** $\sigma_{yz}$ around the crack tip, $l = 0.2a$.

The only non-zero double stress component along the crack line, i.e. $\sigma_{zyx}$ is

$$\sigma_{zyx}^{crack}(x,y)=\frac{a\mu}{2\pi}\int_{-1}^{1}\left(-2\frac{l^2}{X^2}+K_0\left(\frac{\left|X\right|}{l}\right)+2\frac{l}{R}K_1\left(\frac{\left|X\right|}{l}\right)\right)B_z\mathrm{d}t\,,\tag{4.3.5}$$

where $X=x-at$. The other double stress components vanish on the $x$-axis. As expected, the double stress component is singular at the crack tip.

To shed light on the effect of the non-standard (non-classical) terms in the traction within the gradient elasticity theory, the standard (classical) traction-free condition is examined here, i.e.

$$t_z=\sigma_{yz}=0\tag{4.3.6}$$

On using the condition given by Eq. (4.3.6) in the gradient theory, the kernel $K^t$ is simplified to

$$K^t(s,t)=\frac{\sigma_{yz}(X,Y)}{b_z}=\frac{\mu}{2\pi}\frac{X}{R^2}\left[1-\frac{R}{l}K_1\left(R/l\right)\right]\tag{4.3.7}$$

Such approximation was also used previously in [118]. Using the kernel of Eq. (4.3.7), the dislocation density is determined and is compared to the one derived from the non-classical traction boundary condition in Fig. 20. It is observed that at the vicinity of the right (left) crack tips, the dislocation density for the non-classical traction-free boundary condition approaches a positive (negative) singularity, while the situation for the classical traction-free condition is vice versa. Finally, the stress components resulting from classical and nonclassical traction-free boundary conditions are compared in Fig. 21. It is observed that on using the classical traction-free condition in gradient elasticity, the stress field $\sigma_{yz}$ is zero along the crack-faces.

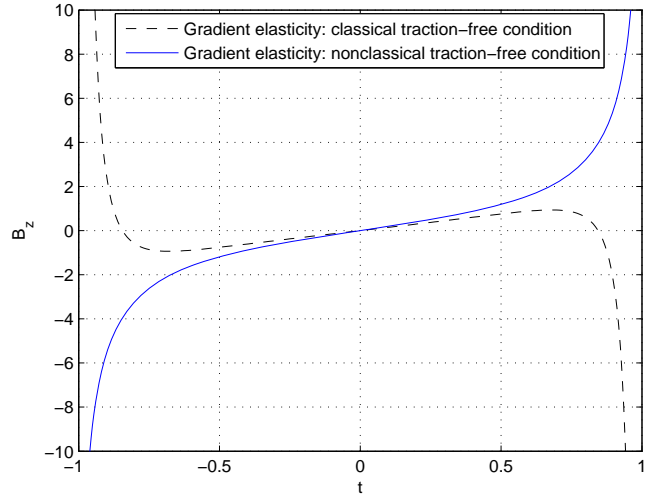

***Figue 20:*** *Dislocation density for standard and non-standard traction-free boundary conditions.*

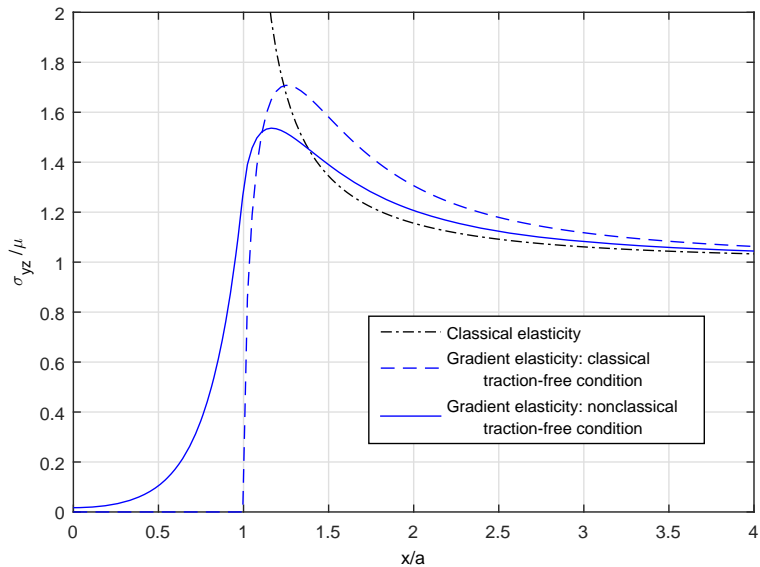

***Figure 21:*** *Stress $\sigma_{yz}$ for the standard and non-standard traction-free boundary conditions in gradient theory compared to the one in classical elasticity for l = 0.2a.*

### 4.4 Gradient Plasticity and Shear Instabilities: Size-Dependent Stability Diagrams

In this section we consider the implications of the ILG framework to the description of shear instabilities in deforming materials undergoing thermal and structural changes. A characteristic example that we focus upon is the technologically important class of bulk metallic glasses (BMGs). We examine the interplay between strain temperature and free volume internal lenghts (ILs) on the onset of instability, within a coupled deterministic ILG multiphysics framework. Then, attention is focused on finite domains and size-dependent stability diagrams are derived.

### 4.4.1 Shear Bands in BMGs for Infinite Domains

The model equations describing the evolution of deformation, temperature and free volume (internal variable) in the shear direction for the physical configuration depicted in Fig. 22 below

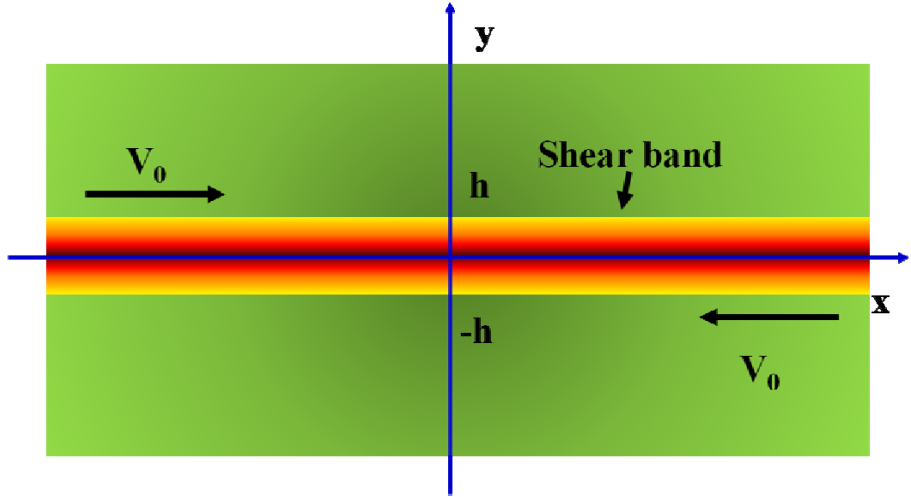

***Figure 22:*** *Illustration of shear banding under simple shear.*

read as follows:

$$\frac{\partial \tau}{\partial y} = 0 \;\; ; \;\; \rho c_v \frac{\partial \theta}{\partial t} = k \frac{\partial^2 \theta}{\partial y^2} + F(\xi, \theta, \tau) \;\; ; \;\; \frac{\partial \xi}{\partial t} = D \frac{\partial^2 \xi}{\partial y^2} + G(\xi, \theta, \tau) \, ,$$

$$\tau = \kappa(\gamma^p, \dot{\gamma}^p, \theta, \xi) - c \frac{\partial^2 \gamma^p}{\partial y^2} \;\; ; \;\; \gamma = \frac{\tau}{\mu} + \gamma^p \, . \tag{4.4.1}$$

The first three are balance equations for the stress (under quasistatic conditions), the internal energy, and the free volume internal variable. The fourth is a gradient-dependent constitutive equation for the flow stress, and the fifth is a standard linear decomposition of the total strain in its elastic and plastic components. Thus, $(\gamma, \dot{\gamma})$ denote the total strain and its rate, while $(\gamma^p, \dot{\gamma}^p)$ denote their plastic counterparts, and $\tau / \mu$ gives the elastic strain with $\mu$ being the elastic shear modulus. The internal variable $\xi$ that obeys the reaction-diffusion equation is a dimensionless measure of the free volume density. In particular, $\xi = \alpha \upsilon_f / \upsilon^*$, where $\upsilon^*$ is a critical volume (hard-sphere volume of an atom), $\upsilon_f$ is the average free volume per atom, and $\alpha$ is a geometrical factor of order 1. The parameters $(\rho, c_v, k, D)$ denote respectively the mass density, the specific heat, the thermal conductivity, and the free volume diffusion coefficient. The source terms $F(\xi, \theta, \tau)$ and $G(\xi, \theta, \tau)$ represent respectively the net generation rates of heat and free volume. Usually, the heat source term has the form $F(\xi, \theta, \tau) = \beta \tau \dot{\gamma}^p$, where $\beta$ is the so-called

Taylor-Quinney coefficient that represents the fraction of the rate of plastic work transformed into heat. Explicit expressions for $G(\xi,\theta,\tau)$ available in the literature are those proposed by Spaepen [119] and Johnson et al [120].

Let $(t_o,\theta_o,\tau_o)$ denote reference values for time, temperature and stress, respectively. These are introduced to normalize the governing equations, and their explicit expressions depend on the material parameters included in the constitutive equations for $\kappa(\gamma^p,\dot{\gamma}^p,\theta,\xi)$ and $G(\xi,\theta,\tau)$. Then, Eqs. (4.4.1) are written in a normalized form as

$$\frac{\partial\tilde{\tau}}{\partial y}=0 \; ; \; \frac{\partial\tilde{\theta}}{\partial\tilde{t}}=\ell_\theta^2\frac{\partial^2\tilde{\theta}}{\partial y^2}+\tilde{\beta}\tilde{\tau}\dot{\tilde{\gamma}}^p \; ; \; \frac{\partial\xi}{\partial\tilde{t}}=\ell_\xi^2\frac{\partial^2\xi}{\partial y^2}+\tilde{G}(\xi,\tilde{\theta},\tilde{\tau}), \tag{4.4.2}$$

$$\tilde{\tau}=\tilde{\kappa}(\gamma^p,\dot{\tilde{\gamma}}^p,\tilde{\theta},\xi)-\ell_\gamma^2\frac{\partial^2\gamma^p}{\partial y^2} \; ; \; \gamma=\frac{\tilde{\tau}}{\tilde{\mu}}+\gamma^p, \tag{4.4.3}$$

where the dimensionless variables are defined as follows: shear stress $\tilde{\tau}=\tau/\tau_o$; temperature $\tilde{\theta}=\theta/\theta_o$; time $\tilde{t}=t/t_o$; plastic shear strain rate $\dot{\tilde{\gamma}}^p=\dot{\gamma}^pt_o$, while $\tilde{\beta}=\beta\tau_o/\rho c_v\theta_o$; $\tilde{\mu}=\mu/\tau_o$; $\tilde{G}=Gt_o$; $\tilde{\kappa}=\kappa/\tau_o$. It is noted that the space variable $y$ remains not normalized, while the parameters $\ell_\theta=\sqrt{kt_o/\rho c_v}$, $\ell_\xi=\sqrt{Dt_o}$ and $\ell_\gamma=\sqrt{c/\tau_o}$ are length scales related with the presence of the three different gradient terms (heat diffusion, free volume diffusion, and strain viscoplasticity) in the constitutive equations. We also introduced the following normalized quantities: plastic strain hardening parameter $\tilde{h}=\partial\tilde{\kappa}/\partial\gamma^p\geq0$, plastic strain rate hardening parameter $\tilde{s}=\partial\tilde{\kappa}/\partial\dot{\tilde{\gamma}}^p>0$, thermal softening parameter $\tilde{\Phi}=-\partial\tilde{\kappa}/\partial\tilde{\theta}>0$, and the free-volume softening parameter $\tilde{\Xi}=-\partial\tilde{\kappa}/\partial\xi>0$. Since $\tilde{G}(\xi,\tilde{\theta},\tilde{\tau})$ is a function of $\xi,\tilde{\theta}$ and $\tilde{\tau}$, we can also define the following parameters: $\tilde{G}_\xi=\partial\tilde{G}/\partial\xi$, $\tilde{G}_{\tilde{\theta}}=\partial\tilde{G}/\partial\tilde{\theta}$, and $\tilde{G}_{\tilde{\tau}}=\partial\tilde{G}/\partial\tilde{\tau}$.

Next, we seek solutions of Eqs.(4.4.2) and (4.4.3) that can be expressed by the time-dependent homogeneous solution $\left(\gamma_h,\gamma_h^p,\tilde{\theta}_h,\xi_h,\tilde{\tau}_h\right)$ and their perturbations $\left(\delta\gamma,\delta\gamma^p,\delta\tilde{\theta},\delta\xi,\delta\tilde{\tau}\right)$ in the form

$$\left(\delta\gamma,\delta\gamma^p,\delta\tilde{\theta},\delta\xi,\delta\tilde{\tau}\right)=\left(\hat{\gamma},\hat{\gamma}^p,\hat{\theta},\hat{\xi},\hat{\tau}\right)\exp(\omega\tilde{t}+iqy) \tag{4.4.4}$$

where $\left(\hat{\gamma},\hat{\gamma}^p,\hat{\theta},\hat{\xi},\hat{\tau}\right)$ are small constants that characterize the initial magnitude of the perturbations, $q$ is the corresponding wave number and $\omega$ denotes the normalized initial rate of growth. Introducing Eqs.(4.4.4) into Eqs.(4.4.2) and (4.4.3), taking into account only terms that are of first order in $\left(\delta\gamma, \delta\gamma^p, \delta\tilde{\theta}, \delta\xi, \delta\tilde{\tau}\right)$, and solving $\hat{\gamma}, \hat{\tau}$ in terms of $\hat{\gamma}^p, \hat{\theta}, \hat{\xi}$, we obtain a set of linear equations of the form $\mathbf{MX} = \mathbf{0}$, with

$$
\mathbf{M} = \begin{bmatrix}
\tilde{s}\omega + \tilde{h} + \ell_\gamma^2 q^2 & -\tilde{\Phi} & -\tilde{\Xi} \\
\tilde{\beta}\tilde{\tau}_h\omega + \tilde{\beta}\tilde{\gamma}_h^p\left(\tilde{s}\omega + \tilde{h} + \ell_\gamma^2 q^2\right) & -\left(\omega + \ell_\theta^2 q^2 + \tilde{\beta}\tilde{\gamma}_h^p\tilde{\Phi}\right) & -\tilde{\beta}\tilde{\Xi}\tilde{\gamma}_h^p \\
\tilde{G}_{\tilde{\tau}}\left(\tilde{s}\omega + \tilde{h} + \ell_\gamma^2 q^2\right) & \tilde{G}_{\tilde{\theta}} - \tilde{G}_{\tilde{\tau}}\tilde{\Phi} & \tilde{G}_\xi - \tilde{G}_{\tilde{\tau}}\tilde{\Xi} - \omega - \ell_\xi^2 q^2
\end{bmatrix} \quad (4.4.5)
$$

with $\mathbf{X} = \begin{bmatrix} \hat{\gamma}^p, & \hat{\theta}, & \hat{\xi} \end{bmatrix}^T$. For a non-trivial solution, the determinant of matrix $\mathbf{M}$ vanishes, leading to the following spectral or dispersion equation for the initial growth rate $\omega$ of the perturbation

$$
\omega^3 + a_2\omega^2 + a_1\omega + a_0 = 0 , \quad (4.4.6)
$$

with the $a$'s defined by the relations

$$
\left.
\begin{aligned}
a_2 &= \frac{q^2\left[\ell_\gamma^2 + \tilde{s}(\ell_\theta^2 + \ell_\xi^2)\right] + \tilde{h} - \tilde{\beta}\tilde{\Phi}\tilde{\tau}_h - \tilde{s}\tilde{G}_\xi}{\tilde{s}}, \\
a_1 &= \frac{\left(q^2\ell_\xi^2 - \tilde{G}_\xi\right)\left[q^2\left(\ell_\gamma^2 + \tilde{s}\ell_\theta^2\right) + \tilde{h} - \tilde{\beta}\tilde{\Phi}\tilde{\tau}_h\right] + q^2\ell_\theta^2\left(\tilde{h} + q^2\ell_\gamma^2\right) - \tilde{\beta}\tilde{\Xi}\tilde{G}_{\tilde{\theta}}\tilde{\tau}_h}{\tilde{s}}, \\
a_0 &= \frac{q^2\ell_\theta^2\left(\tilde{h} + q^2\ell_\gamma^2\right)\left(q^2\ell_\xi^2 - \tilde{G}_\xi\right)}{\tilde{s}}.
\end{aligned}
\right\} \quad (4.4.7)
$$

According to the well-known Routh-Hurwitz (R-H) criterion, all the roots of this polynomial equation have negative real part (indicating stable deformation), if and only if the following conditions hold true

$$
a_2 > 0, \ a_0 > 0 \ , \ a_2 a_1 - a_0 > 0, \quad (4.4.8)
$$

which, in turn, give the inequalities

$$
q^2 > \frac{\tilde{\beta}\tilde{\Phi}\tilde{\tau}_h + \tilde{s}\tilde{G}_\xi - \tilde{h}}{\left[\ell_\gamma^2 + \tilde{s}(\ell_\theta^2 + \ell_\xi^2)\right]}, \ q^2 > \frac{\tilde{G}_\xi}{\ell_\xi^2}, \ b_3 q^6 + b_2 q^4 + b_1 q^2 + b_0 > 0 , \quad (4.4.9)
$$

where $b_3 = \left(\ell_\gamma^2 + \tilde{s}\ell_\theta^2\right)\left(\ell_\theta^2 + \ell_\xi^2\right)\left(\ell_\gamma^2 + \tilde{s}\ell_\xi^2\right) > 0$, while the coefficients $b_2, b_1, b_0$ are not given here due to their lengthy form.

The violation of the second condition in Eqs.(4.4.9) corresponds to a Turing instability (a real eigenvalue crosses zero at marginal stability, i.e. $\omega_c = 0$ with $q_c \neq 0$). In the present case we have $q_c = \sqrt{\tilde{G}_\xi} / \ell_\xi$ and the corresponding symmetry-breaking instability has the characteristic length $l_c = 2\pi / q_c = 2\pi\ell_\xi / \sqrt{\tilde{G}_\xi}$. Moreover, in the original (dimensional) variables, Eq.(4.4.9)$_2$ is equivalent to $t_D < t_G$ , where $t_D = 1 / Dq^2$ is a characteristic time for the free volume perturbation diffusion, and $t_D = 1/G_\xi$ is the characteristic time for the free volume coalescence. Accordingly, Eq.(4.4.9)$_2$ expresses the competition between these two processes. When this condition is violated (i.e. $t_D > t_G$), the coalescence process is faster than diffusion and the perturbation will grow leading to the instability of the uniform deformation. In this connection, it is noted that the same instability criterion also appears when isothermal conditions are considered. As discussed in [121], the same holds also true when the strain gradient term is not included in the flow stress and the inertia effects are not negligible. The violation of the third condition in Eqs.(4.4.9) is associated, in general, to a pair of complex conjugate eigenvalues crossing the imaginary axis at criticality, i.e. $\operatorname{Re}\omega_c = 0$, $\operatorname{Im}\omega_c = \sqrt{a_1} \neq 0$ corresponding to another critical $q_c \neq 0$. This suggests the emergence of a time symmetry-breaking instability representing a Hopf bifurcation.

### 4.4.2 Finite Domains and Size Effects

As it may be deduced from Eqs.(4.4.9), there is a maximum wave number $q_{max}$, such that the homogeneous solution is stable for $q > q_{max}$. This, in turn, implies a size effect. In particular, for finite domains (e.g. for a layer of thickness $L$), the boundary conditions will constrain the set of allowable wave numbers $q$ to $q \sim 1/L$. Considering, for example, that adiabatic conditions and no diffusion flux of the free volume prevail on the boundary along with zero strain gradients, it can be easily shown that the dispersion relation given by Eqs. (4.4.6) and (4.4.7) still holds, but with $q = n\pi/L$. This implies that the homogeneous deformation remains stable to any perturbation when the specimen size is smaller than $L_{min} = \pi/q_{max}$.

For illustration purposes of the aforementioned remarks, we use the following explicit expressions

$$G(\xi,\theta,\tau) = \frac{1}{\alpha} f \exp\left(-\frac{\Delta G^m}{k_B\theta} - \frac{1}{\xi}\right)\left\{\frac{2k_B\theta}{\upsilon^* S \xi}\left[\cosh\left(\frac{\tau\Omega}{2k_B\theta}\right) - 1\right] - \frac{1}{n_D}\right\}, \qquad (4.4.10)$$

$$\kappa(\dot{\gamma}^p,\theta,\xi) = \frac{2k_B\theta}{\Omega}\operatorname{arcsinh}\left[\frac{\dot{\gamma}^p}{2f}\exp\left(\frac{\Delta G^m}{k_B\theta} + \frac{1}{\xi}\right)\right], \qquad (4.4.11)$$

which were introduced by Spaepen [119]: $f$ is the frequency of atomic vibration (~Debye frequency), $\Delta G^m$ is the activation energy of an atom jump, $k_B$ is the Boltzmann constant, $S = 2\mu(1+v)/3(1-v)$ with $v$ the Poisson's ratio, $\Omega$ is the atomic volume, and $n_D$ is the number of atomic jumps necessary to annihilate a free volume equal to $\upsilon^*$ and is usually taken to be 3-10 [122]. It is noted that the homogeneous part of the flow stress may be taken not to depend explicitly on the plastic strain $\gamma^p$, since a typical (compressive) stress–strain curve usually consists of an elastic part followed by yielding and perfect plasticity. While such an absence of plastic strain hardening $\left( \tilde{h} = 0 \right)$ may be attributed to the lack of dislocation plasticity in amorphous materials, there have been some recent experiments suggesting this may not be the always the case (see, for example, [123] and references quoted therein).

Suitable values of the material constants are obtained from the literature [124] and they are given in Table 1 below for a typical $Zr_{41.2}Ti_{13.8}Cu_{12.5}Ni_{10}Be_{22.5}$ bulk metallic glass (a.ka. Vitreloy 1).

**Table 1:** Material Constants Values Used

| $\rho$ | $c_v$ | $k$ | $\beta$ | $D$ | $\mu$ | $v$ | $f$ | $\Delta G^m$ | $\Omega$ | $\overset{*}{v}$ | $n_D$ | $\alpha$ |
|---|---|---|---|---|---|---|---|---|---|---|---|---|
| 6125 kg/m³ | 400 J/(kg K) | 20 W/(m K) | 0.9 | $10^{-16}$ m²/s | 35.3 GPa | 0.36 | $\sim 10^{13}$ s⁻¹ | 0.2–0.5 eV | 20 Å³ | 1.41 Å³ | 3 | 0.15 |

For calculation purposes, we also assume a value for the gradient coefficient $c = 5$ N and compute numerically the homogeneous solution for the stress, temperature, free volume, and plastic strain as a function of the applied shear strain by using Eqs. (4.4.2) and (4.4.3). The details are given in [125], where the instability regime of applied shear vs. the wavelength of the fluctuation is obtained.

For a specimen of thickness $L$, the wave numbers can take only discrete values of the form $q = n\pi / L$. Substituting this relation in Eqs.(4.4.9) and demanding the resulting inequalities to hold for all q's, the following stability diagram of Fig. 23 is obtained (for details the reader is referred to [125]).

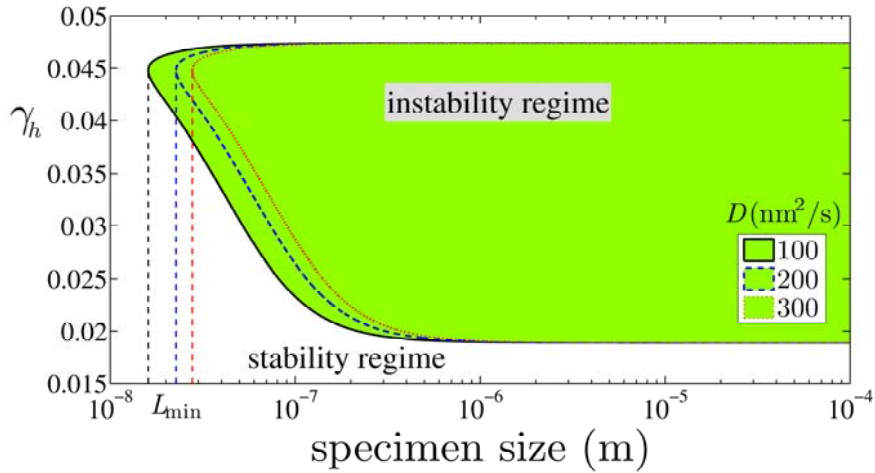

**Figure 23:** *Size effects on the stability-instability regimes of the homogeneous deformation for various values of free volume diffusion coefficient.*

It is seen that smaller specimens can support larger stable homogeneous deformations even in the softening regime (i.e. for $\gamma_h > 0.04$). This effect is more pronounced as the value of $D$ increases, i.e. greater values of $D$ lead to larger stable homogeneous strains. Moreover, for layers smaller than a critical size $L_{\min} \left(= \pi / q_{\max} \right)$, homogeneous deformation remains stable to any perturbation, i.e. the instability onset is suppressed. From a physical point of view this suggests that nanometer-scale glasses are too thin to accommodate a shear band of critical thickness. This response is in accordance with experimental results reported in the literature (see, for example, [126] and references quoted therein).

### *4.5 Combined Gradient-Stochastic Models and Size Effecst in Micro/Nanopillars*

In the previous sections gradient elasticity and gradient plasticity multiphysics problems are considered within an ILG deterministic framework. The effect of stochasticity due to randomly evolving micro/nanostructures was not shown, and the corresponding effect on local instabilities and size-dependence was not illustrated. To account for such effects, combined gradient-stochastic models should be employed. An initial effort along this direction is outlined below to interpret size-dependent serrated stress-strain curves in micro/nano pillars (Section 4.5.1). Statistical characterization of such serrated stress-strain curves is not always possible with standard analyses based on Boltzmann-Gibbs-Shannon (B-G-S) entropy thermodynamics, and corresponding power-law expressions cannot interpret the measured probability density functions (PDFs). In contrast, Tsallis q-statistics based on non-extensive entropy thermodynamis can sufficiently be employed as shown below (Section 4.5.2).

### *4.5.1 Stochasticity-enhanced Gradient Plasticity Model*

For a material characterized by linear hardening after yielding, the following simple gradient plasticity model may be employed [6e, 63a]

$$\overline{\sigma} = \begin{cases} E\varepsilon, & \text{for } \varepsilon \leq \sigma^{y}/E, \\ \sigma^{y} + \beta\varepsilon^{p} - \beta\ell^{2}\dfrac{\partial^{2}\varepsilon^{p}}{\partial x^{2}}, & \text{for } \varepsilon > \sigma^{y}/E, \end{cases} \tag{4.5.1}$$

where $\varepsilon^{p}$ is the plastic strain, $\beta$ is the hardening modulus, $\overline{\sigma}$ and $\sigma^{y}$ are the applied and yield stress, respectively, and $\ell$ is the corresponding internal length.

The discretized version of the constitutive model of Eq. (4.5.1) may then be implemented in a cellular automaton (CA) grid where stochastic material heterogeneity enters the formulation in the form of a fluctuating yield stress. Each cell's yield stress is assumed to follow a specific distribution with a certain mean and variance. In earlier and recent work [127-132] the local (cell) yield stresses was assumed to follow a Weibull distribution, the statistical characteristics of which were either measured experimentally or properly fitted to fit the experimental data. In particular, the local yield stress was assumed to be of the form $\sigma_{w}^{y} = (1+\zeta)\sigma^{y}$, with $\zeta$ being a Weibull distributed random variable. When applying Weibull statistics, the probability density function is defined as

$$\mathrm{PDF}[\delta] = \frac{\kappa}{\lambda}\left(\frac{\delta}{\lambda}\right)^{\kappa-1} e^{-(\delta/\lambda)^\kappa}; \quad \overline{\delta} = \lambda\Gamma\left[1+(1/\kappa)\right], \quad \left\langle\delta^2\right\rangle = \lambda^2\Gamma\left[1+(2/\kappa)\right] - \overline{\delta}^2, \quad (4.5.2)$$

where $\kappa$, $\lambda$ are the so-called shape (or Weibull modulus) and scale parameters of the Weibull distribution, $\overline{\delta}$ is the mean value and $\left\langle\delta^2\right\rangle$ is the variance; $\Gamma$ denotes the usual gamma function.

The CA simulations may be conducted for either load/stress or displacement/strain controlled conditions. Strain-controlled simulations can be performed for modeling stress drops, while stress controlled simulations for modeling strain bursts. The system is slowly loaded by increasing the external driving force (stress/load) from zero to a certain value in small steps. During the simulation a cell is considered "unstable" when the following inequality holds

$$\overline{\sigma} + \sigma_{int} > \sigma_w^y ; \qquad \sigma_{int} = \beta\ell^2\frac{\partial^2\varepsilon^p}{\partial x^2} - \beta\varepsilon^p, \qquad (4.5.3)$$

i.e. when the local (external plus internal) stress exceeded the local yield stress giving rise to instabilities. The local strain at all unstable sites is then increased by a small amount and new internal stresses $\sigma_{int}$ are computed for all cells, with the validity of the inequality in Eq. (4.5.3) being checked again. This process is repeated until a stable configuration is reached, i.e. there are no "unstable" cells. Then the external driving force is increased again, and so on, until a certain criterion is met, e.g. the total strain reaching a certain value, or an avalanche occurs and all cells become "unstable".

Experimental load-displacement curves of $Zr_{50}Ti_{16.5}Cu_{15}Ni_{18.5}$ pillars with diameters of 112nm and 410nm, and a 3:1 height-to-diameter ratio [133], were modeled [132] using cellular automata implementation of the stochasticity-enhanced gradient plasticity model given in Eq. (4.5.1). The comparison between the experimental and simulated results are shown in Fig. 24, and the parameters used are given in Table 2.

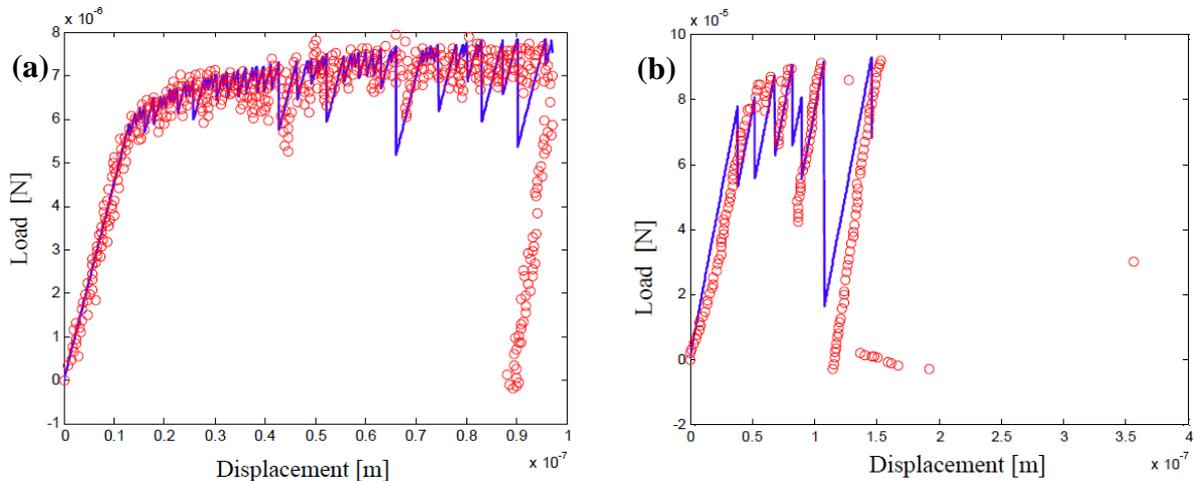

***Figure 24.*** *Comparison between simulations (solid line) with experimental data (dots) for $Zr_{50}Ti_{16.5}Cu_{15}Ni_{18.5}$ pillars with diameters of **(a)** 112nm, and **(b)** 410nm (Reprinted from [132] with permission from Elsevier).*

**Table 2**

| | | | | | |
|---|---|---|---|---|---|
| **Modeling BMG Nanopillars with various diameters** | | | | | |
| Diameter (nm) | $\sigma^y$ (MPa) | $\beta$ (MPa) | E (GPa) | $\ell$ (nm) | ($\lambda$, $\kappa$) |
| 112 | 800 | 500 | 92 | 8.4 | (1.06, 10) |
| 410 | 1000 | 500 | 92 | 8.4 | (1.06, 10) |

For all cases for which the initial heterogeneity statistics were not known [127-129, 131-132] many simulations were performed with different Weibull characteristics (mean and variance), and only the ones approximating the experimental measurements better were kept. This was not the case in [130] where the local yield stress distribution was experimentally measured and its characteristics (which were indeed fitted very accurately by a Weibull distribution) were directly used in the cellular automaton implementation. It follows that experimental or theoretical model information on the distribution characteristics of the local yield stresses is desired for the implementation of the proposed formulation through the above CA procedure.

An effective approach for deducing theoretical expressions for PDFs from a statistical analysis of the experimental data is to employ Tsallis q-generalization of thermodynamics. The q-entropy $S_q$ proposed by Tsallis [29] reads

$$S_q = k \frac{1 - \sum_{i=1}^{W} p_i^q}{q - 1},$$ (4.5.5)

where W is the total number of microstates of the system, $p_i$ are the occupation probabilities and $q$ is a real parameter. The standard Boltzmann-Gibbs-Shannon (B-G-S) entropy is recovered in the limit $q \rightarrow 1$. Based on Eq. (4.5.5), q-exponential, q-Gaussian and q-Weibull distributions can be generated for the PDFs. Such PDFs can be obtained, for example, from solutions of the differential equation

$$\frac{dy}{dx} = py^q.$$ (4.5.6)

When p is constant, the solution of Eq. (4.5.6) is a q-exponential, whereas when if $p \propto x$, the solution is a q-Gaussian. If y is identified with the cumulative distribution function (CDF), and $p \propto x^r$, a q-Weibull CDF is generated. For an exponential relaxation, $\xi$ (~CDF) is obtained as a solution of

$$\frac{d\xi}{dt} = -\lambda \xi \Rightarrow \xi = e^{-\lambda t}; \quad \lambda \geq 0 \ \text{(Lyapunov exponent)},$$ (4.5.7)

and if some fractality is involved, Eq. (4.5.7) is replaced by

$$\frac{d\xi}{dt} = -\lambda_q \xi^q \Rightarrow \xi = \left[1 + (q-1)\lambda_q t\right]^{\frac{1}{1-q}},$$ (4.5.8)

i.e. a q-exponential distribution is generated. The implications of Eq. (4.5.8) in interpreting statistical aspects of plastic deformation in micro/nano pillars and modeling size-dependent stress-strain curves are considered in the next subsection.

### 4.5.2 Analysis of Heterogeneity and Size-dependence through Tsallis q-Statistics

Molybdenum micropillar compression experiments are reported in [134] with pillar diameters ranging between 160 nm to 5.5 μm. The deformation is characterized by an irregular sequence of strain bursts (Fig. 25a) manifesting themselves as steps in the stress-strain curves. The statistics of these strain bursts have been calculated and fitted through a usual power-law relation in [134], as shown in Fig. 25b which, however, does not fit the whole burst size range.

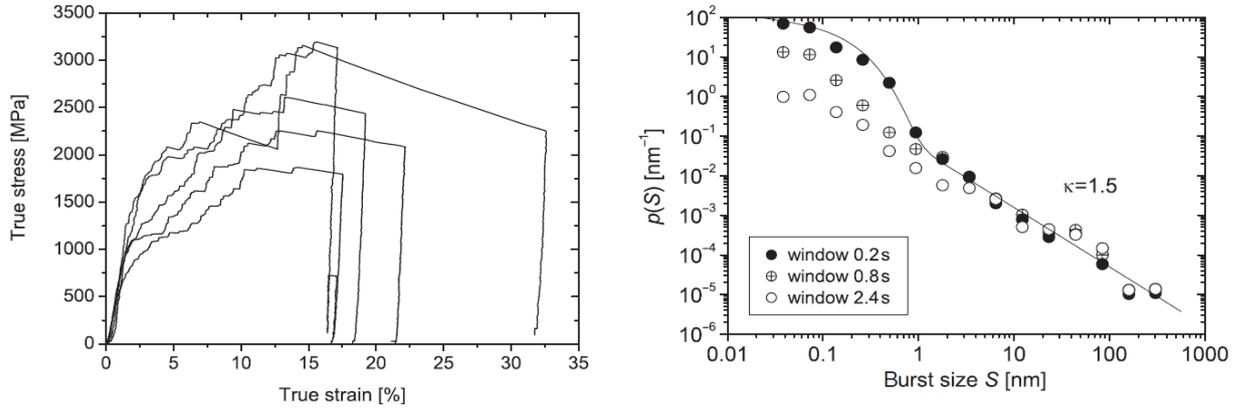

**Figure 25: (a)** *Stress-strain response showing an irregular sequence of strain bursts.* **(b)** *Distribution of strain bursts of size s fitted by a power law (reprinted from [135] with permission from Elsevier).*

In contrast, the whole range of the experimental curve is modeled by using the following generalization of the q-exponential distribution of Eq. (4.5.8) of the form

$$p(s) = A\left[1 + (q-1)Bs\right]^{\frac{1}{1-q}}, \qquad (4.5.9)$$

where $q$ is the q-index and $(A, B)$ are fitting parameters (Fig. 26a). The values of the fitting parameters for Fig. 26a are $A = 1.83745$, $B = 12.031 \, \text{nm}^{-1}$, $q = 1.65$ (with a standard error of 0.5097 and 6.31137 for $A$ and $B$, respectively).

Next, the assumption is made that a material point yields when the stress-strain response departs from the linear behavior, i.e. at the end of the elastic region. In this case the local yield strength $\sigma_y^{loc}$ is related to the local strain at yield $\varepsilon_y^{loc}$ through the relation [135]

$$\sigma_y^{loc} = E\varepsilon_y^{loc}, \qquad (4.5.10)$$

where $E$ is, as usually, the modulus of elasticity. Moreover, the micropillars are divided in a number of slices with size $\ell$, i.e. equal to the internal length of the material. Then, the strain burst sizes $s$ shown in Fig. 25a correspond to the simultaneous or consecutive yielding of an integer number of elements at the same external stress level. It is assumed that the smallest burst size $s^b$ corresponds to the plastic strain of a single element, i.e. $s^b = \varepsilon_y^b \ell = \left(\sigma_y^b / E\right)\ell$, with $\sigma_y^b$ denoting the yield stress of the bulk material. A strain burst of size $s$ is then assumed to be given by $s = n s_b$, where the number $n$ of simultaneously yielding elements is equal to

$$n = int\left[\frac{s}{s_b}\right] = int\left[\frac{\ell \varepsilon_y}{\ell \varepsilon_y^b}\right] = int\left[\frac{\sigma_y/E}{\sigma_y^b/E}\right] = int\left[\frac{\sigma_y}{\sigma_y^b}\right], \qquad (4.5.11)$$

where *int* denotes, as usual, the integer part of the ratio. It may then be argued that the probability of having a burst of size $s$ is equal to the probability of having elements with yield stress equal to $\sigma_y$, i.e.

$$p\left(\sigma_y\right) \equiv p(s) = A\left[1 + (q-1)Bns_b\right]^{\frac{1}{1-q}} = A\left[1 + (q-1)B\ell\frac{\sigma_y}{E}\right]^{1/(1-q)}, \qquad (4.5.12)$$

or, since $E/\ell = \sigma_y^b / s^b$, the above equation takes the form

$$p\left(\sigma_y\right) = A\left[1 + (q-1)Bs^b\left(\frac{\sigma_y}{\sigma_y^b}\right)\right]^{1/(1-q)}. \qquad (4.5.13)$$

Using the fitting parameters $A = 1.83745$, $B = 12.031$ nm$^{-1}$, $q = 1.65$, along with the bulk yield stress and strain burst values $\sigma_y^b = 1.1$ GPa and $s^b = 0.03$ nm (deduced from an inspection of the experimental stress-strain curves) the cumulative distribution function of the yield stress for the data reported in [134] takes the form

$$p\left(\sigma_y\right) = 1.83\left[1 + 0.21\,\sigma_y\right]^{-1.538}, \qquad (4.5.14)$$

and the corresponding plot is depicted in Fig. 26b.

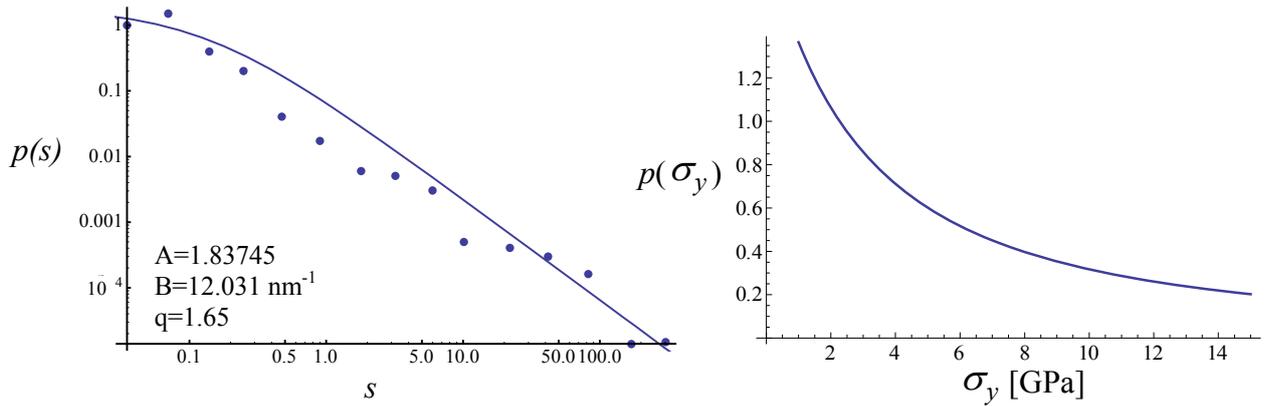

**Figure 26: (a)** *Distribution of strain bursts of size s fitted by the q-exponential distribution of Eq. (4.5.9).* **(b)** *Theoretically deduced yield stress probability distribution (reprinted from [135] with the permission of AIP Publishing).*

By using the statistical characteristics of the probability distribution of Eq. (4.5.14), more precise and physically-based results can be obtained through the cellular automata simulations. To this end the inverse transform sampling [136] method is used for choosing random values from the probability distribution of Eq. (4.5.14), which involves computing the quantile function of the distribution, i.e. computing the cumulative distribution function (CDF) of the distribution (which maps a number in the domain to a probability between 0 and 1) and then inverting that function. Next, by implementing numerically the stochasticity-enhanced gradient plasticity model of Eq. (4.5.1), the stress-strain curves can be obtained, as shown in Fig. 27. It can be seen that these simulation results approximate both qualitatively and quantitatively the experimental measurements, capturing both the size dependence on the micropillar diameter, as well as the observed strain bursts.

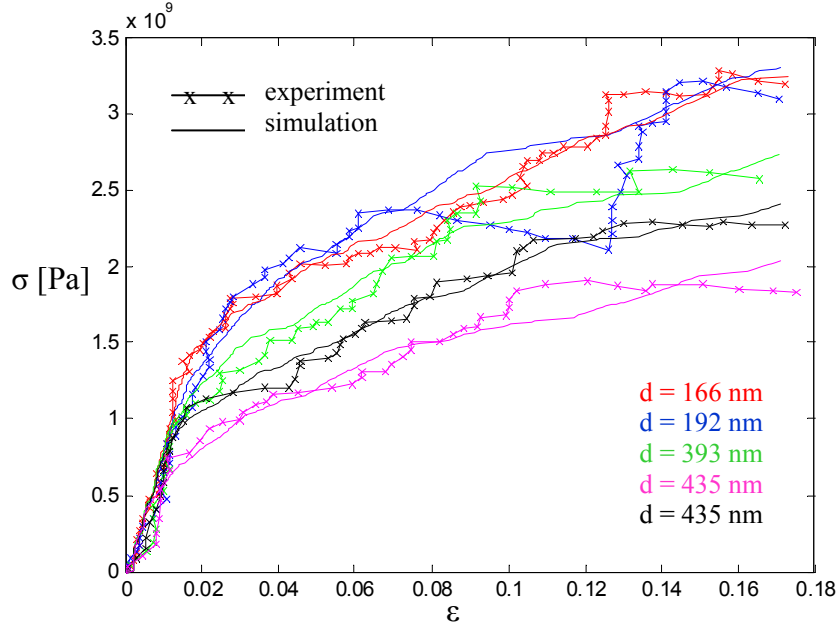

**Figure 27:** *Comparison between experimental and simulation results on the stress-strain behavior of Mo-micropillars with different diameters (d) (reprinted from [135] with the permission of AIP Publishing).*

We conclude this section by elaborating on a 2D counterpart of Eq. (4.5.1) assuming again random variations of cell yield stress according to Weibull distributed random variables with mean $\langle\sigma_y\rangle$. We performed CA simulations where the external stress increased from 0 to 70 MPa while cells yield when

$$\sigma_{EXT} - \beta\varepsilon^p + \beta\ell^2\left(\frac{\partial^2\varepsilon^p}{\partial x^2} + \frac{\partial^2\varepsilon^p}{\partial y^2}\right) > \sigma_y \Rightarrow \sigma_{EXT} + \sigma_{INT} > \sigma_y \ . \tag{4.5.15}$$

A 30 × 30 grid was used with cell dimension of 0,5 μm in order to model a specimen with dimensions 1.5 x 1.5 μm, with the corresponding internal or characteristic length $\ell$ was taken equal to 0.5 μm in both directions. We run the simulation for different values of the gradient coefficient $\beta\ell^2$. By measuring the number $s$ of cells yielding simultaneously we were able to obtain the cumulative distribution of avalanches, and approximate it with the use of Tsallis q-statistics, thus enabling the establishment of a relation between the gradient coefficient and the q-index. For the CA simulations we assumed that the specimens are compressed Ni micropillars for which the material parameters are known. The results are shown in Fig. 28 and the parameters used were E = 200 GPa, $\ell = 0.5\,\mu m$, and $\lambda = 7$ Å, while the interatomic energy of Ni was calculated as $U_{int} = c\lambda\left(\ell/\lambda\right)^2$, assuming that the force-like gradient coefficient c coincides with the interatomic force. For each value of gradient coefficient $c = \beta\ell^2$ we determine the corresponding value of q. The relation between c and q is shown in Fig. 29.

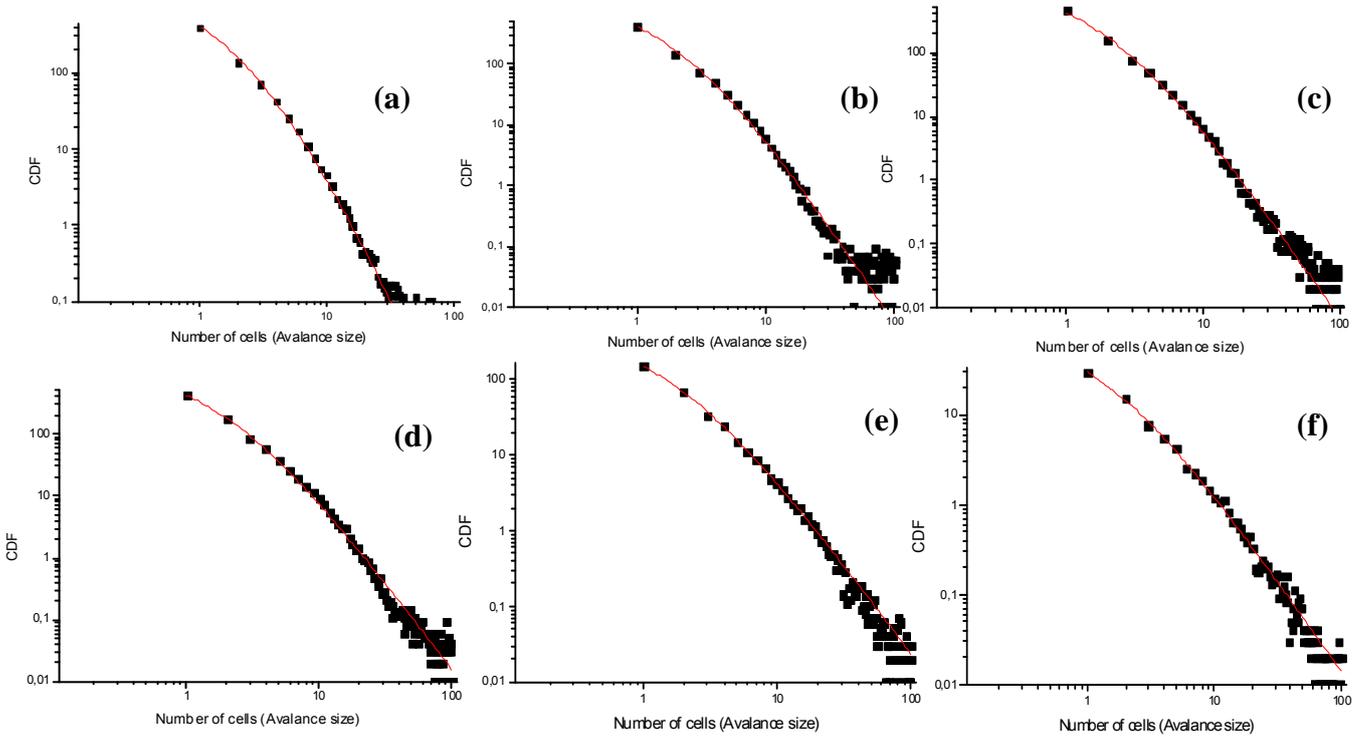

**Figure 28:** *Use of Tsallis q-statistics for modeling the dynamics of avalanches produced in simulations for (a) c = 0.5 mN, β = 2 GPa, q = 1.28, $U_{int}$= 4.475eV; (b) c = 0.75 mN, β = 3 GPa, q = 1.31, $U_{int}$= 6.562eV; (c) c = 0.875 mN, β = 3.5 GPa, q = 1.32, $U_{int}$= 7.656eV; (d) c = 1.75 mN, β = 7 GPa, q = 1.35, $U_{int}$= 6.562eV; (f) c = 3 mN, β = 12 GPa, q = 1.42, $U_{int}$= 26.25eV; (g) c = 5 mN, β = 20 GPa, q = 1.45, $U_{int}$= 43.75eV.*

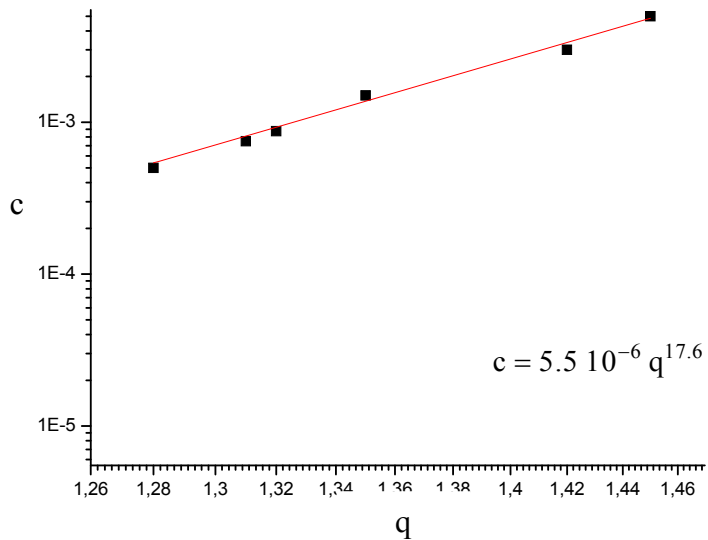

$c = 5.5 \ 10^{-6} \ q^{17.6}$

This initial result (see also [137]) providing a power law relation between the gradient coefficient and Tsallis q-index needs to be further explored, and such work is currently in progress [138]. For q = 1 we find a minimum value of the gradient coefficient (c = 5.5 μN) which indicates short-range weak interactions between Ni atoms (since the interatomic energy $U_{int} = 0.048 eV$ is much smaller than the cohesive energy of Ni). This suggests that at a macroscopic scale the force coefficient c is not large enough to deform plastically the cells of the Ni micropillars.

### 4.6  Further Considerations on Tsallis q-Statistics

As mentioned earlier, NC/UFG and BMG materials exhibit serrated stress-strain curves which can only be captured when a suitable "strain gage" or small-scale resolution device is employed. These serrations in the form of load drops (for constant strain rate tests) or displacement bursts (for constant stress rate tests) are attributed to local instabilities which also manifest through the emergence of multiple shear bands. In principle, the approach of nonlinear physics and dynamical systems may be employed to derive and solve partial differential equations including gradient and stochastic terms. This task, which remains a challenge for the future, may be facilitated by performing a statistical analysis of the available experimental data concerning both serration and shear band characteristics. Such type of analyses for BMGs is provided in this section by using a more sophisticated (than in the previous section) Tsallis q-statistics.

#### 4.6.1  Tsallis q-Statistics for Serrations

Here we analyze in detail the serrated stress-strain curve of $Zr_{64.13}Cu_{15.75}Ni_{10.12}Al_{10}$ [139], while a summary of the results concerning other MGs is given in Table 3. We begin with estimating the non-extensive statistics (i.e. Tsallis $q$-triplet, namely $q_{stat}, q_{sens}, q_{rel}$ [140]) of the stress time series, shown in Fig. 30a. We first remove the linear part and from the resultant serration time series ($s_t$) we remove subsequently the drift (trend) with a second order difference filter [141], namely $\Delta S_t = s_t - 2s_{t-1} + s_{t-2}$. The drift corrected stress time series are shown in Fig. 30b. In general, the Tsallis $q$-entropic indices can be estimated by using the probability density function (PDF) computed from the experimental data $X = \{s_t; t = 1,2,\ldots,N\}$. In particular, the best $q_{stat}$ value corresponds to the best linear fit (maximum correlation coefficient, $cc$) of the graph $\ln_q(p(s_i))$ vs $s_i^2$, where the function $\ln_q(s) = \dfrac{s^{1-q} - 1}{1-q}$ corresponds to the $q$-logarithm (inverse of the $q$-exponential). The statistical analysis is based on the algorithm described in [41].

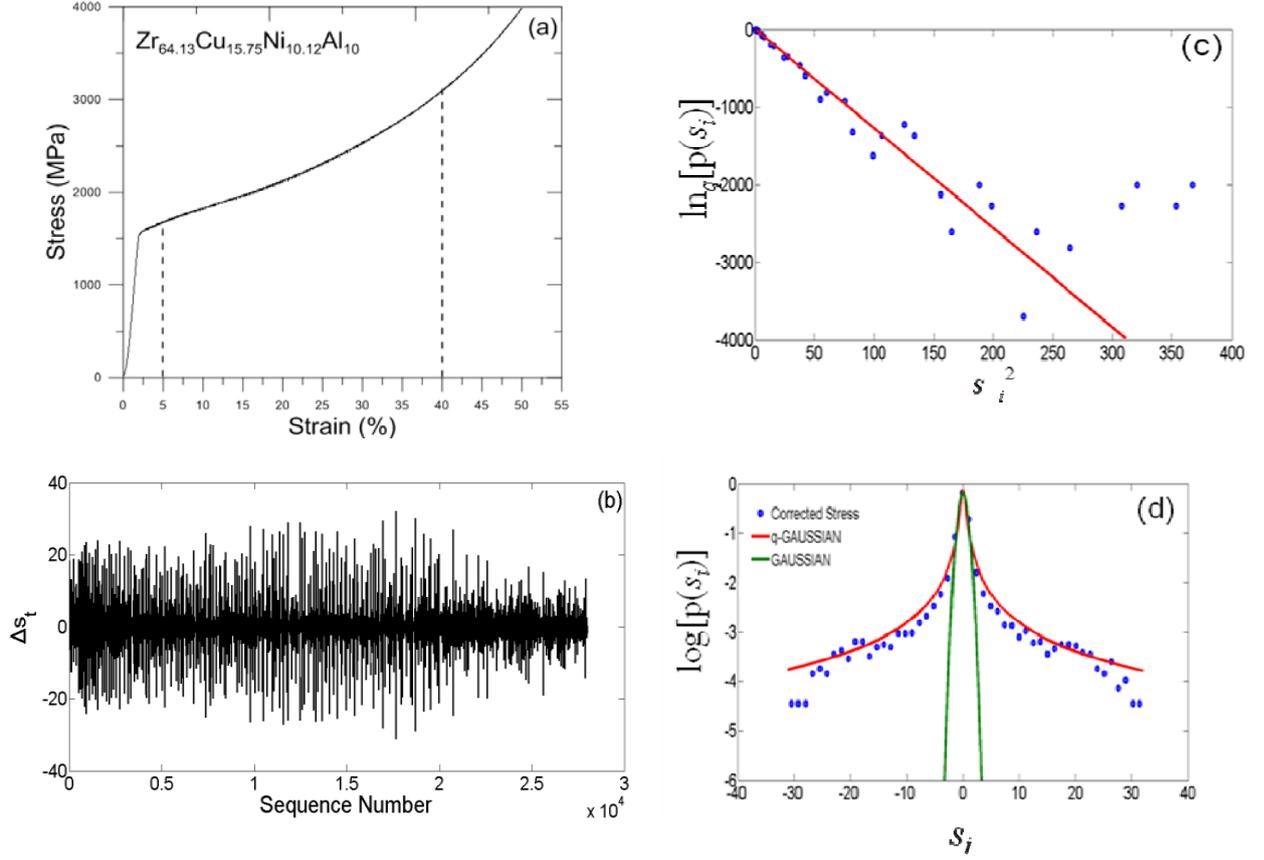

**Figure 30:** **a)** *Stress-strain curve for* $Cu_{47.5}Zr_{47.5}Al_5$ *at* $10^{-1}$ $s^{-1}$(blue). **b)** *Corrected drift stress signals* ($\Delta S$) *corresponding to blue stress-strain of* **(a)**. **c)** *Best linear correlation fitting between* $\ln_q[p(s_i)]$ *and* $(s_i)^2$ *for the detrended stress signal.* **d)** *$\log[(p(s_i)]$ vs $s_i$ for the detrended stress time series of curve A (blue circles), the theoretical q-Gaussian (blue line) and the normal Gaussian (green line).*

In Figs. 30c,d, we present the results for the corrected stress time series shown in Fig. 30b, while in Fig. 30c, we present the best linear correlation between $\ln_q[p(s_i)]$ (open blue circles) and $(s)_i^2$. The best fit was found for the value of $q_{stat} = 1.32 \pm 0.08$, and correlation coefficient (*cc*) $cc$=0.9545±0.034. This value was used to estimate the $q$-Gaussian distribution presented in Fig. 30d by the solid red line. The difference between the $q$-Gaussian and the Gaussian *PDF* (green line) in long tails is clearly shown, in a $\log[p(s_i)]$ vs $s_i$ graph.

Moreover, we estimated Tsallis $q_{sens}$ and $q_{rel}$ indices using the algorithms used in [142]. In particular, Tsallis $q_{sen}$ entropic index ($q$-sensitivity) is correlated with entropy production and is given by

$$q_{sen} = 1 + \frac{a_{max} a_{min}}{a_{max} - a_{min}} \qquad (4.6.1)$$

The $a_{max}, a_{min}$ values correspond to the extremes of multifractal spectrum for which $f(a) = 0$. In addition, the $q_{rel}$ index ($q$-relaxation) is given by $q_{rel} = (k-1)/k$, where $s$ is the slope of the log-log plot for the mutual information $I(\tau)$ given by the relation

$$I(\tau) = -\sum_{s(i)} p(s(i)) \log_2 p(s(i)) - \sum_{s(i-\tau)} p(s(i-\tau)) \log_2 p(s(i-\tau)) +$$
$$\sum_{s(i)} \sum_{s(i-\tau)} p(s(i), s(i-\tau)) \log_2 p(s(i), s(i-\tau))$$
(4.6.2)

The corresponding results are shown in Fig. 31.

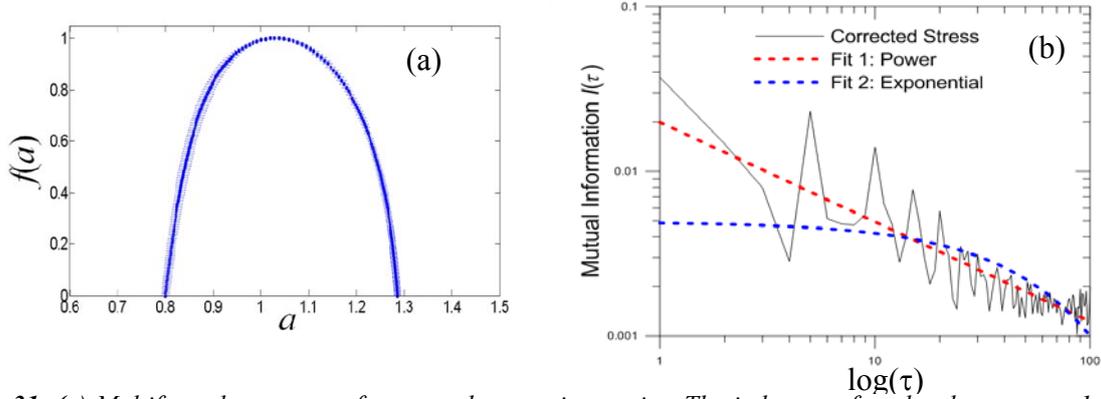

**Figure 31:** (a) *Multifractal spectrum of corrected stress time series. The index was found to be $q_{sens}$ = - 1.1134 ± 0.0324.* (b) *Double logarithmic plot of mutual Information $I(\tau)$ vs $\tau$. The $q_{rel}$ index was found to be $q_{rel}$ = 2.653.*

In particular, the estimation of the multifractal spectrum presented in Fig. 31a, along with the error bars, gives $a_{min} = 0.8 \pm 0.006$, $a_{max} = 1.286 \pm 0.00346$ and, therefore, $q_{sens} = -1.1134 \pm 0.0324 < 1$. In addition, in Fig. 31b we present the best log $I(\tau)$ fitting of the mutual information function for the corrected stress time series. The results suggest that the $q_{rel}$ index is $q_{rel} = 2.653$, indicating a $q_{rel}$ exponential decay relaxation of the system to meta-equilibrium non-extensive stationary states. The above results show that the system is in an off-equilibrium stationary state whose physics is properly described by Tsallis $q$-triplet with values: $\{q_{sens}; q_{rel}; q_{stat}\} = \{-1.1134; 2.653; 1.32\}$.

**Table 3**

| | $Zr_{64.13}Cu_{15.75}Ni_{10.12}Al_{10}$ | $Cu_{47.5}Zr_{47.5}Al_5$ | $Zr_{65}Cu_{15}Ni_{10}Al_{10}$ | $Zr_{52.5}Ti_5Cu_{17.9}Ni_{14.6}Al_{10}$ | $Zr_{51}Cu_{23.25}Ni_{13.5}Al_{12.25}$ |
|---|---|---|---|---|---|
| $q_{stat}$ | 1.338±0.15<br>cc=0.914±0.02 | 1.32±0.08<br>$cc = 0.95 \pm 0.034$ | 1.38±0.057<br>$cc = 0.95 \pm 0.02$ | 1.59±0.103<br>$cc = 0.91 \pm 0.035$ | 1.29±0.06<br>$cc = 0.9313 \pm 0.03$ |
| $q_{sens}$ | -1.0004 ± 0.047 | -0.0168 ± 0.049 | -0.033 ± 0.0013 | -0.36 ± 0.01 | -1.0697 ± 0.153 |
| $q_{rel}$ | 2.953 | 2.162 | 2.12 | 1.8 | 2.552 |

Similar analysis can be carried out for other BMGs [143] and the results for 5 MGs [144] are listed in Table 3. In all cases the Tsallis $q$-indices were found to be different from unity, verifying a possible general scheme, $q_{sens} < 1 < q_{stat} < q_{rel}$ as noted in [145], concluding that nonextensivity is clearly a characteristic of the systems under study.

### 4.6.2 Image Analysis of Multiple Shear Bands

Another aspect of BMGs is the occurrence of multiple shear bands. Figure 32 shows a series of SEM images of shear bands concerning the deformation of various BMGs [139, 146], for which corresponding image analysis is conducted herein to estimate three statistical indexes of the images; i.e. fractal dimension, lacunarity, and Tsallis $q_{sens}$ index.

For the estimation of the fractal dimension ($D$), we used an improved variation of box counting method (box merge), the same used in [141]. Scans are made to the data set with a box sized $1/s$=1/2, 1/4, ..., $1/s_{max}$ of the size of the box containing the data set. In each scan the number $n$ of the non-empty boxes is counted. The fractal dimension is then calculated from the slope of the linear part of the log($n$)-log($s$) plot, i.e. $D = \log(n) / \log(s)$. However, the use of a single fractal dimension has two deficiencies [147]: First, although the fractal dimension describes how much space is filled, this value does not indicate how the space is filled by the object. This can be confronted by measuring the "lacunarity" which is a parameter that describes the distribution of the sizes of gaps surrounding the object within the image. Greater lacunarity reflects a greater size distribution of the gaps and can also be used to distinguish objects with similar fractal dimension. It can be estimated from the relation $\Lambda_r(B) = \dfrac{E[(X_r^B(n))^2]}{(E[X_r^B(n)])^2}$, where $B$ is the binary image, $\Lambda$ is the lacunarity index, $r$ is the size of box, $n$ is the number of mass points (black pixels) falling into the box, and $X$ is the number of the boxes containing $n$ mass points. Lacunarity is routinely measured using a gliding box algorithm [147]. The second problem of standard fractal analysis (assuming that shear band networks can be described by a single fractal dimension alone) is associated with the fact that, in reality, shear band complex structures exist within subsets of regions having different scaling properties. Thus, the use of "multifractal analysis" provides additional information about the space filling properties than the fractal dimension $D$ alone. Multifractal analysis takes intensity variations by measuring pixel density within a box. One can measure the generalized fractal dimensions in order to construct the multifractal spectra by modifying the box-counting algorithm.

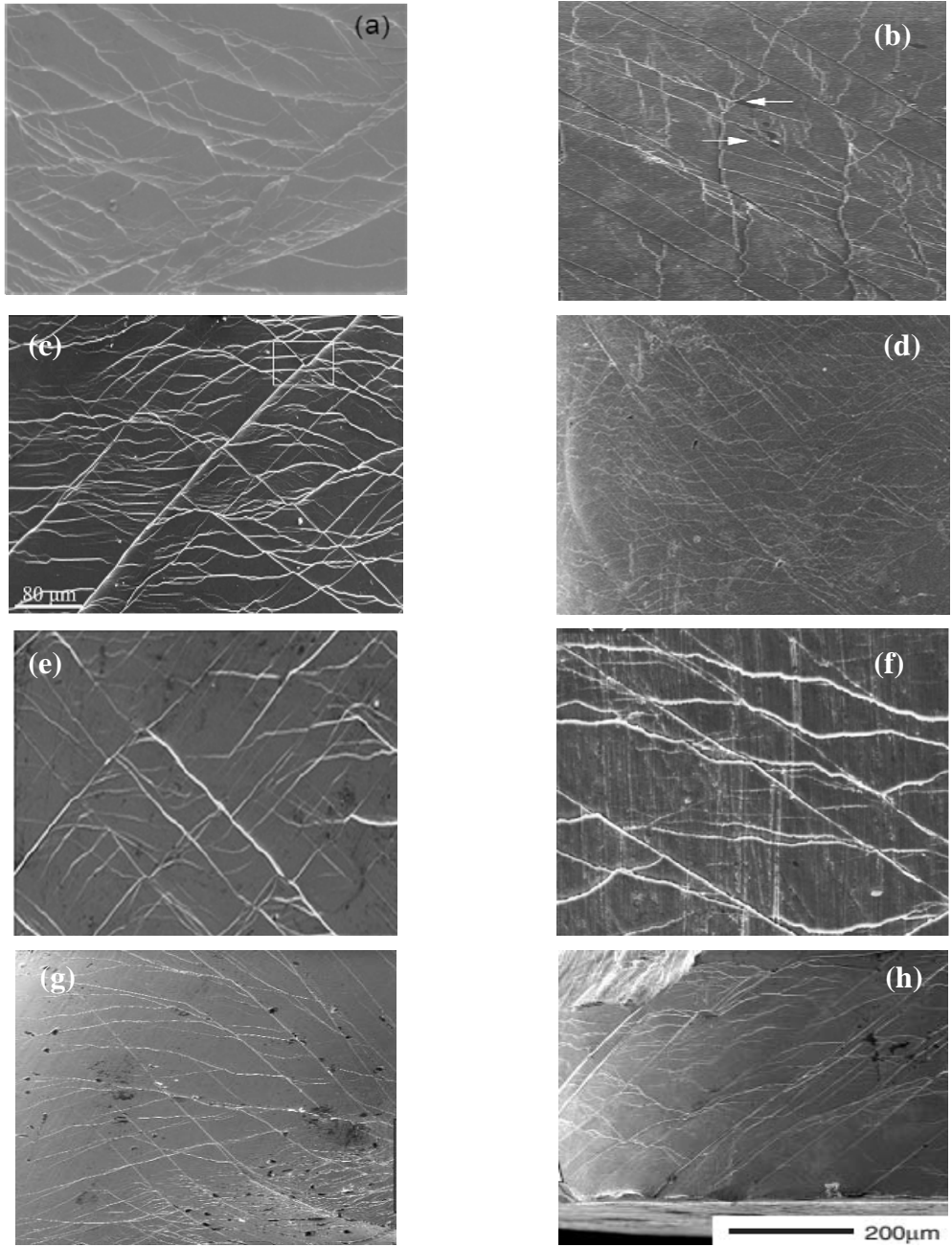

**Figure 32:** *SEM images of shear bands concerning different BMGs such as: a) $Zr_{64.13}Cu_{15.75}Ni_{10.12}Al_{10}$ (reprinted from [139] with the permission of AIP Publishing). b) $Ti_{55}Zr_{10}Cu_9Ni_8Be_{18}$ (reprinted from [146c] with the permission of Elsevier). c) $Cu_{47.5}Zr_{47.5}Al_5$ (reprinted from [146a] with the permission of Cambridge University Press). d,e) $Cu_{46.5}Zr_{47.5}Al_5Co_1$ (reprinted from [146d] with the permission of Elsever). f) $Ti_{40}Zr_{25}Ni_3Cu_{12}Be_{20}$ (reprinted from [146e] with the permission of Elsevier). g) $(Zr_{50}Cu_{50})_{95}Al_5$ BMG (d = 2 mm). h) $(Zr_{50}Cu_{50})_{95}Al_5$ BMG (d = 1.5 mm).*

The multifractality of the images was investigated using the freeware programs ImageJ [148] and its plugin FracLac [149] based on [150]. In particular, according to [150], the multifractal spectrum *f(a)* of an image can be estimated from the generalized dimension spectrum, $D_Q = \lim \dfrac{[\ln I_{Q\varepsilon} / \ln \varepsilon^{-1}]}{1-Q}$ where $I_{Q,\varepsilon} = \Sigma[P_i^Q]$ is the probability distribution which is found from the number of pixels (*M*) that were contained in each *i*th element of a size (*ε*) required to cover an object: $P_{i,\varepsilon} = M_{i,\varepsilon} / \Sigma M_{\varepsilon}$. Then, using the Legendre transformation $f(a) = Qa - (Q-1)D_Q$ we estimate the multifractal spectrum $f(a)$. Finally, from the multifactal spectrum we estimate Tsallis $q_{sens}$ index from Eq. (4.6.1).

**Table 4**

|            | (a)     | (b)     | (c)     | (d)     | (e)     | (f)     | (g)     | (h)     |
|------------|---------|---------|---------|---------|---------|---------|---------|---------|
| $D$        | 2.3364  | 2.192   | 2.3489  | 2.4253  | 2.3629  | 2.5271  | 2.2794  | 2.5695  |
| $\Lambda$  | 1.9085  | 1.3531  | 1.5819  | 1.2319  | 2.2133  | 1.5681  | 1.6579  | 1.5935  |
| $q_{sens}$ | -2.41   | -3.465  | -2.357  | -3.871  | -1.172  | -1.75   | -1.894  | -2.303  |

The results shown in Table 4 strongly suggest that the spatial evolution of multiple shear bands form a complex fractal (multifractal) network. This multifractal geometry indicates processes of far from thermodynamic equilibrium and intense interaction between shear bands. In addition, the use of lacunarity and multifractal analysis can capture differences that fractal dimension alone cannot (e.g. a, c, d). A detailed analysis of Tsallis q-statistics for both serrations and multiple shear bands is provided in a forthcoming article [144].

### 4.7 Fractional Calculus and Fractal Media

An extension of the ILG framework to consider fractional derivatives and fractal media is outlined here, as this topic may evolve into a useful subject of material mechanics research for small scale objects. We focus on the extension of gradient elasticity and gradient plasticity in this direction by elaborating on the fractional form of the extra Laplacian term appearing in these models. First we discuss the fractional counterpart of gradient elasticity and provide a brief account of related developments for fractal elastic materials. Then, we illustrate how these ideas carry on to fractional and fractal considerations for gradient plasticity.

### 4.7.1 Fractional Gradient Elasticity and Fractal Elasticity

**(i) Fractional Gradient Elasticity:** Non-standard generalizations of gradient elasticity theory to account for nonlocality, memory and fractality have recently been proposed in [9,110] by introducing derivatives of non-integer order for both 1D and 3D situations, within a variatonal formulation. For 3D spatial fractional models, the apparatus of fractional vector calculus is also used. A new fractional variational principle for Lagrangians with Riesz fractional derivatives was employed to consider static and dynamic Euler-Bernoulli and Timoshenko beam models for nonlocal fractional and fractal cases, and solutions to some typical boundary value problems were obtained.

To describe complex materials characterized by non-locality of power-law type and long-term memory, the following non-standard generalizations of the GradEla are suggested:

- *Fractional GradEla with power-law non-locality*

$$\sigma_{ij} = (\lambda \varepsilon_{kk} \delta_{ij} + 2\mu \varepsilon_{ij}) - l_s^2(\alpha)(-^R\Delta)^{\alpha/2}(\lambda \varepsilon_{kk} \delta_{ij} + 2\mu \varepsilon_{ij}), \qquad (4.7.1)$$

where $(-^R\Delta)^{\alpha/2}$ is the fractional generalization of the Laplacian in the Riesz form, and

$$\sigma_{ij} = (\lambda \varepsilon_{kk} \delta_{ij} + 2\mu \varepsilon_{ij}) - l_s^2(\alpha)^C\Delta_W^\alpha(\lambda \varepsilon_{kk} \delta_{ij} + 2\mu \varepsilon_{ij}), \qquad (4.7.2)$$

where $^C\Delta_W^\alpha$ is the fractional Laplacian in the Caputo form.

- *Fractional GradEla with power-law memory and non-locality*

$$\sigma_{ij} = (\lambda \varepsilon_{kk} \delta_{ij} + 2\mu \varepsilon_{ij}) - \left[ l_s^2(\alpha)(-^R\Delta)^{\alpha/2} + l_d^2(\beta)(^RD_t^\beta)^2 \right](\lambda \varepsilon_{kk} \delta_{ij} + 2\mu \varepsilon_{ij}), \qquad (4.7.3)$$

where $(^RD_t^\beta)^2$ is the square of the derivative of non-integer order $\beta$ with respect to time $t$. Equations (4.7.1) and (4.7.2) are fractional generalizations of the original GradEla model, while Eq. (4.7.3) is a fractional generalization of its extended version to consider internal inertia (see, for example, the review of [24]).

To shed further light on the above suggested generalization of fractional gradient elasticity, a specific 3D configuration with spherical symmetry is considered below. The model of Eq. (4.7.1) is employed due to the fact that definite results are available for the fractional Laplacian of Riesz type. The corresponding fractional GradEla governing equation is of the form

$$c_\alpha((-\Delta)^{\alpha/2}u)(r) + c_\beta((-\Delta)^{\beta/2}u)(r) = f(r) \quad (\alpha > \beta), \qquad (4.7.4)$$

where $r \in \mathbb{R}^3$ and $r = |r|$ are dimensionless, $(-\Delta)^{\alpha/2}$ is the Riesz fractional Laplacian of order $\alpha$, and the coefficients ($c_\alpha$, $c_\beta$) are material constants related to the Young's modulus and the internal length, respectively. The rest of the symbols have their usual meaning: $u$ denotes displacement and $f(r)$ body force (acceleration is reflected). For $\alpha > 0$ and suitable functions $u(r)$, the Riesz fractional derivative can be defined in terms of the Fourier transform $F$ by

$$((-\Delta)^{\alpha/2}u)(r) = F^{-1}(|k|^\alpha (Fu)(k)), \qquad (4.7.5)$$

where $k$ denotes the wave vector. If $\alpha = 4$ and $\beta = 2$, we have the well-known GradEla equation

$$c_2\Delta u(r) - c_4\Delta^2 u(r) + f(r) = 0, \qquad (4.7.6)$$

where $c_2 = E$, $c_4 = \pm l^2 E$.

Equation (4.7.4) is a fractional partial differential equation with a solution of the form

$$u(r) = \int_{\mathbb{R}^3} G_{\alpha,\beta}^3(r - r')f(r')d^3r', \qquad (4.7.7)$$

with the Green-type function $G_{\alpha,\beta}^3(r)$ given by

$$G^3_{\alpha,\beta}(\boldsymbol{r}) = \int_{\mathbb{R}^3} \frac{1}{c_\alpha \mid \boldsymbol{k} \mid^\alpha + c_\beta \mid \boldsymbol{k} \mid^\beta} e^{+i(\boldsymbol{k} \cdot \boldsymbol{r})} d^3\boldsymbol{k} = \frac{1}{(2\pi)^{3/2} \sqrt{\mid \boldsymbol{r} \mid}} \int_0^\infty \frac{\lambda^{3/2} J_{1/2}(\lambda \mid \boldsymbol{r} \mid)}{c_\alpha \lambda^\alpha + c_\beta \lambda^\beta} d\lambda, \qquad (4.7.8)$$

where $J_{1/2}(z) = \sqrt{2/(\pi z)} \sin(z)$ is the Bessel function of the first kind and the dot denotes inner product.

To proceed further, we consider Thomson's problem of an applied point load $f_0$, i.e.

$$f(\boldsymbol{r}) = f_0 \delta(\boldsymbol{r}) = f_0 \delta(x)\delta(y)\delta(z). \qquad (4.7.9)$$

Then, the displacement field $u(\boldsymbol{r})$ has a simple form given by the particular solution

$$u(\boldsymbol{r}) = f_0 G^3_{\alpha,\beta}(\boldsymbol{r}), \qquad (4.7.10)$$

with the Green's function given by Eq. (4.7.8), i.e.

$$u(\boldsymbol{r}) = \frac{1}{2\pi^2} \frac{f_0}{\mid \boldsymbol{r} \mid} \int_0^\infty \frac{\lambda \sin(\lambda \mid \boldsymbol{r} \mid)}{c_\alpha \lambda^\alpha + c_\beta \lambda^\beta} d\lambda \quad (\alpha > \beta). \qquad (4.7.11)$$

It turns out that the asymptotic form of the solution given by Eq. (4.7.11) for $0 < \beta < 2$, and $\alpha \neq 2$, reads

$$u(\boldsymbol{r}) \approx \frac{f_0 \Gamma(2-\beta)\sin(\pi\beta/2)}{2\pi^2 c_\beta} \frac{1}{\mid \boldsymbol{r} \mid^{3-\beta}} \quad (\mid \boldsymbol{r} \mid \to \infty). \qquad (4.7.12)$$

This asymptotic behavior for $\mid \boldsymbol{r} \mid \to \infty$ does not depend on the parameter $\alpha$, and (as will be seen below) the corresponding asymptotic behavior for $\mid \boldsymbol{r} \mid \to 0$ does not depend on the parameter $\beta$, where $\alpha > \beta$. It follows that the displacement field at large distances from the point of load application is determined only by the term $(-\Delta)^{\beta/2}$, where $\beta < \alpha$. This can be interpreted as a fractional non-local "deformation" counterpart of the classical elasticity result based on Hooke's law. We can also note the existence of a maximum for the quantity $u(\boldsymbol{r})\mid \boldsymbol{r} \mid$ in the case $0 < \beta < 2 < \alpha$. Indeed, these observations become clear by considering in detail the following two special cases that emerge.

*A) Sub-gradient elasticity model: $\alpha = 2$; $0 < \beta < 2$.* In this case Eq. (4.7.4) becomes

$$c_2 \Delta u(\boldsymbol{r}) - c_\beta ((-\Delta)^{\beta/2} u)(\boldsymbol{r}) + f(\boldsymbol{r}) = 0, \quad (0 < \beta < 2). \qquad (4.7.13)$$

The order of the fractional Laplacian $(-\Delta)^{\beta/2}$ is less than the order of the first term related to the usual Hooke's law. For example, one can consider the square of the Laplacian, i.e. $\beta = 1$. In general, the parameter $\beta$ defines the order of the power-law non-locality. The particular solution of Eq. (4.7.13) in the present case, reads

$$u(\boldsymbol{r}) = \frac{1}{2\pi^2} \frac{f_0}{\mid \boldsymbol{r} \mid} \int_0^\infty \frac{\lambda \sin(\lambda \mid \boldsymbol{r} \mid)}{c_2 \lambda^2 + c_\beta \lambda^\beta} d\lambda \quad (0 < \beta < 2). \qquad (4.7.14)$$

The following asymptotic behavior for Eq. (4.7.14) can be derived in the form

$$u(r) = \frac{f_0}{2\pi^2 |r|} \int_0^\infty \frac{\lambda \sin(\lambda |r|)}{c_2 \lambda^2 + c_\beta \lambda^\beta} d\lambda \approx \frac{C_0(\beta)}{|r|^{3-\beta}} + \sum_{k=1}^\infty \frac{C_k(\beta)}{|r|^{(2-\beta)(k+1)+1}}, \quad (|r| \to \infty), \qquad (4.7.15)$$

where

$$C_0(\beta) = \frac{f_0}{2\pi^2 c_\beta} \Gamma(2-\beta) \sin\left(\frac{\pi}{2}\beta\right); \quad C_k(\beta) = -\frac{f_0 c_2^k}{2\pi^2 c_\beta^{k+1}} \int_0^\infty z^{(2-\beta)(k+1)-1} \sin(z) dz. \qquad (4.7.16)$$

As a result, the displacement field generated by the force that is applied at a point in the fractional gradient elastic continuum described by the fractional Laplacian $(-\Delta)^{\beta/2}$ with $0 < \beta < 2$ is given by

$$u(r) \approx \frac{C_0(\beta)}{|r|^{3-\beta}} \qquad (0 < \beta < 2), \qquad (4.7.17)$$

for large distances $|r| \gg 1$.

*B) Super-gradient elasticity model:* $\alpha > 2$ and $\beta = 2$. In this case, Eq. (4.7.4) becomes

$$c_2 \Delta u(r) - c_\alpha((-\Delta)^{\alpha/2} u)(r) + f(r) = 0, \quad (\alpha > 2). \qquad (4.7.18)$$

The order of the fractional Laplacian $(-\Delta)^{\alpha/2}$ is greater than the order of the first term related to the usual Hooke's law. The parameter $\alpha > 2$ defines the order of the power-law non-locality of the elastic continuum. If $\alpha = 4$, Eq. (4.7.18) reduces to Eq. (4.7.6). The case $3 < \alpha < 5$ can be viewed as corresponding as closely as possible ($\alpha \approx 4$) to the usual gradient elasticity model of Eq. (4.7.6). The asymptotic behavior of the displacement field $u(|r|)$ for $|r| \to 0$ in the case of super-gradient elasticity is given by

$$u(r) \approx \begin{cases} \dfrac{f_0 \Gamma((3-\alpha)/2)}{2^\alpha \pi^2 \sqrt{\pi} c_\alpha \Gamma(\alpha/2)} \dfrac{1}{|r|^{3-\alpha}}, & (2 < \alpha < 3), \\[4mm] \dfrac{f_0}{2\pi\alpha c_\beta^{1-3/\alpha} c_\alpha^{3/\alpha} \sin(3\pi/\alpha)}, & (\alpha > 3). \end{cases} \qquad (4.7.19)$$

Note that the above asymptotic behavior does not depend on the parameter $\beta$, and that the corresponding relation of Eq. (4.7.19) does not depend on $c_\beta$. The displacement field $u(r)$ for short distances away from the point of load application is determined only by the term with $(-\Delta)^{\alpha/2}$ ($\alpha > \beta$), i.e. the fractional counterpart of the usual extra non-Hookean term of gradient elasticity. More details for the above results can be found in [9,110].

**(ii)** *Gradient Elasticity for Fractal Materials:* The elasticity of materials with fractal structure can be described by the notion of density of states [110]. As a starting point, we consider the following elasticity model for fractal materials

$$\sigma_{ij} = (\lambda \varepsilon_{kk} \delta_{ij} + 2\mu \varepsilon_{ij}) - l_F^2(D,d)\Delta^{(D,d)}(\lambda \varepsilon_{kk} \delta_{ij} + 2\mu \varepsilon_{ij}), \qquad (4.7.20)$$

where $\Delta^{(D,d)}$ is the "fractal-Laplacian" that takes into account the power-law density of states of the fractal medium under consideration, with $(D,d)$ denoting respectively volumetric and surface fractal dimensions. This constitutive equation along with the governing equilibrium equation may be devired from a variational principle, which can be used to formulate corresponding boundary value problems; for example, generalizations of the Euler-Bernoulli and Timoshenko beam equations for fractal materials. Building on the approach proposed in [110] new non-standard generalizations of gradient elasticity models of fractal materials can be obtained by using the methods of vector calculus for non-integer-dimensional spaces [151-152]. The corresponding governing equations are differential equations with integer-order derivatives that can easily be solved for typical boundary value problems; for example for problems of cylindrical and spherical symmetry, without the appearance of complexities due to fractional derivatives. In fact, as discussed in [110, 151-153], the definitions of vector operators for non-integer dimensional spaces, can be realized for two cases: $d = D-1$ and $d \neq D-1$ where, as already indicated, $D$ is the dimension of the interior of the considered region and $d$ is the dimension of its boundary. Then, the solutions obtained for fractal elasticity [110, 151-152] can be extended to derive solutions of fractional gradient elasticity by using the operator split method; i.e. the fractional counterpart [110] of the Ru-Aifantis theorem [101] earlier used for solving boundary value problems of non-fractional and non-fractal gradient elasticity [24].

We conclude this section by providing the governing equations of fractal gradient elasticity for the displacement for problems of radial symmetry for both cases $a = D-1$ and $d \neq D-1$.

*A) Fractal Gradient Elasticity for $d = D-1$*: Let us assume that the displacement vector $\boldsymbol{u}$ is everywhere radial and it is a function of $r = |\boldsymbol{r}|$ alone, i.e. $u_k = u_k(|\boldsymbol{r}|)$. The corresponding equilibrium equations (vanishing acceleration) for a continuum model with non-integer dimensional space, resulting from a fractal generalization of gradient elasticity, takes the following form for the displacement field

$$(\lambda + 2\mu)(1 \pm l_s^2(D)^V \Delta_r^D)^V \Delta_r^D \boldsymbol{u} + \boldsymbol{f} = 0. \tag{4.7.21}$$

For spherical symmetry with $d = D-1$, the vector Laplacian for the considered non-integer dimensional space reads

$$^V \Delta_r^D \boldsymbol{u}(r) = \left( \frac{\partial^2 u_r(r)}{\partial r^2} + \frac{D-1}{r} \frac{\partial u_r(r)}{\partial r} - \frac{D-1}{r^2} u_r(r) \right) \boldsymbol{e}_r, \; ; \; \boldsymbol{u}(r) = u_r(r) \boldsymbol{e}_r. \tag{4.7.22}$$

For consistency, we are also assuming that $\boldsymbol{f}(r) = f(r)\boldsymbol{e}_r$. Then, the governing differential equation for the radial displacement becomes [151-156]

$$\frac{\partial^4 u_r(r)}{\partial r^4} + \frac{2(D-1)}{r} \frac{\partial^3 u_r(r)}{\partial r^3} + \left( \frac{(D-1)(D-5)}{r^2} - l_s^{-2}(D) \right) \frac{\partial^2 u_r(r)}{\partial r^2} -$$

$$- \left( \frac{3(D-1)(D-3)}{r^3} + l_s^{-2}(D) \frac{D-1}{r} \right) \frac{\partial u_r(r)}{\partial r} + \tag{4.7.23}$$

$$+\left(\frac{3(D-1)(D-3)}{r^4}+l_s^{-2}(D)\frac{D-1}{r^2}\right)u_r(r)-(\lambda+2\mu)^{-1}l_s^{-2}(D)f(r)=0.$$

For cylindrical symmetry with $d=D-1$, the corresponding fractal gradient elasticity governing equation for the displacement, reads

$$\frac{\partial^4 u_r(r)}{\partial r^4}+\frac{2(D-2)}{r}\frac{\partial^3 u_r(r)}{\partial r^3}+\left(\frac{(D-2)(D-6)}{r^2}-l_s^{-2}(D)\right)\frac{\partial^2 u_r(r)}{\partial r^2}-$$

$$-\left(\frac{3(D-2)(D-4)}{r^3}+l_s^{-2}(D)\frac{D-2}{r}\right)\frac{\partial u_r(r)}{\partial r}+\qquad(4.7.24)$$

$$+\left(\frac{3(D-2)(D-4)}{r^4}+l_s^{-2}(D)\frac{D-2}{r^2}\right)u_r(r)-(\lambda+2\mu)^{-1}l_s^{-2}(D)f(r)=0.$$

*B) Fractal Gradient Elasticity for $d\neq D-1$*: Fractal gradient elastic materials with $d\neq D-1$ and spherical symmetry, are described by the governing equilibrium equation

$$(\lambda+2\mu)(1\pm l_s^2(D,d)\,^V\!\Delta_r^{D,d})\,^V\!\Delta_r^{D,d}\,\boldsymbol{u}+\boldsymbol{f}=0\,.\qquad(4.7.25)$$

where $^V\!\Delta_r^{D,d}\boldsymbol{u}$ is the vector Laplacian for $d\neq D-1$ for the displacement field $\boldsymbol{u}=u(r)\boldsymbol{e}_r$ that is defined by the equation [110,151-153]

$$^V\!\Delta_r^{D,d}\boldsymbol{u}=\frac{\Gamma((d+\alpha_r)/2)\Gamma(\alpha_r/2)}{\pi^{\alpha_r-1/2}\Gamma((d+1)/2)}\left(\frac{1}{r^{2\alpha_r-2}}\frac{\partial^2 u_r}{\partial r^2}+\frac{d+1-\alpha_r}{r^{2\alpha_r-1}}\frac{\partial u_r}{\partial r}-\frac{d\alpha_r}{r^{2\alpha_r}}u_r\right)\boldsymbol{e}_r,\quad(4.7.26)$$

where $\alpha_r=D-d$. The vector differential operator given by Eq. (4.7.26) allow us to describe complex fractal materials with dimension $D$ for the interior of the representative elementary volume (RVE) and dimension $d$ for its bounding surface ($d\neq D-1$). Similarly, we can consider the cylindrical symmetry case for $d\neq D-1$. Some special example solutions of the above equations will be provided in a forthcoming article [153] for illustrative purposes.

### 4.7.2 Fractional Gradient Plasticity and Fractal Plasticity

In this section, we briefly consider non-linear field equations with fractional derivatives of non-integer order to describe nonlinear elasticity or deformation theory of plasticity for fractional continua, with power-law non-locality and weak non-linearity. We shall consider a simple model of fractional gradient plasticity of the form $(\alpha>0)$

$$\sigma(\boldsymbol{x})=E\varepsilon(\boldsymbol{x})+c(\alpha)((-\Delta)^{\alpha/2}\varepsilon)(\boldsymbol{x})+\eta K(\varepsilon(\boldsymbol{x})),\qquad(4.7.27)$$

where $K(\varepsilon(\boldsymbol{x}))$ is a nonlinear function which describes the usual (homogenous) part of the flow stress; $c(\alpha)$ is an internal length scale parameter which describes weak nonlocal interactions; $E$ is the Young's modulus; and $\eta$ is a small parameter of non-linearity. Here $(-\Delta)^{\alpha/2}$ is the fractional Laplacian in the Riesz form. As a simple example, we may assume a power-law relationship for the function $K(\varepsilon)$, designating strain hardening, i.e. $K(\varepsilon)=\varepsilon^n(\boldsymbol{x})$, $(n>0)$. It is noted that for the case $n = 3$, Eq. (4.7.27) is the fractional Ginzburg-Landau equation. It should be also noted that for $\eta \equiv 0$, Eq. (4.7.27) may be viewed as a constitutive equation for one-dimensional gradient elasticity or as a constitutive equation for the flow stress of plasticity theory with linear hardening (in that case $E$ should be replaced by the hardening modulus $h$).

Next, we assume that $\varepsilon(\boldsymbol{x}) = \varepsilon_0(\boldsymbol{x})$ is a solution of Eq. (4.7.27) with $\eta = 0$, i.e. $\varepsilon_0(\boldsymbol{x})$ is a solution of the linear fractional differential equation

$$\sigma(\boldsymbol{x}) = E\varepsilon_0(\boldsymbol{x}) + c(\alpha)((-\Delta)^{\alpha/2}\varepsilon_0)(\boldsymbol{x}). \tag{4.7.28}$$

We may seek a solution of the form $\left(\eta \neq 0\right)$

$$\varepsilon(\boldsymbol{x}) = \varepsilon_0(\boldsymbol{x}) + \eta\varepsilon_1(\boldsymbol{x}) + \dots. \tag{4.7.29}$$

This means that we consider perturbations to the strain field $\varepsilon_0(\boldsymbol{x})$ of the fractional gradient elasticity, which are caused by weak plasticity effects. This allows us to use the perturbation methodology as it will be discussed in more detail in a future publication [153].

Fractional gradient plasticity can alternatively be described by Caputo fractional derivatives, as this is more convenient for reasons associated with initial and boundary conditions; i.e. they allow us to use initial and boundary conditions which are of the same form as for integer-order differential equations. For fractional derivatives of other type (for example, the Riemann-Liouville derivatives), the boundary conditions are represented by integrals and derivatives of non-integer order. Thus, for fractional nonlocal plasticity, we propose the following form of constitutive relation $(\alpha > 0)$

$$\sigma(\boldsymbol{x}) = c(\alpha)(^C\Delta_W^\alpha\varepsilon)(\boldsymbol{x}) + K[\boldsymbol{x}, \varepsilon(\boldsymbol{x})], \tag{4.7.30}$$

where the fractional Laplacian of the Caputo type was used. Example problems based on the above constitutive equation will be discussed in [153], along with their physical implications.

In concluding this section, it is pointed out that a similar formulation as the one discussed earlier for fractal elasticity, can also be adopted for fractal plasticity. This will result to a new plasticity framework for fractal materials, based on non-integer dimensional spaces. Typical examples will be provided in a future publication.

## 5. Concluding Remarks

As this chapter was essentially completed and first submitted in June 2015, there are several articles that came to our attention since then and more progress was made on the topic of gradient mechanics by our Lab collaborators and other authors. While it is impossible to expand in detail on all these developments, it is only fair to provide a brief account on various issues that have not been elaborated or touched upon here.

## 5.1  Generalized Continuum Mechanics Aspects

Even though not explicitly stated by a number of prominent continuum mechanics authors in the recent literature, the revival of interest on second deformation gradient materials (i.e. the early contributions of Mindlin, Toupin, Rivlin, Eringen, etc. in 1960's) was directly motivated by the author's simple gradient elasticity and plasticity models (in mid 1980's and early 1990's). These simple but robust non-classical elasticity, plasticity and damage models have clearly shown the ability of gradients to stabilize the behavior in the material softening regime, to predict widths and spacings of shear bands, to interpret size effects, and eliminate unphysical singularities in dislocation lines and crack tips. The same was true for dislocation evolution gradient models proposed by the author and his co-workers in the early 1980's, such as the Walgraef-Aifantis (W-A) model for the development of persistent slip bands in fatigued crystals and dislocation patterning, as already pointed out in previous sections [6a-d].

In fact, it may not be an exaggeration to claim that the simple Laplacian-based models for gradient elasticity and plasticity have resulted to a plethora of very interesting publications worldwide for nanoscale materials and nanotechnology components. The same holds for the W-A model which, despite early criticisms, has resulted to an overwhelming activity of discrete dislocation dynamics simulations initially, followed up by refined continuum dislocation models which lately have also resorted to gradient considerations for dislocation patterning interpretations. On the strain gradient front, the reader may consult recent works [154] where issues of nonlinear strain measures and symmetry properties along with variational principles are considered. On the dislocation gradient or gradient defect kinetics front, generalized gradient-dependent continuum dislocation theories for pattern formation can be found in [155].

A most recent excellent overview on the application of gradient theory to model a large number of mechanical characteristics of nanoscopic materials and objects can be found in [156] where an extensive list of references on the topic is also provided. It is interesting to note that in this overview – with a variety of new results and nanotechnology applications – both stress gradients and strain gradients in the form of Laplacians are used according to the author's earlier suggestion [6a].

## 5.2  Extensions Beyond Nanotechnology

It recently came to our attention that stress gradients in the form of Laplacians have also been used by the rheology community to model shear banding phenomena in complex fluids [157]. An overwhelming number of works exist in this topic based on the introduction of a diffusive-like stress term in the standard Johnson-Segalman model. It is remarkable to note, however, that while the ideas of non-monotonous stress vs. strain (or strain rate) curves and the introduction of Laplacian terms in the rheology literature are exactly the same as in the solid mechanics literature, as well as the corresponding treatment of localized deformation zones in metals [38a-b], no cross-references exist. In view of this, it is recommended that a closer interaction and knowledge transfer between the rheology and solid mechanics communities is needed in this area.

A similar situation exists for fluid turbulence models recently elaborated upon in [158] which are based on an extension of the Navier-Stokes equations to include a Laplacian of the symmetric part of the velocity gradient tensor. This generalization is precisely of the same form as the simple Laplacian modification of Hooke's law for elasticity and the Mises flow stress expression for plasticity, as suggested earlier by the author [32a, 38a-b].

In contrast to the aforementioned cases in the complex fluid and turbulence literature, recent work in geology and earth-system science [159] employs the formal structure of the W-A model [6b] to interpret crack patterns in our planet. Two families of cracks (aged and newborn) are introduced, the densities of which obey the evolution equations of the W-A model. A physical justification for the analogy between crack and dislocation families is given supplemented by discrete element simulations. Further work in this area seems to be quite promising, as concepts from multiscale nonequilibrium thermodynamics have not been sufficiently utilized for modeling pattern formation at earth scales.

### 5.3  Extensions to Biomedicine

The initial gradient models of the author for deformation and defect kinetics were developed in analogy to existing models in biology and population dynamics. The Laplacians of strain introduced for modeling elastic and plastic deformation, as well as the Laplacians of dislocation densities introduced for modeling pattern formation, strongly remind the formalism of reaction-diffusion (R-D) systems. A basic difference, however, lies on the fact that the intrinsic lengths or gradient parameters in front of the Laplacian terms are not simply scale-independent material coefficients (like the diffusion or reaction constants) but they depend on the topological configuration and size of the elementary volume at hand. In general, the internal lengths (as well as the internal times in respective delay differential equation/DDE formulations) may depend on strain or stress applied at the macroscale or experienced at the microscale. While in standard treatments macroscopic and microscopic strains are related through sophisticated averaging or homogenization procedures, a more flexible approach may be to treat these quantities as independent variables following separate evolution equations. This is the case, in particular, for living systems (tissues, cells) where internal (micro) strains and stresses arise; for example, during the crawling of cells and their interaction with the extracellular matrix (ECM).

Strain and strain gradient effects in the form of Laplacians used by the author for technological materials [38a-b] were also used by Murray and co-workers [71] – his treatise on mathematical biology, Chapter 6 of Vol. II on the mechanical theory for generating pattern and form development. Interestingly, similar to the case of complex fluids, there has not been so far cross-referencing between the articles on pattern formation between manmade technological and naturemade living materials. The connection between the two approaches needs to be addressed in the future.

The situation is similar to some recent models proposed for cancer. These "Go or Grow" cancer models for motile and immotile cells [160] are similar in form with the W-A model (motile and immotile cells correspond to immobile and mobile dislocations, respectively), but internal strain and strain gradient effects have not been included. This topic is currently considered in [161].

Other direct analogies between models for technological nanostructured materials and living micro/nanoscopic systems exist in the area of neural transmission. An additional current term due to flexoelectric effects may be added to reduced Fitzhugh-Nagumo equations enriched with a time delay term associated, for example, with the influence of alcohol or other drugs on neurons. Some preliminary results in this direction have been obtained [162]. Such analogies are more transparent for deformation wave propagation along a microtubule. In fact, it can readily be shown that the governing equation for signal propagation across a microtubule [163] for the displacement is similar to the governing equation for strain propagation along a one-dimensional nanostructured solid or polymeric fiber [164]. Soliton-like kink and periodic solutions are possible where travelling wave speed and amplitude are interrelated.

Linearized versions of such wave propagation theories involving both internal lengths and internal times have also been used to model vibration and dispersion properties of nanotubes [165], and more recently of metamaterials [166].

In view of all the above developments, it follows that the field of internal length-internal time (IL-IT) gradient mechanics is still an area for multidisciplinary research activity with important applications to new technology, environment, biology and medicine. As a notable example we refer to "nanofracture" of materials ranging from metals/polymers/ceramics and their composites to nanocrystalline/nanoglass and amorphous solids, and from bone and tissue materials to cell membranes and mitochondria. As a first step, classical fracture mechanics should be revisited by adopting IL-IT considerations. Higher-order strain/stress gradients including $\nabla^2$ and $\nabla^4$ operators can conveniently be introduced to eliminate singularities that cannot be treated or be dispensed with for small nanoscale volumes. Stochastic effects are also important to include and their interplay with the deterministic gradients has to be evaluated by also resorting to solutions of stochastic differential equations. Some work along these directions, i.e. $\nabla^4$ - fracture mechanics and stochastic differential equations for strain have been presented in [167] and related publications are in preparation. In this connection, it is pointed out that some very interesting recent work on nanofracture mechanics similar to the author's formulation has been reported in [168].

## Acknowledgments

This chapter is an edited version of a final report prepared last year for the projects *Hellenic ERC-13* and ARISTEIA II funded by the General Secretariat of Research and Technology (GSRT) of Greece. Discussions and contributions from my former students A. Konstantinidis and I. Tsagrakis, as well as my recent collaborators Y. Chen, Y. Yue, A. Iliopoulos, C. Bagni, M. Mousavi and V. Tarasov are acknowledged. The particular examples fo the ILG approach treated here were benchmark problems addressed within the aforementioned pojects. Further elaboration based on these initial results will follow in related future journal publications. I am especially indebted to my daughter Katerina Aifantis for stimulating my interest to Li-ion batteries (as well as to neuromechanics), as well as to Stephane Bordas for inviting this contribution and Daniel Balint for his useful editing comments.